\documentclass[10pt,letterpaper]{article}

\usepackage[margin=0.74in]{geometry}
\usepackage[T1]{fontenc}
\usepackage{amsmath}
\usepackage{amssymb}
\usepackage{amsthm}
\usepackage{booktabs}
\usepackage{multirow}
\usepackage{array}
\usepackage{tabularx}
\usepackage{longtable}
\usepackage{pdflscape}
\usepackage{makecell}
\usepackage{graphicx}
\usepackage[font=small,labelfont=bf]{caption}
\usepackage{subcaption}
\usepackage{algorithm}
\usepackage{algpseudocode}
\usepackage{listings}
\usepackage{xcolor}
\usepackage[numbers,sort&compress]{natbib}
\usepackage{microtype}
\usepackage{placeins}
\usepackage[colorlinks=true, linkcolor=blue, citecolor=blue, urlcolor=blue]{hyperref}
\hypersetup{
  pdftitle={Knowledge Boundary Probing and Demand-Guided Intervention for LLM-Based Power System Code Generation},
  pdfauthor={Hui Wu, Xiaoyang Wang, Zhong Fan},
  pdfkeywords={large language models; power-system code generation; API knowledge probing; retrieval-augmented generation; on-premise AI deployment}
}

\newcommand{\mM}{\mathcal{M}}
\newcommand{\mF}{\mathcal{F}}
\newcommand{\mD}{\mathcal{D}}

\newcommand{\pp}{\texttt{pandapower}}

\title{Knowledge Boundary Probing and Demand-Guided Intervention\\
for LLM-Based Power System Code Generation}

\author{
Hui Wu\thanks{Corresponding authors: Hui Wu
(\texttt{hw865@exeter.ac.uk}) and Zhong Fan
(\texttt{z.fan@exeter.ac.uk}).}\\
Department of Engineering, University of Exeter\\
\texttt{hw865@exeter.ac.uk}
\and
Xiaoyang Wang\\
Department of Computer Science, University of Exeter\\
\texttt{x.wang7@exeter.ac.uk}
\and
Zhong Fan\\
Department of Engineering, University of Exeter\\
\texttt{z.fan@exeter.ac.uk}
}

\date{Author version, October 2026}

\begin{document}
\maketitle
\begin{center}
\small Published in \emph{Advanced Engineering Informatics} 77 (2027), 105328.\\
Version of record: \url{https://doi.org/10.1016/j.aei.2026.105328}
\end{center}

\begin{abstract}
Large language models (LLMs) can turn grid-analysis requests into
executable programs for power-system simulation, but utilities and
research laboratories often require on-premise deployment.  In this setting,
first-pass failures frequently arise at an \emph{API-knowledge
boundary}, through hallucinated functions, misused parameters, and
mishandled result tables.

We present \emph{PowerCodeBench}, a parameterised benchmark generator
released as a frozen $2{,}000$-task suite for \pp{}, and a
deployment-time workflow that requires no weight updates.
Documentation-driven L0--L3 probes produce per-model API profiles for
diagnosis, model comparison, documentation allocation, and backend
calibration.  A query-side demand estimator selects layered API
evidence before generation, while execution feedback routes targeted
repair.

Across ten open-weight LLMs ($1.5$B--$480$B) and four mid-tier APIs,
the validation-enabled workflow raises scalar-match accuracy by
$32$--$56$ percentage points after up to three repair rounds relative
to an unassisted first pass, for every model of at least $7$B and
every API.  Open-weight models in the $70$B--$120$B range reach the four-vendor mid-tier
accuracy range under matched no-tool conditions.  Selective injection
approaches the full-layer reference using $41\%$ of its prompt tokens.
Among model--item pairs passing numerical checks under both workflows,
engineering review confirms the requested analysis in $88\%$ of
full-workflow outputs versus $66\%$ under plain repair.  Round-0 pilots
on OpenDSS and PyPSA motivate staged onboarding from broad retrieval at
cold start to calibrated selective injection.  Measured throughput,
latency, energy, and allocated GPU memory establish a practical
on-premise serving envelope.
\end{abstract}

\noindent\textbf{Keywords:}
Large language models;
power-system code generation;
power-system simulation;
API knowledge probing;
retrieval-augmented generation;
on-premise AI deployment.

\section{Introduction}
\label{sec:intro}

The transition to high-penetration renewable and integrated energy
systems is increasing the volume, diversity, and complexity of
analyses that grid operators and energy-research labs must perform:
contingency screening under stochastic renewable injection, optimal power-flow (OPF)
coordination, time-series simulation across variable distributed
resources, post-fault state reconstruction~\cite{donti2021mlenergy,
heymann2024ai40years}.  Large language models (LLMs) are therefore
emerging as an engineering-informatics interface for modern energy
systems~\cite{huang2024foundationpower,majumder2024electricenergy}:
they translate operational questions phrased in natural language into
executable programs over domain simulation libraries such as
\pp{}~\cite{thurner2018pandapower}, OpenDSS~\cite{dugan2011opendss},
and PyPSA~\cite{brown2018pypsa}.  This places LLMs between
engineers, domain software APIs, and executable simulation workflows,
and raises a deployment question that is specific to energy
engineering practice: which model, at what scale, should be trusted
for which grid-analysis workload, and where can it be served?

Energy-sector practitioners (utilities, system operators, research
labs) typically require on-premise serving for structural reasons:
data confidentiality, regulatory compliance, reproducibility, and
operating cost~\cite{henao2025concerns,cheng2026secureenergy}.
Hosted closed-source APIs typically do not meet these constraints,
since data leaves the operator's control.
Locally deployable open-weight LLMs can, but their useful range is not
a single ``small-model'' category.  Whether one open-weight tier is universally viable is therefore not
the practical question; the practical question is how deployment-time
assistance can map each model tier to the grid-analysis workloads it
can reliably support, without the recurring cost of fine-tuning or of
multi-agent orchestration around a stronger model.

First-pass failures of LLM assistance on power-system code generation
are dominated not by reasoning limits alone but by \emph{API-driven
non-execution}: hallucinated function names, misused parameters,
mishandled result tables, and broken multi-step workflows in a
stateful, version-evolving
library~\cite{kuhar2025libevolutioneval,zhuo2025apimisuse,
chen2025apihallucination,patil2023gorilla}.  The same broad failure
mode appears across our panel ($1.5$B lower-bound reference,
$7$B--$480$B open-weight cohort, four mid-tier closed-source APIs);
what differs across models is the \emph{location} of each model's
API-knowledge boundary, not its existence.  We formalise this as a layered
\emph{knowledge boundary} (recognition, recall, comprehension,
application), loosely inspired by revised Bloom-style cognitive
levels~\cite{anderson2001taxonomy} and adapted to executable API use
(Sec.~\ref{sec:probing}).

Existing method families each supply part of what this deployment
scenario needs (from execution-scored code benchmarks to
retrieval-augmented generation, API-retrieval tooling, and
self-debugging loops).  Within these closest comparison classes, we
found no approach that combines a persistent model-side knowledge
profile with query-side demand estimation and layered,
profile-conditioned documentation delivery
(Sec.~\ref{sec:related:gaps}).

This setting raises three research questions.  \emph{RQ1:} at which
cognitive layer (recognition, recall, comprehension, application)
does each model's knowledge of a stateful simulation API break down?
\emph{RQ2:} can model-side weakness and query-side API demand be
reconciled at deployment time, so that documentation is injected and
repair routed only where the serving model needs it?  \emph{RQ3:}
with such assistance, which locally deployable open-weight tiers
reach hosted mid-tier-API accuracy, and on which grid-analysis
workloads?

\begin{figure}[!htbp]
  \centering
  \includegraphics[width=\linewidth]{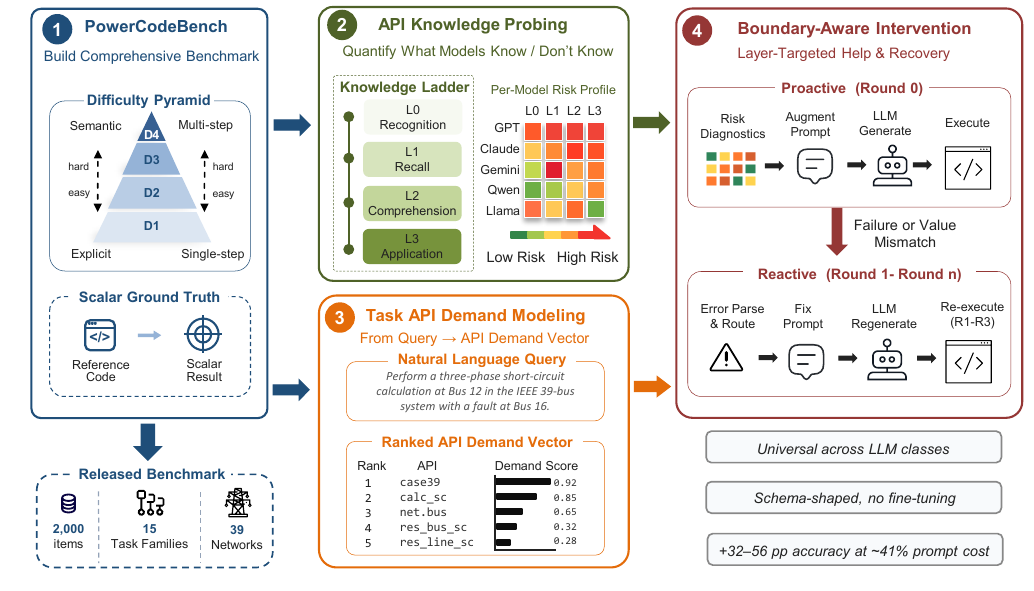}
  \caption{Reliability workflow for LLM-based power-system code
    generation: a natural-language grid-analysis request becomes an
    LLM-generated program, executed on the target grid case to yield
    a scalar engineering output.  PowerCodeBench supplies
    execution-validated tasks; L0--L3 probing converts structured API
    documentation into per-model knowledge profiles; demand modelling
    estimates query-side API requirements; the boundary-aware
    intervention combines per-model risk and per-query demand for
    proactive injection and reactive repair.}
  \label{fig:system_overview}
\end{figure}

We answer these questions with the deployment-time reliability
workflow in Fig.~\ref{fig:system_overview}.  Its evaluation instrument,
\emph{PowerCodeBench}, pairs natural-language grid-analysis requests
with reference \pp{} programs and scalar engineering ground truth; the
frozen release contains $2{,}000$ execution-validated tasks from 15
power-system families and 39 grid networks.  A documentation-driven
L0--L3 probe generator (four graded documentation layers) then
converts the API specification into a
per-model risk profile $\rho_\mM(f)$, a probe-measured risk score
per API function, locating API-knowledge failures
and providing a diagnostic signal for model selection and
model-specific documentation allocation.

At inference time, a query-side demand predictor identifies the API
functions implicated by the request.  The intervention combines this
demand with the model profile and library-derived intent, boundary,
and output contracts to select the documentation supplied before
generation and to route execution failures afterwards.  It requires
only structured documentation and chat access to the model, changes no
weights, and is instantiated here on \pp{} with Round-0 pilots of the
probing and proactive-allocation path on OpenDSS and PyPSA.  Unlike
fine-tuning against an evolving
library~\cite{cheng2025advanceddispatch} or multi-agent orchestration
around a stronger model~\cite{jin2025gridmind,zhang2025poweragent}, it
adds reliability at deployment time.

To the grid operator, this machinery is invisible: the interaction
remains a conventional analysis request, not a prompt-engineering
task.  An engineer states, say, an $N\!-\!1$ contingency screen;
behind the request, the workflow selects the documentation that
screen implicates, executes the generated program on the target grid
case, gates the output for validity, and routes any failure to
targeted repair.  Persistent failures
are returned for engineering review rather than silently retried, so
the operator neither tunes the model nor inspects its internal
knowledge profile.

Across ten open-weight LLMs ($1.5$B--$480$B) and four commercial
mid-tier APIs on the full benchmark, the validation-enabled workflow
raises accuracy after up to three repair rounds by $32$--$56$
percentage points relative to an unassisted first pass for every
evaluated $\geq\!7$B model and every API.
The $70$B--$120$B open-weight class enters the mid-tier API accuracy
range, while the largest open-weight configurations reach the top of
the matched no-tool panel and remain competitive when stronger API
reasoning modes are enabled.  The validity audit further shows that
the gain reflects more faithful engineering workflows, not merely more
scalar matches.  Together with measured throughput, latency, energy,
and allocated GPU memory, these results support a workload-aware policy: select the
smallest local tier that covers the target analyses, then use the
complete selective pipeline to approach the full-layer reference
accuracy without fine-tuning or cloud-resident inference.

\paragraph{Contributions}
The resulting contributions are as follows.
(1)~PowerCodeBench, a parameterised, execution-validated
natural-language-to-code benchmark generator for power-system
simulation.
(2)~An L0--L3 documentation-driven probe generator that turns an API
specification into per-model profiles for library-specific diagnosis,
model comparison, documentation allocation, and calibration of
evidence allocation for each backend.
(3)~A complete selective proactive pipeline (query-side demand
estimation, layered documentation artifacts, profile-conditioned
function--layer allocation, and budgeted packing) that improves
first-pass generation while controlling prompt length.
(4)~Reactive repair that routes runtime failures to targeted fixes
and, when a validation signal is available, also handles code that
runs but returns the wrong value.
(5)~The systems finding that first-pass failures in this setting
concentrate at the versioned API boundary before successful execution, corroborated
by an engineering-validity audit of the evaluation protocol itself.
\section{Related Work}
\label{sec:related}

Table~\ref{tab:related_positioning} positions this work against the
six method families closest to it, along the six capabilities that
our framework combines.
The following review traces these families and then distils the gaps
addressed by the combined workflow.
\begin{table}[!htbp]
\centering
\footnotesize
\setlength{\tabcolsep}{4pt}
\caption{Positioning of this work against the closest method
  families.  Capabilities: execution-based scoring of generated
  programs; stateful domain simulator with engineering ground truth;
  model-side knowledge profiling; query-side API-demand modelling;
  layered documentation injection (L0--L3); routed reactive repair.
  \checkmark{} = central to the family; $(\checkmark)$ = partial (see
  notes); $\times$ = absent.  Each row cites representative systems.}
\label{tab:related_positioning}
\resizebox{\textwidth}{!}{%
\begin{tabular}{@{}p{4.2cm}cccccc@{}}
\toprule
\textbf{Method family} &
\makecell{\textbf{Execution}\\\textbf{scoring}} &
\makecell{\textbf{Stateful}\\\textbf{simulator}} &
\makecell{\textbf{Model-side}\\\textbf{profile}} &
\makecell{\textbf{Query-side}\\\textbf{demand}} &
\makecell{\textbf{Layered}\\\textbf{injection}} &
\makecell{\textbf{Routed}\\\textbf{repair}} \\
\midrule
General code benchmarks~\cite{chen2021evaluating,austin2021programsynthesis,hendrycks2021apps} &
\checkmark & $\times$ & $\times$ & $\times$ & $\times$ & $\times$ \\
\addlinespace[2pt]
Scientific / library-grounded benchmarks~\cite{lai2023ds1000,tian2024scicode,zhuo2025bigcodebench} &
\checkmark & $(\checkmark)^{a}$ & $\times$ & $\times$ & $\times$ & $\times$ \\
\addlinespace[2pt]
RAG-augmented code generation~\cite{lewis2020rag,zhang2023repocoder,jeong2024adaptiverag} &
$(\checkmark)^{b}$ & $\times$ & $\times$ & $(\checkmark)^{c}$ & $\times$ & $\times$ \\
\addlinespace[2pt]
Uncertainty-gated adaptive retrieval~\cite{jiang2023flare,asai2024selfrag} &
$\times$ & $\times$ & $(\checkmark)^{g}$ & $(\checkmark)^{c}$ & $\times$ & $\times$ \\
\addlinespace[2pt]
API-retrieval and tool-use methods~\cite{patil2023gorilla,qin2023toolllm,li2023apibank} &
$(\checkmark)^{d}$ & $\times$ & $\times$ & $(\checkmark)^{c}$ & $\times$ & $\times$ \\
\addlinespace[2pt]
Self-debugging / execution-feedback methods~\cite{chen2023selfdebug,madaan2023selfrefine,jia2024feedbacksim} &
\checkmark & $(\checkmark)^{e}$ & $\times$ & $\times$ & $\times$ & $(\checkmark)^{f}$ \\
\addlinespace[2pt]
\textbf{This work} &
\checkmark & \checkmark & \checkmark & \checkmark & \checkmark & \checkmark \\
\bottomrule
\end{tabular}%
}

\smallskip
\begin{minipage}{0.97\linewidth}
\footnotesize
$^{a}$~Real scientific libraries, but predominantly stateless
single-shot tasks without persistent simulator state or engineering
ground truth.
$^{b}$~Execution scoring appears in code-completion variants only.
$^{c}$~Retrieval is keyed on surface similarity to the query, not on
an estimate of which API functions the request requires.
$^{d}$~Scored by call-level correctness (API name and
abstract-syntax-tree match), without end-to-end program execution.
$^{e}$~Power-system simulators appear in agent case
studies~\cite{jia2024feedbacksim} rather than as a controlled
benchmark substrate.
$^{f}$~Uniform re-prompting with raw error text; no error-type
routing and no output-table contracts.
$^{g}$~Retrieval is gated by transient generation-time confidence
signals; no persistent, per-function, layer-resolved profile is
measured or retained.
\end{minipage}
\end{table}

\subsection{Execution-based and library-grounded code benchmarks}
\label{sec:related:codegen}

Execution-based code-generation benchmarks judge model outputs by
running generated programs: HumanEval/MBPP/APPS for function synthesis
and SWE-bench for repository-level
maintenance~\cite{chen2021evaluating,austin2021programsynthesis,
hendrycks2021apps,jimenez2024swebench}; scientific and library-grounded
variants (DS-1000, SciCode, BigCodeBench, LibEvolutionEval) move
closer to our setting by requiring real-library use, scientific
computation, or version-specific code
generation~\cite{lai2023ds1000,tian2024scicode,zhuo2025bigcodebench,
kuhar2025libevolutioneval}.  PowerCodeBench
(Sec.~\ref{sec:benchmark}) inherits this execution-grounded
lineage; the capabilities it adds beyond it, including the paired
probe suite, are summarised in the positioning table.

\subsection{LLMs for power systems and energy AI}
\label{sec:related:power}

Existing LLM work in power systems explores complementary directions:
dispatch reasoning (ElecBench), fine-tuned dispatch (GAIA),
natural-language grid visualisation (ChatGrid), and agent-orchestration
of solver workflows (GridMind, PowerAgent)~\cite{zhou2024elecbench,
cheng2025advanceddispatch,jin2024chatgrid,jin2025gridmind,
zhang2025poweragent}.  The closest neighbours ask LLMs to use
power-system simulation libraries directly: ChatGPT for OpenDSS, the
Daline toolbox, and a feedback-driven multi-agent framework spanning
Daline and MATPOWER~\cite{bonadia2023opendss,jia2024daline,
jia2024feedbacksim}.  These show that domain-library use is a real
bottleneck, but they remain case studies or tool/agent frameworks and stop
short of systematic executable code-generation evaluation paired with
API-level diagnosis.

LLM assistance also enters a field with a mature
computational-intelligence reliability stack.  Its applications
include uncertainty-aware fault diagnosis, statistical and
entropy-based~\cite{abbasi2025statistical,abbasi2025winding} or
wavelet--CNN-based methods in microgrids~\cite{arevalo2025microgrid}, and
vision-based substation inspection~\cite{jiang2025yoloiase}.  The
diagnostic methods consume simulation and measurement outputs of the
kind the benchmark's tasks compute.
Simulation-backed optimisation likewise covers dynamic-security
assessment~\cite{li2026dsa}, scenario-decoupled $N\!-\!1$-secure cost
control~\cite{wang2026scenario}, and renewable- and EV-aware economic
dispatch~\cite{zadehbagheri2023energy}.  Beyond energy systems,
generative circuit synthesis illustrates the wider move towards
engineering-design automation~\cite{zhang2026siwfilter}.  Together,
these studies illustrate the downstream engineering context; none
addresses the reliability of LLM-generated simulation code itself.

On the serving side, a systems literature is making inference on
resource-constrained local hardware increasingly practical, from
collaborative inference across heterogeneous edge
LLMs~\cite{jin2026moe2} to memory- and scheduling-level optimisation
of integrated edge devices~\cite{zhang2025breakingedge}.  This line
strengthens the paper's on-premise premise:
serving open-weight models locally is increasingly practical.  Our
question is complementary: not how to serve a given
model efficiently, but which model tier suffices for which
grid-analysis workload and how far deployment-time intervention can
raise it.

The broader energy-AI literature positions foundation models as a
layer in the intelligent-energy-system
stack~\cite{antonesi2025transformersenergy,zhang2026llmenergy};
within that framing our contribution is a concrete
reliability-and-evaluation building block.

\subsection{API knowledge, tool use, and retrieval-augmented generation}
\label{sec:related:api}

Tool-use research treats API invocation as a distinct capability
(Toolformer, ReAct, API-Bank, Gorilla, ToolLLM,
ToolACE)~\cite{schick2023toolformer,yao2023react,li2023apibank,
patil2023gorilla,qin2023toolllm,liu2025toolace}, and code-specific
analyses establish hallucinated APIs and parameter misuse as
recurrent LLM failure modes~\cite{zhuo2025apimisuse,
chen2025apihallucination,liu2024codeupdatearena}, motivating
inference-time mechanisms over weight edits.  Retrieval-augmented
generation (RAG), which prepends retrieved text to the prompt, provides
the natural comparison class for proactive injection (RepoCoder,
Adaptive-RAG)~\cite{lewis2020rag,zhang2023repocoder,
jeong2024adaptiverag} and execution-feedback methods for reactive
correction (self-debug, self-refine)~\cite{chen2023selfdebug,
madaan2023selfrefine}.  Related model-internal methods organise
linguistic inputs at multiple granularities in neural
translation~\cite{lv2024dmgts} or suppress irrelevant multimodal
evidence through knowledge-guided modulation~\cite{shi2024ktmn}; our
method instead selects explicit, versioned API documentation at
deployment time.  Uncertainty-gated adaptive retrieval (FLARE,
Self-RAG) also conditions on the model side, but from transient
decode-time confidence or learned reflection
tokens~\cite{jiang2023flare,asai2024selfrag}; the contrast with our
persistent, offline-measured profile recurs as the first gap below.  Our
intervention sits in both code-oriented families but
conditions its evidence differently; the next subsection makes the
differences explicit.

\subsection{Limitations of existing studies}
\label{sec:related:gaps}

Three gaps recur across the method families compared in
Table~\ref{tab:related_positioning}.  First, benchmarks score models,
retrieval systems condition on the query, and uncertainty-gated
methods react to transient decode-time confidence; the compared
systems do not retain a per-function, layer-resolved profile as
inspectable allocation state.  Second, API-retrieval methods estimate
query-side relevance, but the compared systems do not reconcile those
estimates with persistent, layer-resolved measurements of the serving
model.  Third, evidence delivery and post-execution correction are not
jointly \emph{layered and profile-conditioned}: standard retrieval
provides flat context rather than selecting among signature, semantic,
and example layers, while post-execution repair generally re-prompts
without routing by failure type.  The L0--L3 probe profiles of
Sec.~\ref{sec:probing} address the first gap, the demand estimator of
Sec.~\ref{sec:demand} the second, and layered, profile-conditioned
injection with routed repair
(Secs.~\ref{sec:interv:proactive}--\ref{sec:interv:reactive}) the
third.
\section{PowerCodeBench: Task, Protocol, and Benchmark}
\label{sec:benchmark}

PowerCodeBench is not a fixed hand-written dataset but a configurable
benchmark generator: it produces (query, reference code, scalar
ground-truth) triples through a parameterised pipeline over grid
networks, task families, modification operators, and natural-language
templates.  The 2{,}000-item suite used throughout this paper is one
frozen release of that generator, retained for reproducibility;
because every ground truth is derived from reference-code execution
instead of manual annotation, both the configuration pool and the
sampling budget can be extended (and the suite regenerated
end-to-end when \pp{} is upgraded) without changing the rest of the
evaluation pipeline.  Below we define the task and scoring contract,
the difficulty axes the generator controls, the coverage of its
current configuration, and the reference execution that produces
ground truth.

\subsection{Task and Scoring Contract}
\label{sec:problem}

We consider the task of \emph{scientific code generation for power system
analysis}.  Given a natural-language query $q$ and a target library
(here, \pp{}), a large language model $\mM$ generates Python
code $C = \mM(q)$.  Unlike short, stateless function synthesis, each
task requires constructing a multi-step stateful computation over a
network object (network loading, parameter modification, simulation
execution, and result extraction, all mediated by a specific,
versioned API).  This places the benchmark closer to realistic
library-use settings than to isolated algorithmic function synthesis,
while adding a power-system simulation layer absent from existing
scientific-code suites~\cite{chen2021evaluating,
austin2021programsynthesis,hendrycks2021apps,jimenez2024swebench,
lai2023ds1000,tian2024scicode,zhuo2025bigcodebench}.
The generated code is executed in a controlled
environment and evaluated against a scalar reference output $g^*$ produced
by verified reference code.  A generated answer is counted as successful only
when two conditions hold:
\begin{enumerate}
  \item \textbf{Executability}: $C$ runs to completion with no runtime
    exception (\texttt{AttributeError}, \texttt{ImportError},
    \texttt{TypeError}, \mbox{\texttt{SyntaxError}}, etc.); and
  \item \textbf{Scalar correctness}: the result $r(C)$ extracted from $C$
    matches $g^*$ under the benchmark's type-specific matching rule: exact
    match for integer and Boolean outputs, and mixed absolute/relative
    tolerance for floating-point outputs (concrete thresholds are specified
    with the reference-validation protocol in
    Sec.~\ref{sec:bench:gt}).
\end{enumerate}

We report accuracy at two endpoints: \textbf{R0 accuracy}
(\textbf{Pass@1}, the fraction of tasks solved by the first generation,
before reactive repair) and \textbf{accuracy after $K$ fix rounds}
(the cumulative
fraction of tasks solved on the first attempt or after any of up to
$K$ successive error-feedback fix rounds).  We set $K=3$ throughout
this paper and refer to this endpoint as R3 below; the protocol
itself is parametric in $K$.  Both endpoints are primary metrics:
R0 isolates proactive injection at generation time, R3 captures the
full reactive correction loop.  Alongside accuracy we report
\textbf{prompt-token cost} (per item and per successful answer),
because the method is designed to trade off accuracy and token
consumption jointly.  This protocol follows the
execution-based evaluation tradition in code-generation benchmarks and the
later use of execution feedback for self-debugging and iterative repair
\cite{chen2021evaluating,hendrycks2021apps,chen2023selfdebug,
madaan2023selfrefine}.

We distinguish this \emph{model-scoring} protocol from the
\emph{benchmark-construction} protocol.  During construction, the
reference code for each item must pass the domain validity checks of
the hard admission filter (described below) before its
scalar output is accepted as $g^*$; these checks
keep the benchmark instances physically meaningful at construction time.
They are not applied as additional filters on free-form model
outputs, which are scored only by execution and scalar-result matching,
keeping scoring objective and reproducible.
This separation between reference-solution validation and model-output scoring
is consistent with executable scientific and library-grounded code benchmarks,
where ground truth is obtained by running trusted reference programs
\cite{lai2023ds1000,tian2024scicode,zhuo2025bigcodebench}.

\subsection{Difficulty Taxonomy}
\label{sec:bench:taxonomy}

Benchmark items are not sampled randomly: the generator imposes
systematic control over two pressure sources: \emph{workflow
complexity} (single-step analysis versus multi-step or conditional
execution) and \emph{semantic grounding} (whether the modification
step states parameters explicitly or requires resolving them from
the network state).  This $2\times2$ design isolates the failure
modes that empirical power-system code generation actually faces;
it draws on power-domain natural-language evaluation and
simulation-assistant work but makes the semantic-grounding and
workflow-composition axes explicit for executable code
generation~\cite{zhou2024elecbench,
cheng2025advanceddispatch,bonadia2023opendss,jia2024daline,
jia2024feedbacksim}.  The four difficulty levels D1--D4 instantiate
this design (Table~\ref{tab:difficulty}): comparing
(D1, D3) against (D2, D4) isolates the effect of workflow complexity,
while comparing (D1, D2) against (D3, D4) isolates the effect of
semantic grounding; D4 stacks both.

``Semantic'' here refers specifically to the setup/modification step:
the target element is specified by an operating rule
(\emph{``the most heavily loaded line''}, \emph{``the highest-voltage
bus with load connected''}) that must be resolved against the current
network state before the modification is applied.  It does not mean
every entity in the query is implicit; downstream analysis routines
and final extraction indices may still be stated explicitly.
The same table pairs each level with a benchmark item as
the model receives it, so the two axes can be read off the query text
itself.  D4's split
into a semantic setup phase followed by a deterministic multi-step
skeleton preserves executable-reference scoring: a fully semantic
multi-step pipeline would require the model's own intermediate
decisions to feed downstream steps, breaking deterministic ground
truth.
\begin{table}[!htbp]
\centering
\footnotesize
\setlength{\tabcolsep}{4pt}
\caption{The D1--D4 difficulty taxonomy: the workflow-complexity
  $\times$ semantic-grounding design, with one benchmark item per level.
  Queries are shown as delivered to the model (lightly abridged);
  \emph{GT} is the scalar ground truth produced by executing the
  reference solution.  Emphasis marks the semantic setup instruction
  that D3 and D4 must resolve against the network state before the
  modification is applied.}
\label{tab:difficulty}
\footnotesize
\renewcommand{\arraystretch}{0.95}
\begin{tabular}{@{}>{\raggedright\arraybackslash}p{0.15\linewidth}p{0.815\linewidth}@{}}
\toprule
\textbf{Level} & \textbf{Representative benchmark item} \\
\midrule
\textbf{D1} \newline single-step \newline explicit &
  ``Set transformer 51 to 2 parallel units, disconnect line 54, execute a
  power flow on the Illinois 200-bus system, and report the minimum bus
  voltage (p.u.).'' \hfill GT: 0.9861 \\
\addlinespace[2pt]
\textbf{D2} \newline multi-step \newline explicit &
  ``Perform N-1 contingency analysis on the 30-bus case and find the most
  loaded line.  Add a 14~MW generator at bus~9, re-run power flow with that
  line out of service, and report the new maximum line loading (\%).''
  \hfill GT: 137.875 \\
\addlinespace[2pt]
\textbf{D3} \newline single-step \newline semantic &
  ``Using the CIGRE HV benchmark, \emph{identify the most heavily loaded
  line under the current operating point} and reinforce it by adding one
  parallel circuit.  Run power flow and report the loading (\%) of
  transformer~5.'' \hfill GT: 68.062 \\
\addlinespace[2pt]
\textbf{D4} \newline multi-step \newline semantic &
  ``On the PEGASE 1354-bus case, \emph{find the online generator with the
  highest active dispatch} and trip it offline.  Run power flow; if any
  line loading exceeds 110.1\%, disconnect line~1066 and re-run, reporting
  the minimum bus voltage (p.u.).'' \hfill GT: 0.9820 \\
\bottomrule
\end{tabular}
\end{table}

\subsection{Generator Coverage and Frozen Release}
\label{sec:bench:coverage}

The generator's current configuration samples from a pool of
\textbf{39 grid networks} (ranging from 4-bus examples to PEGASE
9241-bus, RTE 6470-bus, Polish 3120-bus, and GB 2224-bus
transmission cases), \textbf{15 task families} spanning steady-state
power flow, DC power flow, OPF, short-circuit studies, contingency
analysis, time-series simulation, state estimation, and composed
multi-step workflows such as sequential diagnosis,
power-flow--short-circuit composition, and contingency repair, and \textbf{17 query-target patterns}
(point values, extrema, which-element-is-largest (argmax) queries,
threshold checks, time-series point queries).  These analyses are part of the modelling and simulation
scope supported by \pp{}~\cite{thurner2018pandapower}.  The
semantic setup rules and feasibility filters are summarised in the
generation protocol of
SM Table~S1; the modification operators, D4 compound wrappers,
and the full item-level
inventory are in the PowerCodeBench repository
(\url{https://github.com/huiwxing/PowerCodeBench}).

The frozen release used throughout this paper instantiates this
configuration as \textbf{2{,}000 items}, sampled with a fixed seed
and a 30\%/30\%/20\%/20\% D1--D4 split.  The items derive from
$54$ operative query-template keys ($47$ single-task and $7$
multi-task-family keys; cluster sizes $3$--$171$, median $22$).  The per-level item counts
and coverage diagnostics of the release are tabulated alongside the
protocol, and the per-family and per-network composition
plots, including the 39 instantiated networks (from a 40-network
generator registry), are released with the benchmark
repository.  D3 draws
on the same 8 single-step task families as D1, and D4 on the same
7 multi-step families as D2: semantic levels are not separate task
sets, only controlled rewrites of the explicit levels in which the
setup instruction is replaced by a rule-grounded operating
description.

\subsection{Reference Execution and Validation}
\label{sec:bench:gt}

For each item, the ground truth $g^*$ is produced by executing the
verified Python reference solution against a deterministic network
snapshot and extracting a single scalar value (float, int, or bool)
specified by a structured result-table query (e.g.\ $\max$ of
\texttt{res\_line.loading\_percent}); for specialised tasks such as
time-series simulation and state estimation, dedicated
reference-code generators construct the required controllers,
measurements, and result-extraction logic.  The model-scoring protocol is
the task-and-scoring contract introduced above; throughout the benchmark,
integer and Boolean outputs require exact match, while floating-point
outputs use mixed absolute/relative tolerance (absolute tolerance
$10^{-3}$, relative tolerance $10^{-2}$).  No manual answer annotation
enters the loop.  We audit the numerical criterion in both directions,
including whether programs that pass it actually perform the requested
physical analysis (Sec.~\ref{sec:exp:validity_audit}).

Each candidate item passes a \emph{hard admission filter} before
joining the release: the reference code must run to completion, the
underlying simulation must converge or be feasible (power-flow
convergence, feasible OPF solution), voltage and topology sanity
checks must pass, and the extracted scalar must lie
in a plausible range for its task family.  Items that fail any of
these checks are rejected and regenerated.  The uniqueness,
ground-truth-type, and semantic-coverage diagnostics of the frozen
release are recorded alongside the released generation protocol.  Separately, for D3 and D4
items, a lexical diagnostic flags semantic-setup instructions that
would inadvertently reveal explicit parameter values or indices; this
is a release-quality \emph{diagnostic} rather than a model-scoring
input.  The five-stage construction pipeline
(template design, controlled instantiation, reference execution,
validation and rejection, semantic diagnostics) is documented in
SM Sec.~S1 (Fig.~S1).
\section{Boundary Probing and Demand-Guided Intervention}
\label{sec:method}

\begin{figure}[!t]
  \centering
  \begin{subfigure}{\linewidth}
    \centering
    \includegraphics[width=\linewidth]{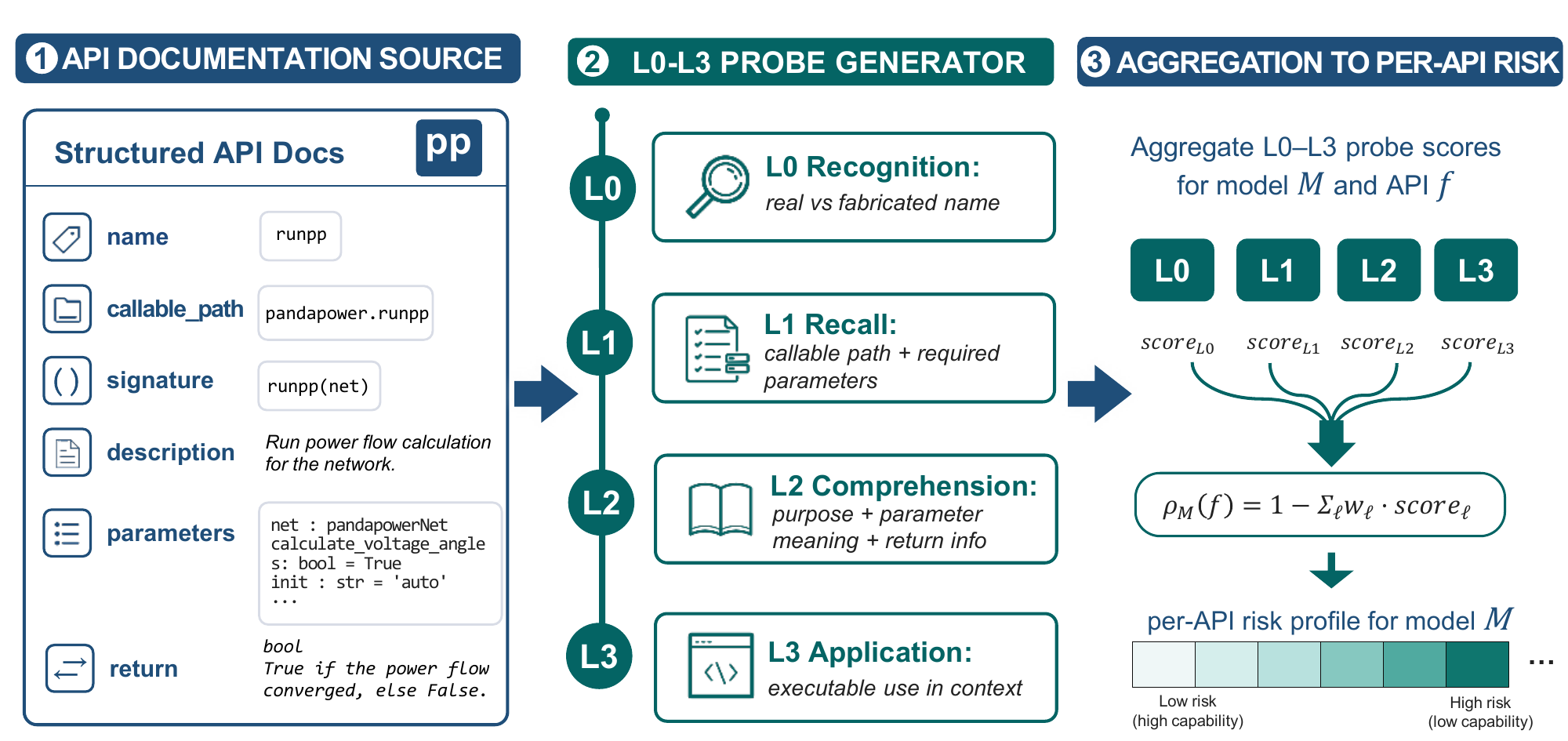}
    \caption{}
    \label{fig:probe_generator_concept}
  \end{subfigure}\\[2pt]
  \begin{subfigure}{0.95\linewidth}
    \centering
    \includegraphics[width=\linewidth]{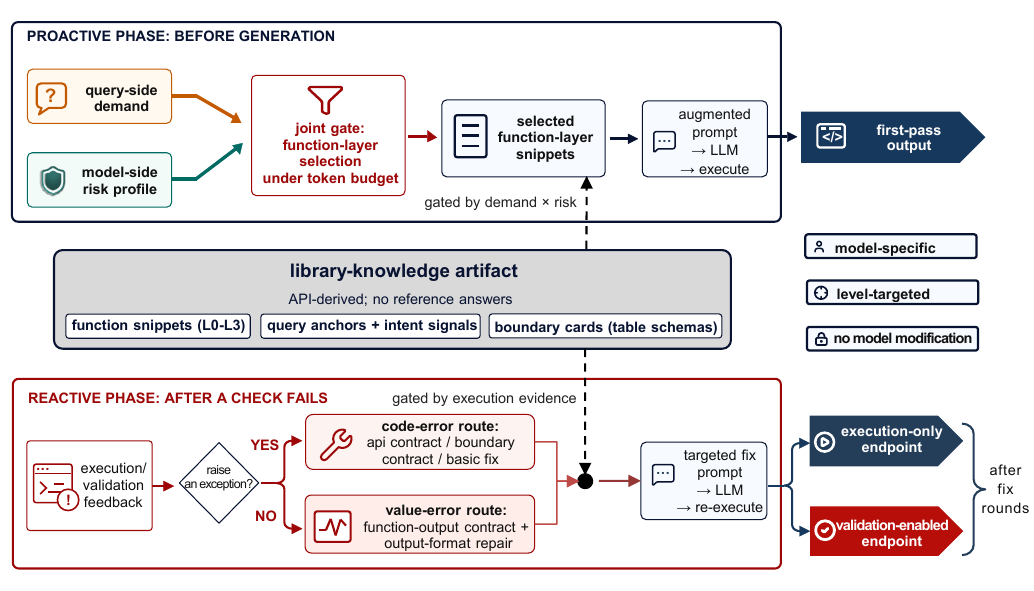}
    \caption{}
    \label{fig:intervention_pipeline}
  \end{subfigure}
  \caption{Method architecture.
    (a)~The L0--L3 probe generator turns structured API
    documentation into per-layer probes, aggregated into a
    model-specific knowledge-risk profile $\rho_\mM(f)$
    (Eq.~\ref{eq:risk}).
    (b)~The two-phase intervention: Phase~1 (proactive) injects
    layer-specific documentation before generation, guided by task
    demand $\hat{d}(f,q)$ (Sec.~\ref{sec:demand}) and the profile;
    Phase~2 (reactive) routes execution feedback to targeted repair
    before re-execution.}
  \label{fig:method_architecture}
\end{figure}

Boundary-aware intervention combines two signals derived from the
same structured API specification.  A documentation-driven
\emph{probe generator} (like PowerCodeBench, a generator rather
than a static question set) emits layered probes whose scores
aggregate into a per-model \emph{knowledge profile}, recording its
measured deficits by API entry and documentation layer.  A query-side \emph{demand
estimator} predicts which entries a natural-language query requires
at inference time, when no reference code is available.  Neither
signal suffices alone: query
similarity cannot tell whether the chosen model already knows the
needed layer, and a model-side profile cannot tell whether the task
needs that API.  The two signals are therefore combined to rank the
candidate documentation before generation.  After a failed attempt,
execution or validation evidence guides targeted repair.  The
subsections below follow this order, and
Fig.~\ref{fig:method_architecture} shows both stages.

\subsection{Design Principles}
\label{sec:probing:principles}
\label{sec:interv:overview}

Three principles govern probe construction and injection alike.
\textbf{Specification-driven construction, no model modification:}
the probe corpus and function-layer snippets are generated from a
structured API specification (entry name, callable path, category,
signature, parameter metadata, descriptions, return information, and
optional examples), without \pp{}-specific synonym tables, entity
lists, or hand-crafted distractor vocabularies.  Runtime interfaces
that are not reliably recoverable from this structure use a compact,
versioned backend artifact layer: workflow-intent rules, boundary
cards, and the curated function-output mapping used by value-error
repair.  The specification-derived core can therefore be regenerated
for a documented API once these backend contracts are supplied.  The
cross-backend pilots below evaluate only probing and Round-0 proactive
allocation, not reactive repair.  No stage requires
fine-tuning, weight updates, or access to model internals, and every
model in the panel is processed by the same procedure.

\textbf{Layered decomposition, layer-targeted correction:} probes
are organised into four layers matching the natural granularity of
API documentation (L0 recognition through L3 application, defined
in Sec.~\ref{sec:probing}), loosely along revised Bloom-style
cognitive levels~\cite{anderson2001taxonomy}, with L3 scored by
executing the model's generated code.  The layer scores provide
allocation state for selecting a signature, contract, or example
so that the whole API entry need not be injected.

\textbf{Model-aware evidence allocation:} query relevance determines
which API entries enter consideration, while the serving model's
per-layer profile determines which facts receive priority.  The same
engineering request can therefore receive different function--layer
evidence under different measured knowledge profiles.  This joint
model--query conditioning distinguishes the allocator from standard
query-only code and documentation retrieval
methods~\cite{lewis2020rag,zhang2023repocoder,
patil2023gorilla,jeong2024adaptiverag}.

\subsection{Layered Probing and Knowledge Profiles}
\label{sec:probing}
\label{sec:probing:l0l3}
\label{sec:probing:coverage}
\label{sec:probing:profile}

\textbf{L0 -- Recognition} tests whether the model distinguishes a
real API name from plausible fabrications, scored by mean accuracy
across paired four-choice \texttt{find\_real} and \texttt{find\_fake}
probes,
directly targeting the
fabricated-API and wrong-method failure modes reported in API-call
and API-misuse studies~\cite{patil2023gorilla,li2023apibank,zhuo2025apimisuse,
chen2025apihallucination}.
\textbf{L1 -- Recall} prompts the model to produce the full API path
and required parameters of $f$.  Its score is the equally
weighted mean of path correctness and required-parameter Jaccard
similarity (overlap over union); the two components are also retained
as separate diagnostics.
\textbf{L2 -- Comprehension} is a multiple-choice test on function
purpose, parameter semantics, and return information, instantiated
from the corresponding documentation fields.
\textbf{L3 -- Application} asks the model to write minimal
executable code invoking the target function; execution success is
the primary score, with API validity and target-function usage as
auxiliary diagnostics.

The frozen corpus covers the complete documented public API rather
than a selected subset: all $275$ callables in the \pp{}
specification ($262$ functions and $13$ callable classes).  These
entries span analysis and simulation routines, network construction,
controllers, time series, result tables, and visualisation.  The
corpus contains \textbf{$2{,}080$ probes} (L0 $550$, L1 $275$, L2
$980$, L3 $275$) and a companion bank of \textbf{$1{,}100$
documentation snippets}; $274$ entries include an executable example
for L3-oriented injection.  An abstract-syntax-tree (AST) audit
confirms that the corpus covers the core \pp{} calls in the benchmark
reference programs.  Calls outside it are generic
Python/NumPy/Pandas operations or network-fixture loaders (SM
Table~S2).  The probes are used only for diagnosis, whereas the
snippets are used only by the intervention.  A newly documented
callable would be incorporated by refreshing both artifacts through
the same generation pipeline.

For each model $\mM$ and each function $f \in \mF$ (the API-entry
corpus), we aggregate the L0--L3 probe scores into a per-function
risk score as a weighted average of the per-level deficits
(Fig.~\ref{fig:probe_generator_concept}):
\begin{equation}
  \rho_\mM(f) \;=\; \sum_{\ell \in \{0,1,2,3\}}
    \tilde{w}_\ell \left(1 - s^{(\ell)}_\mM(f)\right),
  \qquad
  \tilde{w}_\ell \;=\; \frac{w_\ell}{\sum_{\ell' :\, s^{(\ell')}_\mM(f)
    \text{ defined}} w_{\ell'}},
  \label{eq:risk}
\end{equation}
where $s^{(\ell)}_\mM(f) \in [0, 1]$ is the probe score at level $\ell$,
$w_\ell \geq 0$ is the layer weight, and the weights are renormalised
over the levels at which the probe score is defined (the sum over
$\ell$ likewise skips undefined levels).
A high $\rho_\mM(f)$ denotes a larger measured L0--L3 deficit for
function $f$.  The collection $\{\rho_\mM(f)\}_{f \in \mF}$
constitutes the \emph{knowledge profile} of $\mM$.

The probe data is retained at two granularities.  The
\textbf{per-function aggregate} $\rho_\mM(f)$ provides a readable
diagnostic (which API entries exhibit the largest measured
deficits) for cross-model comparison and library-specific model
selection.  The \textbf{per-layer scores} $s^{(l)}_\mM(f)$ provide
the runtime control state: they determine which documentation layer
competes for allocation through the injection score of the proactive
phase below.

The deployed pipeline uses the mildly application-weighted setting
$w\!=\!(0.20, 0.25, 0.20, 0.35)$, which is the configuration behind
the main intervention results in this paper; the same layer weights $w_\ell$ enter
the injection score below.  Accuracy
is insensitive to this choice across the tested simplex directions:
over four weightings spanning uniform, L3-heavy $(0, 0.2, 0.2,
0.6)$, and inverted L0-heavy $(0.4, 0.3, 0.2, 0.1)$ settings on five
models at full $n\!=\!2{,}000$, the maximum pairwise accuracy
difference is $1.3$pp (SM Fig.~S2(a)).

\begin{figure}[!htbp]
  \centering
  \includegraphics[width=\linewidth]{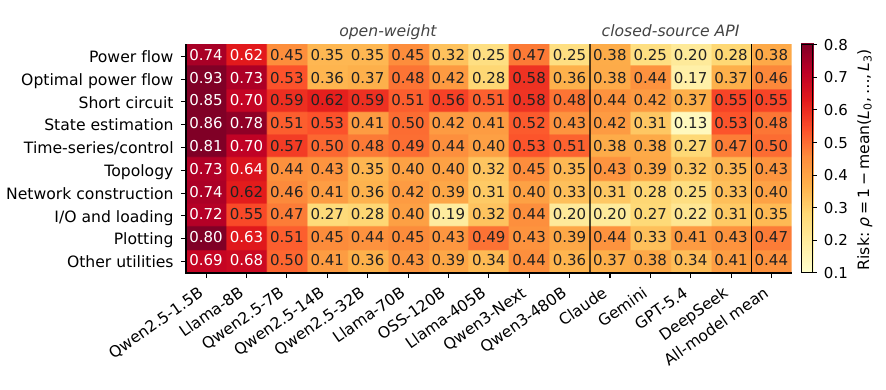}
  \caption{Category-averaged knowledge-risk profile derived from
    $\rho_\mM(f)\!=\!1-\overline{L_0..L_3}$ (the equal-weight
    rendering of Eq.~(\ref{eq:risk})) across the
    open-weight panel (10 models) and 4 closed-source APIs; warmer
    colours indicate higher risk.  Vertical separators delimit the
    two groups; the right-most column is the all-model mean.}
  \label{fig:knowledge_profile_heatmap}
\end{figure}

\paragraph{What the profiles reveal}
\label{sec:probing:results}
The profiles reveal a consistent split.  Deployable models generally
understand what \pp{} functions do, but are less reliable at
recognising valid API names and turning that knowledge into runnable
calls.  Eight of the nine deployable open-weight models score above
$0.90$ on L2, yet no open-weight model reaches $0.50$ on L3, and the
best open-weight L0 score is $0.536$.  Because L0 also tests rejection
of plausible distractors, we treat it as a directional recognition
diagnostic rather than a pure measure of stored API knowledge.
Fig.~\ref{fig:knowledge_profile_heatmap} shows how this pattern varies
across API categories.  It uses the equal-weight rendering
$\rho_\mM(f)\!=\!1-\overline{L_0..L_3}$ so that no layer is emphasised;
the full per-model scores are reported in SM Table~S3.

The four layers provide complementary information.  L2 contributes
parameter and return-value contracts even though it separates models
only weakly.  L1 is the layer most closely associated with unassisted
first-pass accuracy across models ($\rho_s\!=\!0.93$, the Spearman
rank correlation between a layer's mean probe score and that accuracy
over the ten open-weight models).  L0 and L3
instead indicate where name-level guidance or executable examples may
be useful.  Application skill also does not increase monotonically
with model size: Qwen2.5-Coder-32B scores $0.425$ on L3, compared with
$0.200$ for Llama-3.1-70B.  The profiles therefore complement
parameter count as a library-specific diagnostic and provide the
state used to allocate documentation.

The closed-source APIs show the same broad separation between strong
comprehension and less consistent recognition and application.  Their
profiles enter the same documentation-allocation procedure, while
failures that become visible only after execution are handled by the
reactive routes described below.

\subsection{Query-Side Demand Estimation}
\label{sec:demand}

The demand estimator has a deliberately narrow role.  It does not
predict the complete program; it retains a small, high-recall shortlist
of API functions that the query may require.  The model-side profile then
determines which documentation layers to inject for those APIs.  More
formally, let $\mD^*(q) \subseteq \mF$ denote the functions required to
solve query $q$.  We learn a relevance score
$\hat d(f,q) \in [0,1]$ whose top-$k$ set approximates $\mD^*(q)$.
Although $\mD^*(q)$ can be derived from reference code for evaluation,
inference uses only the query text, never reference code, benchmark
ground truth, or execution feedback.

The training data and PowerCodeBench cover some of the same API
functions but differ in query wording.  The downstream evaluation
therefore measures whether ranking transfers to unseen wording.  We train
on $6{,}312$ (query, function-list) pairs
in $511$ base groups covering $125$ API functions.  The pairs are
generated by API traversal, labelled through AST verification, and
split by group to prevent paraphrase leakage.  Of the $50$ functions
used by benchmark reference solutions, $35$ ($70\%$) also occur in
training, mainly analysis routines.  In an overlap audit comparing
$500$ sampled benchmark queries with $5{,}000$ training queries, the
maximum word-level Jaccard similarity is $0.41$, and none exceeds $0.5$
(corpus construction and overlap checks are in SM Sec.~S4.1).

\label{sec:demand:models}\label{sec:demand:eval}
We compare six variants spanning lexical term-frequency--inverse
document-frequency (TF--IDF) retrieval, dense Sentence-BERT (SBERT)
retrieval, and cross-encoder reranking, in unsupervised and
supervised-pairwise forms~\cite{robertson2009bm25,reimers2019sentencebert}.
We set $k=10$, where recall@$k$, the fraction of required functions in
the top-$k$ set, begins to flatten.  The
deployed hybrid TF--IDF ranker combines zero-shot lexical evidence
with a supervised pairwise scorer; its mixing weight is selected for
recall@$10$ on a held-out validation split.  This combination is the
most robust of the six under the shift in query style.  Section~\ref{sec:exp:demand}
and SM Sec.~S4.2 report the comparative results and full formulations.

\label{sec:demand:adaptation}
The remaining shift is in API-role mix: the source corpus is
construction-heavy, whereas the target workload emphasises operational
and analysis functions.  Adaptation first applies the unadapted ranker
to an unlabelled batch of target queries and predicts the top-$k$
functions for each query.  It then aggregates the roles of those
functions to estimate target role shares, reweights source examples by
the target-to-source role ratio, and refits the supervised
component~\cite{huang2007kmm}.  This procedure needs no target function
labels, but it does require a batch of target queries in advance.  This
makes the adaptation transductive and workload specific.  We report
both rankers.  The adapted ranker reaches $71.8\%$ recall@$10$
on the benchmark-reference diagnostic split and supplies
$\hat d(\cdot,q)$ to Eq.~(\ref{eq:intervention_score}) below.  In a
separate end-to-end comparison, the unadapted ranker still retains a
$+27.79$pp accuracy lift (Sec.~\ref{sec:exp:robustness}).

\subsection{Phase~1: Proactive Layer-Wise Injection}
\label{sec:intervention}
\label{sec:interv:proactive}

In Phase~1 (Fig.~\ref{fig:intervention_pipeline}), the model
profile and demand estimate jointly select a small layer-targeted
documentation set that is prepended to the generation prompt; in
Phase~2 the demand estimate is replaced
by evidence from the run (the exception class, failing source line,
or parsed-payload mismatch), which already pinpoints the implicated
APIs.  Both phases draw from the same versioned library-knowledge
artifact.  It stores \textbf{function snippets} generated from the
structured API specification, one rendering per
documentation layer (L0 short name/purpose hint, L1 call signature,
L2 parameter/return contract, L3 compact usage example);
\textbf{anchors and intent signals}, compact backend-facing lexical patterns
tying a query phrase to a specific API or to one of the mutually
exclusive time-series, short-circuit, contingency, OPF, and
DC-power-flow workflow executors; and \textbf{boundary cards}, short
DataFrame schema contracts for recurrent table conventions that span
multiple APIs (e.g.\ result-table columns or OPF cost extraction).
None reads or conditions on the reference answer (construction rules
in SM Sec.~S3.1); the artifact defines \emph{what} library facts are
available, while the probe profile and demand estimator decide
\emph{whether} and \emph{at which layer} to inject them.

Phase~1 treats intervention as a set of function--layer choices: for
each candidate function $f$ and layer $\ell \in
\{\mathrm{L0,L1,L2,L3}\}$, the selector scores whether the query
needs that function and the model needs that layer, then fills a
fixed token budget with the highest-value candidates.

Each candidate function--layer pair is scored by a single
\emph{injection score}:
\begin{equation}
  I_\ell(f, q, \mM) \;=\; \hat{d}(f, q) \cdot
  \max\!\left(r_{\mM,\ell}(f), \tau_\ell(f,q)\right) \cdot
  a_\ell(q) \cdot w_\ell,
  \label{eq:intervention_score}
\end{equation}
where $\hat{d}(f, q) \in [0, 1]$ is the demand estimate of
the previous subsection;
$r_{\mM,\ell}(f) = \max(0, (\tau - s^{(\ell)}_\mM(f)) / \tau)
\in [0,1]$ is the layer-wise probe deficit: how far the per-layer probe
score $s^{(\ell)}_\mM(f)$ falls below the reliability threshold
$\tau\!=\!0.5$, clipped at zero and normalised by $\tau$.  The risk
profile $\rho_\mM$ of Eq.~(\ref{eq:risk}) combines the same scores by a
weighted average of the deficits; here they are instead thresholded.

The deterministic keep-alive floor $\tau_\ell(f,q)$ keeps a candidate in
contention in three high-confidence cases: $f$ matches a
high-precision query anchor, $\hat d(f,q)$ clears a confidence
threshold, or $f$ is tied to a boundary-card contract triggered by
$q$.  The depth multiplier $a_\ell(q)$ adds small layer-specific
increments for procedural-complexity signals in the query, such as
workflow intent, sequential or conditional connectives, aggregation
phrasing, and multi-modification verbs.  Finally, $w_\ell$ is the
layer weight of Eq.~(\ref{eq:risk}) (deployed mild-L3 setting
$(0.20, 0.25, 0.20, 0.35)$).  The numerical constants, the
$\tau_\ell$ and $a_\ell$ rule tables, the intent signals, and the
anchors and boundary cards are all frozen as part of the versioned
artifact specification.  At runtime,
these quantities depend only on the query text and the
versioned library artifact.  They encode fixed layer-role priors and
token-budget arithmetic and are not tuned against PowerCodeBench
accuracy.

Given the scored candidates, we fill the token budget greedily:
candidates are ranked in decreasing order of
$I_\ell(f,q,\mM)\,/\,|\text{doc}(f,\ell)|^{1/2}$ (a token-aware
ranking rule that prefers short, high-value snippets when the budget
is tight) and added one at a time until the cumulative
documentation length reaches the token budget $B$.  A
query-conditioned cap on L3 examples keeps a few procedural
snippets from exhausting the budget on procedurally light queries.
Higher-layer snippets suppress redundant lower-layer ones for the
same function, so an L2 contract makes a separate L0 hint unnecessary.
L2 and L3 can still appear together because a contract and an example
address different needs.  An intent-consistency filter prevents
evidence for one workflow, such as OPF, from pulling in an unrelated
executor.  The selected snippets are prepended to the generation
prompt as a targeted API-reference section.

This phase targets \textbf{R0 (first-pass)} accuracy.  Two
model-agnostic comparisons separate context volume from profile
conditioning: a \emph{demand-hit full-layer reference} that exposes all
available layers for the demand-hit functions under the full-context
budget, and therefore retains any recall errors made by the demand
predictor,
and a same-nominal-budget isolation that keeps the boundary-guided
condition's demand estimates, artifacts, renderer, layer pruning,
and packer unchanged but replaces the model-specific deficit
$r_{\mM,\ell}(f)$ with the constant $1$ (maximal deficit).
Sec.~\ref{sec:exp:intervention}
reports both comparisons.

\subsection{Phase~2: Reactive Correction and Operating Regimes}
\label{sec:interv:reactive}

When generated code raises an exception, the run supplies direct
evidence through the exception class and failing source line.  Code
that executes enters value-side repair only when a validation signal
flags its output.  The reactive phase uses this evidence instead of
re-applying the demand score in Eq.~(\ref{eq:intervention_score}).
Execution-feedback prompting and iterative self-refinement provide
the comparison class for this phase~\cite{chen2023selfdebug,
madaan2023selfrefine}.  Our router first distinguishes code errors
from value errors, then selects the library facts needed for repair.

\paragraph{Reactive routes}
Phase~2 handles two failure classes.  Runtime exceptions are mapped
from exception class and failing source line to one of three
policies: basic fix (local exceptions with no API symbol), API
contract (L2 call contract for an implicated function), or boundary
card (DataFrame/time-series/OPF/short-circuit boundary failures).
Executed-but-wrong outputs are split into numerical mismatches,
handled by extracting an AST trace of the generated code and
injecting interface contracts plus function-output
\texttt{net.res\_*} contracts for each analysis routine encountered
(exception-class clusters, AST-trace structure, and the eight-routine
function-output mapping in SM Sec.~S3.2),
and output-format failures, handled by a focused format instruction
without altering analysis logic.  Runtime and output-format routes
use only execution and parsing evidence.  Numerical-mismatch repair
requires the validation signal described next; after it is triggered,
the router selects contracts from the generated code rather than from
the reference solution.

\paragraph{Operating regimes: reference-free vs.\ validation-enabled}
The two reactive endpoints correspond to distinct operating regimes.
The default reactive endpoint is reference-free, consuming only
execution-side evidence (exception class, failing source line, parse
failure on the printed payload), and is therefore applicable wherever
the generated code is run against the target library.  The
validation-enabled endpoint additionally activates the value-error branch
whenever an external validation signal flags an executed-but-wrong
output.  In the benchmark this signal is provided by scalar matching
against the reference.  In deployment it may come from an acceptance
test, disagreement with supervisory control and data acquisition
(SCADA) records, phasor measurement unit (PMU) records, or historical
cases, a physical-invariant check, an independent solver, or operator
feedback.  The signal triggers repair but is not included
as a target value in the repair prompt.  Without such a signal, repair
is limited to runtime and output-format failures.  The experiments
report both endpoints.

The reactive fix prompt is a structured user-turn template
preserving the original task, the previous code, the failure
description, and any router-selected documentation, with the system
prompt retained unchanged across fix rounds.  When the router returns
a basic-fix decision, the relevant-doc block is empty and the prompt
reduces to the standard error-feedback baseline.  This phase targets the post-first-pass fix
trajectory: reaching a requested validation endpoint with fewer fix
rounds and lower total token consumption than an always-document
baseline, by injecting only the minimum library context that the
failure evidence implicates.  The number of permitted fix rounds is
an evaluation setting; the experimental setup instantiates
it as $K\!=\!3$ and reports the resulting R0$\to$R3 trajectory.
\section{Experiments}
\label{sec:experiments}

\subsection{Experimental Setup}
\label{sec:exp:setup}

\textbf{Models.}
We evaluate ten open-weight models from $1.5$B to $480$B
parameters: Qwen2.5-Coder-\{1.5B, 7B, 14B, 32B\}, Qwen3-Coder-Next
(a mixture-of-experts model, MoE, with a mid-tier number of active
parameters), Qwen3-Coder-480B, Llama-3.1-\{8B, 70B,
405B\}, and GPT-OSS-120B.  They run alongside four closed-source
mid-tier API models from four different vendors: Anthropic
Claude-Haiku-4-5, Google Gemini-2.5-Flash, OpenAI GPT-5.4-mini, and
DeepSeek-V4-Flash.  For
readability, subsequent prose and tables use Qwen2.5-\{1.5B, 7B, 14B,
32B\}, Qwen3-Next, Qwen3-480B, Llama-\{8B, 70B, 405B\}, GPT-OSS-120B,
Claude-Haiku, Gemini-Flash, GPT-5.4-mini, and DeepSeek-Flash as short
labels.  All fourteen models see identical inputs on the full $2{,}000$-item
PowerCodeBench benchmark, with no subset filtering.  The $1.5$B variant
is included as a low-end reference.  Throughout this section,
``deployable tier'' refers to the $\geq\!7$B open-weight cohort and the
four APIs.  The panel reflects the early-2026 generation of mid-tier
APIs.  The same probing and intervention pipelines can be applied to
newer open-weight families.

\textbf{Decoding and protocol.}
Probe generation, first-pass code generation, and fix-round generation
all use greedy decoding (temperature $0$).  Closed-source API calls use
the same \emph{matched no-tool setting}: temperature $0$, no tool or
function calling, and no explicit reasoning mode where the vendor
exposes these controls.  Reasoning-tier variants are evaluated
separately in Sec.~\ref{sec:exp:robustness}.  Each benchmark item
is evaluated under the two-phase protocol, with up to $K\!=\!3$ fix
rounds.  FDR uses
execution status and exception traces.  FDRS also
uses a value-mismatch notice when an acceptance test or equivalent
validation signal is available.  This signal triggers repair against
the relevant output contract without supplying the reference scalar.
Float results use the matching tolerance of Sec.~\ref{sec:bench:gt}.  Injection
conditions use the scenario-adapted (transductive) demand predictor
(Sec.~\ref{sec:demand:adaptation}).  The unadapted variant is reported
in Sec.~\ref{sec:exp:robustness}.  Exact
model identifiers, parameter counts, seeds, token budgets, execution
timeouts, hardware, and serving precision are in SM Sec.~S6.

\textbf{Metrics.}
We report three basic per-model quantities.  \emph{ExecRate} is the
fraction of generated programs that run to completion without raising
an exception.  \emph{Acc} is the fraction whose printed scalar matches
the reference under the matching tolerance.  \emph{Acc$|$Exec} is
accuracy conditioned on execution.  By construction
$\text{Acc}\!\le\!\text{ExecRate}$, and Acc$|$Exec separates value-side
errors from execution-side ones.  At Round~$0$ (R0), Acc measures
first-pass quality and is the target of proactive injection.  The
$R0\!\to\!R3$ Acc trajectory and round counts evaluate reactive
correction under the $K\!=\!3$ setting.  Prompt-token cost is
reported for both phases: first-pass prompt cost for proactive
injection and cumulative fix-round prompt cost for reactive
correction.

\textbf{Variance.}
To bound seed effects, we re-ran every open-weight panel cell on a
stratified $400$-item subset across $10$ vLLM generation seeds.  The
panel-median per-cell standard deviation is $\pm 0.20$pp.  The
cross-condition lifts reported below sit well above this band.
Near-band cross-vendor gaps are reported with paired-bootstrap confidence intervals
(resampling items with replacement to bound noise) alongside the
point estimates.  Because items instantiate the $54$ template keys
of Sec.~\ref{sec:bench:coverage}, the principal contrasts were also
recomputed under a template-cluster bootstrap, with task families as
a sensitivity analysis.  The signs and significance conclusions are
unchanged: the intervals are $[+25.31,+37.64]$pp for panel C--A and
$[+0.66,+3.85]$pp for FDRS--FDR, both excluding zero.  The $\pm5$pp
equivalence certifications in Table~\ref{tab:cross_vendor}(b), however,
hold under item-level resampling only
(Sec.~\ref{sec:exp:cross_vendor}).

\paragraph{Proactive and reactive condition codes}
\label{sec:exp:conditions}
\textbf{Proactive injection} (applied at Round~0):
\begin{description}\setlength{\itemsep}{0.2em}
  \item[A] no proactive injection (baseline).
  \item[B] function-name-only injection (L0-only ablation).
  \item[C] profile-conditioned, demand-aware layer allocation: the
    full proactive method.
  \item[X] demand-only allocation of all available L0--L3 layers under
    the full-context budget: the model-agnostic demand-hit full-layer
    reference (it inherits the demand predictor's recall errors,
    Sec.~\ref{sec:interv:proactive}).
  \item[R] vanilla BM25 retrieval (a standard keyword-frequency ranker)
    over the raw API
    documentation~\cite{robertson2009bm25,lewis2020rag}: off-the-shelf
    RAG baseline.
  \item[R$_{\text{sem}}$] vanilla dense retrieval: identical to R except
    that the ranker is SBERT cosine similarity over the same corpus,
    with the same renderer, packer, and token budget: the dense-RAG
    counterpart (Sec.~\ref{sec:exp:proactive}).
\end{description}
\textbf{Reactive correction} (Rounds~1--3, composed with a proactive
condition; mechanism in Sec.~\ref{sec:interv:reactive}):
\begin{description}\setlength{\itemsep}{0.2em}
  \item[FX] plain fix loop with execution feedback only, no
    documentation: standard self-debug
    baseline~\cite{chen2023selfdebug,madaan2023selfrefine}.
  \item[FD] always-document fix at static depth (no routing).
  \item[FDR] reference-free routed repair from execution
    feedback (no value-error branch).
  \item[FDRS] FDR with validation-triggered value-error repair and
    function-output contracts: the full reactive method when an
    external acceptance signal is available.
\end{description}
Proactive--reactive combinations are written ``proactive+reactive'',
e.g.\ \emph{C+FDRS} for the layer-wise proactive condition composed
with the full reactive method.  R$_{\text{semB}}$ denotes dense
retrieval at C-matched measured cost (budget calibration in
SM Sec.~S5.4).  The transfer pilot
(Sec.~\ref{sec:exp:transfer}) additionally uses C-recal, a
single-change recalibration of C defined there.

The experiments begin with cross-model results and failure
localisation (Secs.~\ref{sec:exp:cross_vendor}--\ref{sec:exp:r0}).
Component and cost ablations follow (Sec.~\ref{sec:exp:intervention}),
then robustness, Round-0 onboarding pilots on OpenDSS and PyPSA, and an audit of whether
numerical matches reflect the requested analysis
(Secs.~\ref{sec:exp:robustness}--\ref{sec:exp:validity_audit}).
\subsection{Cross-Model Gains and the Local Deployment Range}
\label{sec:exp:cross_vendor}

\begin{figure}[!htbp]
  \centering
  \includegraphics[width=\linewidth]{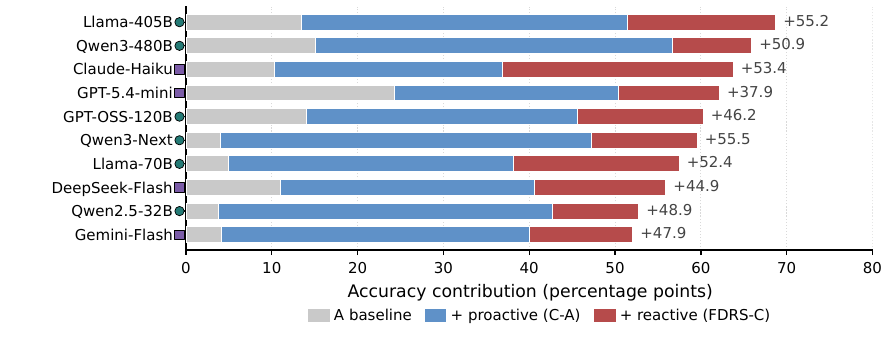}
  \caption{Per-model decomposition of the C+FDRS endpoint for the
    ten cross-vendor configurations of Table~\ref{tab:cross_vendor} (circles:
    open-weight; squares: mid-tier APIs) into three stacked
    contributions: the A baseline (R0),
    the proactive increment ($\text{C}\!-\!\text{A}$), and the
    reactive increment ($\text{C+FDRS}\!-\!\text{C}$).  The total
    lift $\Delta$(FDRS--A) is annotated to the right of each bar.}
  \label{fig:cross_vendor_headline}
\end{figure}

Assistance improves every deployable model in the panel.  Across the
nine open-weight models of at least $7$B and the four mid-tier APIs,
C+FDRS raises accuracy by $32$--$56$ percentage points over the
unassisted first pass on the full $2{,}000$-item benchmark, after up
to three repair rounds.  Table~\ref{tab:cross_vendor} reports the
exact endpoints for the six strongest open-weight configurations and
the four APIs.  Fig.~\ref{fig:cross_vendor_headline} separates the
proactive and reactive contributions.  For the API models,
validation-triggered value repair adds a further $1.0$--$8.6$ points
beyond the reference-free endpoint; per-model repair endpoints for
the open-weight models are reported in SM Table~S8.

\begin{table}[!htbp]
\centering
\caption{Cross-vendor evaluation on the full $2{,}000$-item PowerCodeBench
  benchmark.  The open-weight section lists the six highest-scoring
  configurations; the full $1.5$B--$480$B R0 panel appears in
  Table~\ref{tab:proactive_main}, and open-weight per-model repair
  endpoints in SM Table~S8.  The
  closed-source panel covers four mid-tier APIs; per-cell seed
  variance is in SM Fig.~S2(b).  Condition codes:
  Sec.~\ref{sec:exp:conditions}.}
\label{tab:cross_vendor}
\footnotesize
\setlength{\tabcolsep}{4pt}
\textbf{(a) Endpoint accuracies}\\[3pt]
\begin{tabular}{lrrrrr|r}
\toprule
Model & A (R0) & C (R0) & C+FX & C+FDR & C+FDRS &
\makecell{$\Delta$\\FDRS--A} \\
\midrule
\multicolumn{7}{l}{\textit{Closed-source API}} \\
GPT-5.4-mini          & 24.25 & 50.45 & 61.25 & 61.15 & \textbf{62.15} & $+37.90$ \\
Claude-Haiku          & 10.35 & 36.90 & 61.10 & 61.10 & \textbf{63.75}\,${\scriptstyle[61.70,\,65.85]}$ & $+53.40$ \\
DeepSeek-Flash        & 11.05 & 40.65 & 47.30 & 47.30 & \textbf{55.90} & $+44.85$ \\
Gemini-Flash          &  4.10 & 40.00 & 48.65 & 49.35 & \textbf{52.05} & $+47.95$ \\
\midrule
\multicolumn{7}{l}{\textit{Open-weight (locally deployable)}} \\
Llama-405B            & 13.45 & 51.45 & 64.75 & 64.75 & \textbf{68.70}\,${\scriptstyle[66.80,\,71.00]}$ & $+55.25$ \\
Qwen3-480B            & 15.05 & 56.65 & 63.80 & 63.75 & \textbf{65.90} & $+50.85$ \\
GPT-OSS-120B          & 14.05 & 45.65 & 58.40 & 58.35 & \textbf{60.25} & $+46.20$ \\
Qwen3-Next            &  4.00 & 47.20 & 55.00 & 57.95 & \textbf{59.55} & $+55.55$ \\
Llama-70B             &  5.00 & 38.15 & 49.00 & 51.45 & \textbf{57.45} & $+52.45$ \\
Qwen2.5-32B           &  3.80 & 42.70 & 50.75 & 52.75 & \textbf{52.75} & $+48.95$ \\
\bottomrule
\end{tabular}

\vspace{6pt}
\textbf{(b) Paired endpoint contrasts}\\[3pt]
\begin{tabular}{@{}lrrc@{}}
\toprule
Pair & $\Delta$ (pp) & $95\%$ CI & TOST ($\pm5$pp) \\
\midrule
Llama-70B vs.\ DeepSeek-Flash & $+1.55$ & $[-0.60, +3.75]$ & equiv \\
GPT-OSS-120B vs.\ GPT-5.4-mini & $-1.90$ & $[-3.90, +0.15]$ & equiv \\
GPT-OSS-120B vs.\ Claude-Haiku & $-3.50$ & $[-5.55, -1.45]$ & not equiv \\
Llama-405B vs.\ Claude-Haiku (upper tier) & $+4.95$ & $[+2.95, +6.95]$ & --- \\
\bottomrule
\end{tabular}

\smallskip
\begin{minipage}{\linewidth}
\footnotesize
\emph{Note}: paired item-level bootstrap on the C+FDRS endpoints,
$10{,}000$ resamples ($1{,}000$ upper tier); two one-sided tests
(TOST) equivalence margin $\pm5$pp.  Qwen3-480B--Claude-Haiku gap $+2.15$pp.
\end{minipage}
\end{table}

The direction is consistent across model families, scales, and
serving routes (Fig.~\ref{fig:cross_vendor_headline}).  Every
configuration improves after proactive guidance, and subsequent
repair preserves or extends that gain.  The result is therefore not
driven by one model family or one serving interface.

For local deployment, the central result is overlap at the tier
level.  Using the pre-set five-point margin for a practically relevant
difference, item-level comparisons place Llama-70B alongside
DeepSeek-Flash and GPT-OSS-120B alongside GPT-5.4-mini
(Table~\ref{tab:cross_vendor}b).  This does not imply parity with every
hosted model: both remain below Claude-Haiku, and Qwen2.5-32B lies only
at the lower edge of the API range.  Template-cluster resampling also
does not establish equivalence (Sec.~\ref{sec:exp:setup}).  The
evidence therefore supports overlap between the $70$B--$120$B local
tier and the observed mid-tier API range under matched no-tool
conditions, and stops short of a claim of universal parity.  Section~\ref{sec:exp:per_task}
examines where smaller models remain useful.

The largest open-weight models reach the top of the tested panel.
Llama-405B and Qwen3-480B both finish at or above the strongest
mid-tier API.  The Llama-405B--Claude-Haiku difference is $+4.95$pp
and remains positive under template clustering
($[+0.16,+10.34]$pp), although the task-family variant is marginal
($[-0.10,+10.63]$pp).  These results show that the intervention
continues to add accuracy at the largest scales.  For utilities,
however, the $70$B--$120$B tier remains the more relevant local
deployment range.
\subsection{Failure Mechanisms and the Remaining Workload Boundary}
\label{sec:exp:r0}

The intervention changes both how often programs succeed and what
remains to be fixed.  Without assistance, failures concentrate at the
API surface and many programs never complete a simulation.  As
documentation removes that barrier, workflow sequencing and semantic
grounding become the more visible limits.  This section traces that
shift from failure mechanism to workload.

\paragraph{Most first-pass failures occur before an analysis runs}
On the unassisted first pass, even the strongest open-weight model
executes successfully on fewer than one in three items
(Table~\ref{tab:proactive_main}).  Among models of $32$B or larger,
roughly half of the programs that execute return the accepted value,
yet only $4$--$15\%$ of all first attempts do so.  Variation in
first-pass accuracy therefore comes mainly from reaching execution,
not from value matching after execution.  Proactive documentation is
well aligned with this bottleneck because it can correct an API call
before any analysis runs.

A stratified, double-reviewed sample of $100$ failures shows what
prevents execution or correct analysis.  About three in five
classifiable cases involve the API surface.  Nearly half call a
nonexistent function, attribute, or import path (C1, $47$--$48$
cases), and a further $11$--$13$ misuse a real parameter or result
table (C2).  Workflow and domain-logic errors account for another $28$
cases (C3).  These include omitted analyses, use of the wrong network,
and time-series programs that replace a requested
\texttt{ConstControl} workflow with ad hoc updates.  The much smaller
remainder comprises format or environment failures (C5, $9$--$10$),
numerical extraction errors (C4, $1$--$2$), and one or two ambiguous
query--reference conflicts retained as task defects.

\begin{figure}[!htbp]
  \centering
  \includegraphics[width=\linewidth]{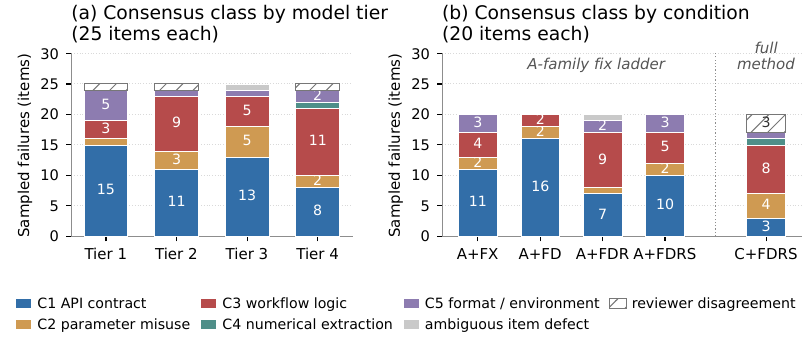}
  \caption{Consensus failure classes of the stratified $100$-failure
    double-review sample by (a)~model tier and (b)~prompt condition.
    T1 is sub-$7$B, T2 is $7$B--$32$B, T3 is $70$B--$120$B, and T4
    comprises Llama-405B, Qwen3-480B, and Qwen3-Next.
    Hatched segments are the three reviewer disagreements, reported
    rather than adjudicated.  Condition codes:
    Sec.~\ref{sec:exp:conditions}.}
  \label{fig:failure_taxonomy}
\end{figure}

The composition of the remaining failures changes as the API barrier
is reduced (Fig.~\ref{fig:failure_taxonomy}).  Among the $20$ sampled
residual failures under A+FDRS, $10$ are API-contract errors; under
C+FDRS, only $3$--$4$ fall in that class, and workflow errors
become the largest group.  The same reordering appears at the highest
model tier.  This is a change in the composition of residual failures,
not evidence that injection creates workflow errors or a direct
estimate of population failure rates.  The intervention helps a model
cross the API boundary, while the residual domain-logic errors call
for workflow-level methods.

The sample was drawn deterministically from $89{,}420$ recorded
failures and stratified across four model tiers, five prompt
conditions, and all $15$ task families.  Two practising power-systems
engineers labelled every record independently and agreed on $97$ of
$100$ cases (Cohen's $\kappa=0.955$).  The reported ranges preserve
the three disagreements.  Because each tier--condition cell contains
only five records, the within-cell shifts are interpreted
qualitatively.
\paragraph{Which workloads remain difficult}
\label{sec:exp:per_task}

The same pattern appears across workloads.  Routine power-flow,
DC-power-flow, and basic comparison tasks show the broadest first-pass
coverage across the panel.  Fault-current screening, SCADA-based
state estimation, post-contingency fault analysis, and parallel
scenario sweeps are close to zero for most unassisted models.  These
tasks depend on specialised calls such as
\texttt{pp.shortcircuit.calc\_sc} and
\texttt{pp.estimation.estimate}, which also receive low recognition
and application scores.  The task results and probe profiles therefore
locate the same weak parts of the API.

Signature recall is the probe layer most closely associated with
unassisted first-pass accuracy across models ($\rho_s\!=\!0.93$).
This association is descriptive rather than causal: the profile
identifies where a model struggles, while the component experiments
determine what the intervention changes.

\begin{figure}[!htbp]
  \centering
  \includegraphics[width=\linewidth]{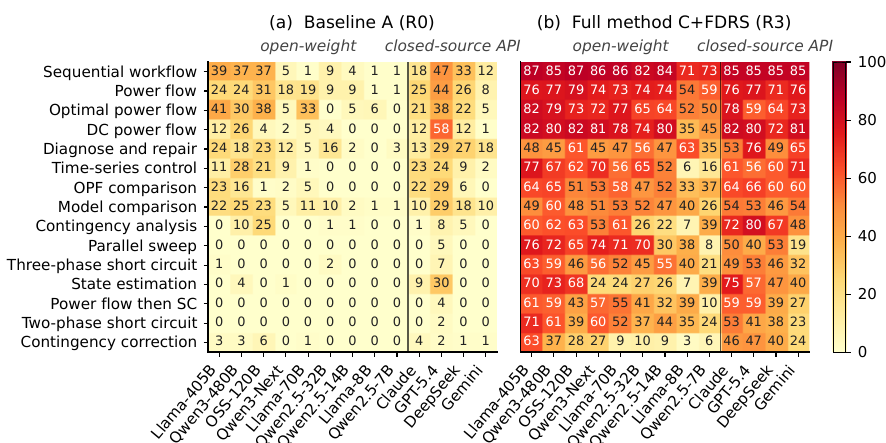}
  \caption{Per-task accuracy heatmap with identical row and column
    ordering in both panels for direct side-by-side comparison.
    \textbf{(a)} Baseline A (no proactive injection, R0).
    \textbf{(b)} Full method C+FDRS after $3$ fix rounds.  Rows are
    \pp{}-derived task families; columns split into open-weight models
    (left) and closed-source APIs (right).}
  \label{fig:per_task_heatmap}
\end{figure}

Fig.~\ref{fig:per_task_heatmap} shows the operational consequence by
comparing the unassisted baseline with the full method.  Short-circuit
and parallel-sweep families account for much of the panel-wide gain.
The difficulty-level gap is a semantic one.  Under
proactive condition C, every model performs better on the
explicit-parameter levels D1 and D2 than on the semantically grounded
levels D3 and D4.  Documentation can expose the relevant API contract,
but it cannot fully translate loosely specified engineering intent
into concrete networks, parameters, and analysis settings.  Semantic
grounding therefore remains a major constraint once the API
surface is supplied (full difficulty breakdown in SM Table~S9).

With injection, mid-size local models become useful for
explicit-parameter workflows.  Rare specialised and semantically grounded
requests still favour larger tiers or closer engineering review.  This
workload envelope informs the deployment guidance in
Sec.~\ref{sec:discussion:implications} and
Table~\ref{tab:deployment_guidance}.
\subsection{Component Contributions and Prompt Cost}
\label{sec:exp:intervention}

Most of the accuracy gain comes from selecting useful API evidence
before the first execution.  On the established \pp{} configuration,
the selective pipeline approaches the demand-hit full-layer reference
with less than half of its prompt tokens.  After execution, routing preserves
the accuracy of always-on documentation with about one fifth of its
documentation volume, while an independent validation signal recovers
additional wrong-value cases.  The component analyses below examine
these effects and their prompt cost.

\paragraph{Selective evidence supplies most of the first-pass gain}
\label{sec:exp:proactive}
Table~\ref{tab:proactive_main} compares the proactive conditions
before any repair occurs, so it isolates the evidence available to the
first program.  The comparison uses the full $2{,}000$-item benchmark
and the ten-model open-weight panel; the corresponding API results
appear in Table~\ref{tab:cross_vendor}.

\begin{table}[!htbp]
\centering
\caption{Round-0 result accuracy ($\%$) under all five proactive
  conditions and the dense vanilla-retrieval baseline
  R$_{\text{sem}}$ on the full $2{,}000$-item PowerCodeBench release,
  with the no-injection execution anatomy: ExecRate (execution rate)
  and Acc$|$Exec (accuracy conditioned on execution) of the
  condition-A first pass, as fractions (metrics defined in
  Sec.~\ref{sec:exp:setup}).
  Condition codes: Sec.~\ref{sec:exp:conditions} (retrieval
  protocols in SM Secs.~S5.4 and~S6.2).}
\label{tab:proactive_main}
\footnotesize
\setlength{\tabcolsep}{2.6pt}
\begin{tabular}{lrr|rrrrrr}
\toprule
Model & \makecell{ExecRate\\(A)} & \makecell{Acc$|$Exec\\(A)} & A & B & C & X & R & R$_{\text{sem}}$ \\
\midrule
Qwen2.5-1.5B         & 0.000 & ---   &  0.00 &  0.00 &  8.10 &  1.75 &  0.20 &  0.10 \\
Qwen2.5-7B           & 0.035 & 0.116 &  0.40 &  1.70 & 23.15 & 24.15 &  3.75 &  6.00 \\
Qwen2.5-14B          & 0.040 & 0.400 &  1.60 & 15.60 & 34.20 & 39.35 &  6.85 &  9.30 \\
Qwen2.5-32B          & 0.079 & 0.484 &  3.80 & 14.90 & 42.70 & 46.05 & 11.10 & 10.50 \\
Llama-8B             & 0.017 & 0.294 &  0.50 &  1.30 & 20.10 & 20.40 &  1.20 &  1.30 \\
Llama-70B            & 0.095 & 0.526 &  5.00 &  5.20 & 38.15 & 41.60 & 10.45 & 11.10 \\
Llama-405B           & 0.262 & 0.513 & 13.45 &  4.20 & 51.45 & 47.45 & 20.85 & 23.05 \\
Qwen3-Next           & 0.076 & 0.530 &  4.00 & 15.60 & 47.20 & 52.15 & 11.15 & 14.20 \\
Qwen3-480B           & 0.290 & 0.520 & 15.05 & 21.55 & 56.65 & 59.45 & 23.10 & 22.75 \\
GPT-OSS-120B         & 0.265 & 0.531 & 14.05 &  9.65 & 45.65 & 48.95 & 21.20 & 22.20 \\
\midrule
\textit{Mean}        & ---   & ---   &  5.79 &  8.97 & 36.74 & 38.13 & 10.99 & 12.05 \\
\bottomrule
\end{tabular}
\end{table}

Selective layer-wise injection raises panel-mean first-pass accuracy
from $5.8\%$ to $36.7\%$, a gain of $31.0$ percentage points
($95\%$ CI $29.8$--$32.1$), and every model improves.
Function names alone are not a reliable substitute: they help several
mid-size Qwen models but reduce accuracy on the strongest
general-purpose models.  Names without semantic context can therefore
do harm.  Condition C finishes within $1.4$ points
of the demand-hit full-layer reference X, whose advantage changes sign
across models.  X is consequently a high-context reference rather
than a strict upper bound.

\label{sec:exp:bm25}
Standard retrieval over the same documentation is less effective.
The BM25 and dense semantic baselines average $11.0\%$ and $12.1\%$
accuracy, respectively, compared with $36.7\%$ for C.  Both use the
same documentation corpus and rendering path but omit the complete
demand- and profile-guided selection process.  On the unrounded panel
means, C leads BM25 by $25.75$ percentage points.  The difference is
not a prompt-budget artifact: when dense retrieval is calibrated to
C's realised token cost, it falls further on seven of the ten models
(SM Sec.~S5.4).

\paragraph{The profile supports diagnosis and evidence allocation}
Signature recall has the strongest rank association with bare
first-pass accuracy across the panel ($\rho_s=0.93$), whereas
application-level competence does not increase monotonically with
model size.  The profile therefore supports model selection alongside
parameter count and provides an inspectable, model-specific allocation
state.  In a same-nominal-budget check across three deployment-scale
models, profile conditioning changed realised prompt length on
$33$--$64\%$ of items.  The panel-mean accuracy difference from the
constant-deficit control ($r_{\mM,\ell}(f)=1$, maximal deficit) was at
most $1.2$ points, and every
paired confidence interval included zero.  On the established \pp{}
configuration, the measured contribution of profile conditioning is
therefore diagnosis and evidence allocation rather than a separately
detected accuracy increment; the first-pass gain and the $41\%$
prompt-cost result belong to the complete selective pipeline.  On
OpenDSS, the initial profile-to-allocation mapping under-supplied name
and signature evidence; a same-pilot priority-floor adjustment
corrected that allocation mismatch (Sec.~\ref{sec:exp:transfer}).
Complete isolation results are reported in SM Sec.~S5.3.

\begin{figure}[!htbp]
  \centering
  \includegraphics[width=\linewidth]{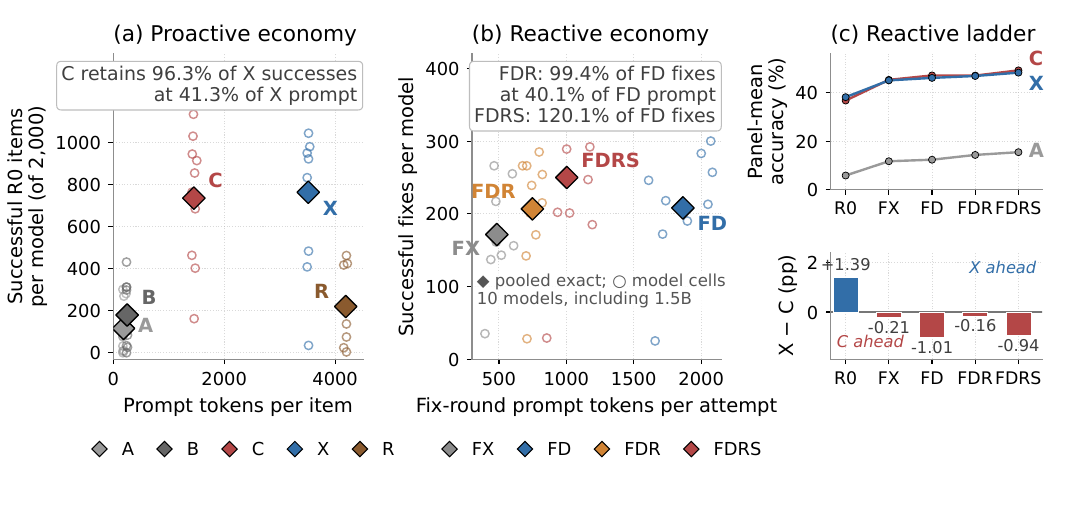}
  \caption{Cost--yield planes behind Table~\ref{tab:token_economy}:
    (a) proactive Round-0 and (b) reactive fix-round economy
    (diamonds: pooled panel values; circles: per-model cells);
    (c) reactive accuracy ladder (Table~\ref{tab:reactive_master})
    and the per-round $\Delta$(X--C) gap.  Condition codes:
    Sec.~\ref{sec:exp:conditions}.}
  \label{fig:token_economy_compact}
\end{figure}

\label{sec:exp:token}
\paragraph{Prompt cost is driven by evidence selection}
C remains within about $1.4$ accuracy points of X while using $1{,}451$
prompt tokens on average against X's $3{,}513$, or $41\%$ of its prompt
cost, a saving of $2{,}062$ tokens per item.
BM25 occupies the opposite corner of Table~\ref{tab:token_economy}:
it uses more context than X but returns fewer matches than the
selective condition C.  Dense retrieval also remains behind C when its
measured cost is matched.  Table~\ref{tab:token_economy} and
Fig.~\ref{fig:token_economy_compact}a therefore locate the economy in
selecting useful evidence, not simply in shortening the prompt.
\paragraph{Execution feedback cannot supply an unfamiliar API contract}
\label{sec:exp:fix_trajectory}

Self-debugging repairs straightforward code errors, but for most
models the largest gain arrives in the first retry.  Models of $70$B and above gain
$3$--$15$ points over three rounds, whereas the $7$B model barely
changes and the $1.5$B reference remains at zero.  Execution rates
also rise faster than result accuracy: many repaired programs run
without returning the requested quantity.  Useful repair therefore
presupposes enough library knowledge to interpret the failing code
(SM Table~S7).

\paragraph{Reactive routing separates two repair needs}
\label{sec:exp:reactive}

The reactive conditions separate two questions: whether another
attempt needs documentation, and whether an independent validation
signal has exposed a wrong value.  FDR answers the first question from
execution evidence alone; FDRS adds the second route when such a
signal is available.  Table~\ref{tab:reactive_master} compares these
choices under each proactive base.  SM Table~S8 gives the open-weight
per-model breakdown for the deployed combination C+FDRS.

\begin{table}[!htbp]
\centering
\caption{Mean accuracy ($\%$) at Round~0 (standalone proactive runs) and after up to three repair rounds over the ten-model open-weight
  panel (1.5B--480B) on the full $2{,}000$-item benchmark.  Rows
  compare proactive bases; the three rightmost columns isolate
  always-on documentation, reference-free routing, and the
  validation-triggered value-error branch.  The bottom rows track the
  C--A and X--C gaps across the repair ladder.  Condition codes:
  Sec.~\ref{sec:exp:conditions}.}
\label{tab:reactive_master}
\footnotesize
\setlength{\tabcolsep}{4pt}
\resizebox{\linewidth}{!}{%
\begin{tabular}{lrrrrr|rrr}
\toprule
Proactive base & R0 & FX & FD & FDR & FDRS &
\makecell{$\Delta$\\FD--FX} & \makecell{$\Delta$\\FDR--FD} & \makecell{$\Delta$\\FDRS--FDR} \\
\midrule
A (no proactive)   &  5.79 & 11.67 & 12.31 & 14.28 & 15.42 & $+0.64$ & $+1.97$ & $+1.14$ \\
C (layer-wise)     & 36.74 & 45.27 & 47.10 & 47.03 & 49.19 & $+1.83$ & $-0.07$ & $+2.16$ \\
X (full-layer reference)  & 38.13 & 45.06 & 46.08 & 46.87 & 48.25 & $+1.02$ & $+0.79$ & $+1.38$ \\
\midrule
$\Delta$(C--A)       & $+30.95$ & $+33.60$ & $+34.79$ & $+32.75$ & $+33.77$ &  &  &  \\
$\Delta$(X--C)       &  $+1.39$ & $-0.21$ & $-1.02$ & $-0.16$ & $-0.94$ &  &  &  \\
\bottomrule
\end{tabular}
}
\end{table}

Plain error feedback raises accuracy by $5.9$--$8.5$ points because
many syntax and local code errors need no additional documentation.
Supplying documentation on every retry adds comparatively little.  On
base C, reactive routing matches always-on documentation
($\Delta$(FDR--FD)$=-0.07$ points; $95\%$ CI $[-0.44,+0.32]$),
showing that documentation can be withheld from local failures without
reducing accuracy.  When independent validation is available, FDRS
adds $2.16$ points over FDR ($95\%$ CI $[+1.82,+2.52]$) by targeting
executed programs with wrong values.  The same direction appears
across the four API models, so the value-error branch is not specific
to one serving route.

The routing logs explain this pattern.  Once proactive evidence has
established the library context, a substantial share of repair
attempts can be handled without further documentation, and this share
generally rises among the stronger models.  Under FDRS, the
validation-triggered route accounts for a larger share of fix attempts
on capable models and carries most of their additional gain (SM
Table~S11).

\begin{table}[!htbp]
\centering
\caption{Token economy in one view: proactive Round-0 (top) and reactive
  fix-round (bottom) conditions over the ten-model open-weight panel
  (column conventions in the note below).  Condition codes:
  Sec.~\ref{sec:exp:conditions}.}
\label{tab:token_economy}
\footnotesize
\setlength{\tabcolsep}{4pt}
\begin{tabular}{lrrrrr}
\toprule
Condition & \makecell{Added/docs\\(avg)} & \makecell{Prompt\\(avg)} & Successes & Tok./succ. & vs X \\
\midrule
\multicolumn{6}{l}{\textit{Round-0 (proactive injection)}} \\
A (no injection)        &     0 &   182 & 115.7 / 2k & 3{,}149 &   5\% \\
B (names only)          &    65 &   248 & 179.4 / 2k & 2{,}760 &   7\% \\
  \textbf{C (full proactive)} & \textbf{1{,}269} & \textbf{1{,}451} & \textbf{734.7 / 2k} & \textbf{3{,}950} & \textbf{41\%} \\
X (full-layer reference)    & 3{,}331 & 3{,}513 & 762.6 / 2k & 9{,}214 & 100\% \\
R (BM25 RAG)            & 4{,}007 & 4{,}189 & 219.7 / 2k & 38{,}133 & 119\% \\
\midrule
\multicolumn{6}{l}{\textit{Round 1--3 (reactive correction, base C)}} \\
C+FX (self-debug)       &     0 &   484 &  171.4 fixes &  9{,}881 &  14\% \\
C+FD (always-doc)       & 1{,}375 & 1{,}864 &  208.0 fixes & 30{,}918 &  53\% \\
C+FDR (routed)          &   262 &   748 &  206.7 fixes & 12{,}427 &  21\% \\
\textbf{C+FDRS}          & \textbf{518} & \textbf{1{,}003} & \textbf{249.9 fixes} & \textbf{13{,}551} & \textbf{29\%} \\
\bottomrule
\end{tabular}

\smallskip
\begin{minipage}{\linewidth}
\footnotesize
\emph{Note}: exact token totals are pooled across the panel:
averages divide by the corresponding items or fix attempts, and
``Tok./succ.'' divides by the pooled matched items (proactive) or
successful fixes (reactive).  ``Successes'' is that pooled count
divided by ten, giving an interpretable per-model scale.  Reactive successes are measured relative to each repair run's own Round-0 snapshot, which can differ slightly from the standalone proactive runs.  At R0,
``Added/docs'' is the prompt increase over A (documentation plus its
framing); in fix rounds it is the recorded retrieved-document volume.
``vs X'' is the prompt-token ratio to X's R0 mean of $3{,}513$.
Boldface marks the deployed configurations; displayed values are
rounded only after these pooled calculations.
\end{minipage}
\end{table}

Routed repair converts this diagnosis into lower context cost.  FDR
uses $19\%$ of FD's documentation tokens while reaching essentially
the same final accuracy.  FDRS uses $38\%$ and exceeds FD by $2.1$
points.  Plain self-debugging is cheaper, but it produces fewer
successful fixes and lower final accuracy, so it is not a
quality-equivalent substitute (Fig.~\ref{fig:token_economy_compact}b--c).
The proactive advantage of C over A persists under every repair
condition; after repair, C also slightly exceeds X while retaining its
lower prompt cost.

\paragraph{Demand adaptation follows the workload mix}
\label{sec:exp:demand}
The demand model must rank useful API evidence for queries whose wording it has
never seen.  Hybrid TF--IDF is the only tested estimator that
improves on zero-shot TF--IDF for benchmark-reference queries while
remaining strong on the augmented held-out set.  Cross-encoder and
pairwise reranking transfer less well under this wording shift.
Role-frequency reweighting then moves attention from construction
towards analysis, gaining $20.7$ points of benchmark-reference
recall@10 at a cost of $5.1$ points on source-style queries (SM
Tables~S4(a)--(b)).  Because this trade depends on the workload's role
mix, the next section tests the adapted and unadapted operating points
on human-written and external queries.

\subsection{Robustness to Phrasing, Workload Mix, and Reasoning Modes}
\label{sec:exp:robustness}

Most of the proactive gain is present without workload-specific demand
adaptation, and the full C+FDRS gain persists when benchmark templates
are replaced by human operator language.  A separate test shows that
the demand estimator also recognises useful APIs in queries
collected outside the benchmark.  Paid reasoning leaves a large
intervention benefit as well.  The main qualification concerns
scenario adaptation: it helps analysis-heavy workloads but can hurt
construction-heavy traffic, so it is best treated as a tunable deployment
setting.

\label{sec:exp:demand_robustness}
Without role-frequency adaptation, condition C still gains $27.8$
percentage points over no injection.  Adaptation adds only a further
$3.2$ points, even though benchmark-reference demand recall rises
from $51.2\%$ to $71.8\%$.  The intervention's main benefit therefore
does not depend on target-workload reweighting.  The per-model results
and a separate cross-style check support the same
reading (SM Fig.~S2(c) and Sec.~S4.3).

\paragraph{The end-to-end gain survives human operator phrasing}
\label{sec:exp:naturalistic_holdout}
Human-written phrasing preserves the same improvement.  On an
independent $80$-item holdout over the same underlying scenarios,
C+FDRS improves every open-weight model of at least $7$B, with a
panel-mean gain of $34.6$ points.  The queries preserve the task,
difficulty, and ground-truth quantity but reuse none of the template
wording; semantic-intent items describe the intended operation rather
than its literal modification.  On average, C+FDRS endpoints are
$6.8$ points lower than on the full benchmark and bare-prompt
endpoints $5.8$ points higher, with the largest shifts on small or
code-specialised models.  The broad model-tier ordering is preserved.
Holding the scenarios fixed makes this a direct test of phrasing
robustness (SM Table~S10).

\label{sec:exp:external_queries}
\paragraph{The demand model transfers, but adaptation follows workload mix}
The demand model also identifies useful API calls outside the
benchmark query pool.  Without role adaptation, its top-ten
suggestions contain $49$--$65\%$ of the required calls across the
three external sets and contain at least one required call for
$83$--$100\%$ of the queries.  These levels bracket the internal
holdout, where the corresponding values are $56\%$ and $93\%$
(Table~\ref{tab:external_queries}).

The three external sets were built to sit as far from the benchmark
as practitioner data allows.  The first contains $84$ controlled
rewrites of real requests posted on the pandapower issue tracker and
forum, screened independently by two practising power engineers under
frozen inclusion rules; $65\%$ of these requests target non-standard
or custom-built networks outside the benchmark's case pool.  The
second is a $10$-query supplement drawn from a separately screened
pool of six \pp{} GitHub Discussions, two Stack Overflow questions,
and two openmod-forum topics; the Stack Overflow and openmod items
come from platforms outside the first set's sources.  The third comprises $20$
requests written from scratch by the same two engineers (a separate
pair from the failure-taxonomy annotators), whose approved edition is
the authoritative text.  Every item carries an executable reference
solution with oracle API labels extracted by the same engine as the
benchmark, so the recall metric applies unchanged.  The estimator and
its adaptation target were fixed before any external query was
collected, and each set was frozen before evaluation (construction
protocols and screening ledgers in SM Sec.~S4.4, Tables~S5--S6).

\begin{table}[!htbp]
\centering
\footnotesize
\caption{External-query evaluation of the demand model, fixed before
  any external query was scored.  Each
  entry reports the fraction of required calls retrieved in the top
  ten / the fraction of queries with at least one required call
  retrieved ($\%$).  The three external sets contain $114$ queries;
  the $80$-item internal holdout tests the same metrics on
  human-written benchmark scenarios.  $\Delta$: adapted $-$
  unadapted top-ten recall (pp), computed from unrounded values.}
\label{tab:external_queries}
\setlength{\tabcolsep}{4pt}
\begin{tabular}{@{}lrccr@{}}
\toprule
Query set & $n$ & Unadapted & Adapted & $\Delta$ (pp) \\
\midrule
Internal holdout (Sec.~\ref{sec:exp:naturalistic_holdout}) & 80 & 55.5 / 92.5 & 68.9 / 98.8 & $+13.5$ \\
Community-sourced rewrites & 84 & 55.1 / 83.3 & 53.4 / 97.6 & $-1.6$ \\
Cross-platform supplement & 10 & 64.7 / 100.0 & 50.0 / 100.0 & $-14.7$ \\
Engineer-written & 20 & 48.7 / 90.0 & 64.5 / 95.0 & $+15.8$ \\
\bottomrule
\end{tabular}
\end{table}

Adaptation is not uniformly beneficial.  It raises recall@10 by
$15.8$ points on the analysis-heavy engineer-written set, but lowers
recall slightly on the construction-heavy community set, even as the
chance of retrieving at least one required call increases.  The
$10$-item supplement points in the same direction but is too small for
a precise estimate.  In the fitted workload profile, reweighting
towards analysis downweights construction APIs.  Within these sets, a
provisional operating rule is to enable adaptation when expected
traffic resembles the analysis-heavy profile and leave it off for
construction-heavy traffic.  A query-conditional
selector is promising on the mixed-role community set, but remains
exploratory pending validation on new mixed-role distributions (SM
Table~S6(a)).

\paragraph{Paid reasoning complements API evidence}
\label{sec:exp:cross_vendor_reasoning}
Additional internal reasoning can raise absolute accuracy, but it does
not replace library evidence.  We test two reasoning-enabled APIs on
a paired $200$-item subset stratified across task families and compare
them with their vanilla endpoints and the strongest open-weight
reference (Table~\ref{tab:reasoning_tier}).

\begin{table}[!htbp]
\centering
\footnotesize
\caption{Reasoning-tier check on the paired $200$-item stratified
  subset: C+FDRS accuracy and R0$\to$C+FDRS lift for two
  reasoning-tier endpoints, their vanilla counterparts, and the
  strongest open-weight reference.
  Condition codes: Sec.~\ref{sec:exp:conditions}.}
\label{tab:reasoning_tier}
\begin{tabular}{@{}lcc@{}}
\toprule
Endpoint ($200$-item subset) & C+FDRS (\%) & R0$\to$C+FDRS (pp) \\
\midrule
GPT-5.4-mini (\texttt{reasoning\_effort=high}) & $68.00$ & $+27.50$ \\
GPT-5.4-mini (vanilla)                         & $59.50$ & $+34.50$ \\
DeepSeek-Reasoner                              & $56.00$ & $+44.00$ \\
DeepSeek-Flash (vanilla)                       & $53.00$ & $+42.00$ \\
Llama-405B (open-weight reference)             & $65.50$ & --       \\
\bottomrule
\end{tabular}
\end{table}

Both reasoning endpoints still receive large gains from C+FDRS.
GPT-5.4-mini-high falls within the observed range of the Llama-405B
reference ($\Delta=+2.5$ points; $95\%$ CI $[-4.0,+9.0]$), whereas
DeepSeek-Reasoner remains $9.5$ points lower.  Within this two-vendor,
$200$-item check, with tools disabled, reasoning and boundary-aware
intervention are complementary.  The transfer pilot next changes the
library, documentation, and probe corpus.
\subsection{Onboarding New Power-System Libraries}
\label{sec:exp:transfer}

The Round-0 probing and proactive-allocation path ran on both OpenDSS
and PyPSA after their backend-facing materials were regenerated.
Selective injection improved on a bare prompt in five of the six
library--model combinations, but it did not outperform broad retrieval
at cold start.  These pilots support beginning with broad retrieval
and switching to selective injection only after the allocation mapping
has been checked against complete programs on the new library.

OpenDSS covers three-phase distribution analysis, whereas
PyPSA~\cite{brown2018pypsa} covers transmission-level modelling with
native optimisation.  We accessed OpenDSS through OpenDSSDirect.  Its
Python binding retains the stateful, namespace-style getter and setter
organisation of the OpenDSS Component Object Model (COM) API.  For example,
\texttt{RegControls.Name} selects the active regulator,
\texttt{RegControls.ForwardVreg} reads or changes its voltage set
point, and \mbox{\texttt{RegControls.}}\allowbreak\mbox{\texttt{TapNumber}} returns its tap position.
We refer to this design as the OpenDSS accessor API.

For each library, we regenerated a compact documentation
specification, its probes and snippets, a frozen $90$-item
mini-benchmark, and a zero-shot demand ranker.  The intervention
policy, packer, and benchmark pipeline were unchanged (SM Sec.~S9).
Qwen2.5-32B, Llama-70B, and DeepSeek-Flash represent a local mid-tier
model, a larger local model, and a hosted API.  Each was compared under
the bare condition A, selective injection C, and broad dense retrieval
R$_{\text{sem}}$ over the same documentation and renderer.

\begin{table}[!htbp]
\centering
\footnotesize
\caption{Cross-backend Round-0 (first-pass) results on the frozen
  $90$-item mini-benchmarks.  At cold start, selective injection improves on the
  bare prompt in five of six cells but does not outperform broad
  retrieval in this pilot.  Panel~(b) reports a targeted OpenDSS
  recalibration that gives the name and signature documentation of
  requested functions a priority floor in the budgeted packer.
  Differences are paired
  item-level percentage points with $95\%$ bootstrap CIs
  ($10{,}000$ resamples; $^{*}$: CI excludes zero).  Condition codes:
  Sec.~\ref{sec:exp:conditions}; full $24$-cell matrix in SM
  Table~S18.}
\label{tab:transfer_pilot}
\setlength{\tabcolsep}{3pt}

\textbf{(a) Cold-start instantiation}\\[3pt]
\begin{tabular}{@{}llrrr@{\hspace{6pt}}rr@{}}
\toprule
Backend & Model & A & C & $\Delta$(C--A) &
R$_{\text{sem}}$ & $\Delta$(C--R$_{\text{sem}}$) \\
\midrule
\multirow{3}{*}{OpenDSS}
 & Qwen2.5-32B    & 17.8 & 15.6 &
 $-2.2\,{\scriptstyle[-10.0,+5.6]}$ &
 66.7 & $-51.1\,{\scriptstyle[-63.3,-38.9]}^{*}$ \\
 & Llama-70B      &  3.3 & 36.7 &
 $+33.3\,{\scriptstyle[+23.3,+43.3]}^{*}$ &
 63.3 & $-26.7\,{\scriptstyle[-37.8,-15.6]}^{*}$ \\
 & DeepSeek-Flash &  6.7 & 37.8 &
 $+31.1\,{\scriptstyle[+21.1,+41.1]}^{*}$ &
 75.6 & $-37.8\,{\scriptstyle[-50.0,-24.4]}^{*}$ \\
\midrule
\multirow{3}{*}{PyPSA}
 & Qwen2.5-32B    & 40.0 & 57.8 &
 $+17.8\,{\scriptstyle[+3.3,+32.2]}^{*}$ &
 54.4 & $+3.3\,{\scriptstyle[-7.8,+14.4]}$ \\
 & Llama-70B      &  5.6 & 50.0 &
 $+44.4\,{\scriptstyle[+33.3,+55.6]}^{*}$ &
 60.0 & $-10.0\,{\scriptstyle[-23.3,+2.2]}$ \\
 & DeepSeek-Flash & 68.9 & 81.1 &
 $+12.2\,{\scriptstyle[+1.1,+23.3]}^{*}$ &
 94.4 & $-13.3\,{\scriptstyle[-22.2,-5.6]}^{*}$ \\
\bottomrule
\end{tabular}

\vspace{6pt}
\textbf{(b) OpenDSS name and signature recalibration}\\[3pt]
\begin{tabular}{@{}lrrrrr@{}}
\toprule
Model & C & C-recal & $\Delta$(C-recal--C) &
R$_{\text{sem}}$ & $\Delta$(C-recal--R$_{\text{sem}}$) \\
\midrule
Qwen2.5-32B
 & 15.6 & 41.1
 & $+25.6\,{\scriptstyle[+14.4,+36.7]}^{*}$
 & 66.7 & $-25.6\,{\scriptstyle[-38.9,-12.2]}^{*}$ \\
Llama-70B
 & 36.7 & 44.4
 & $+7.8\,{\scriptstyle[-1.1,+16.7]}$
 & 63.3 & $-18.9\,{\scriptstyle[-31.1,-6.7]}^{*}$ \\
DeepSeek-Flash
 & 37.8 & 68.9
 & $+31.1\,{\scriptstyle[+17.8,+44.4]}^{*}$
 & 75.6 & $-6.7\,{\scriptstyle[-15.6,+2.2]}$ \\
\bottomrule
\end{tabular}
\end{table}

Table~\ref{tab:transfer_pilot} separates portability from policy
choice.  After the backend-facing materials were regenerated,
probing, profile construction, and proactive allocation ran on
both libraries.  Against the bare prompt, selective injection gains
$12$--$44$ percentage points in five cells.  Against broad retrieval,
it is lower in four cells, while the other two comparisons remain
unresolved at $n=90$.  Broad retrieval supplies much broader coverage
at cold start, whereas selective injection uses roughly one tenth as
many documentation tokens.

Giving retrieval the same prompt budget as C does not reverse this
pattern.  Four
comparisons remain unresolved and two OpenDSS cells still favour
retrieval.  The cold-start deficit therefore depends on which evidence
is selected, not only on how much context is supplied.  On \pp{}, by
contrast, a larger specification, an in-domain demand predictor, and
a calibrated link between profiles and documentation allow C to lead
the same retrieval baseline by $8$--$34$ points
(Table~\ref{tab:proactive_main}).  Broad retrieval is the safer
starting policy while a model's needs on a new API are still being
learned.

OpenDSS shows why the allocation mapping requires calibration.
Qwen2.5-32B reproduces
$72$ of the $76$ accessor paths in isolated probes, so its profile
calls for little name and signature documentation.  In complete
control-device programs, however, $17$ of $18$ selectively assisted
outputs fail because they call members that do not exist, compared
with six under broad retrieval.  Examples include \texttt{Tap} in
place of \mbox{\texttt{TapNumber}}, and \texttt{Bus.kvar} in place of
\texttt{Circuit.TotalPower}.  Isolated path recall therefore does not
show that a model will select and reproduce the right accessor inside
a complete program.  Llama-70B shows the reverse pattern: its weaker
profile triggers more documentation and reduces
the same failures.

We used the frozen base results to define one correction before
generating any recalibrated outputs: for every function a query needs,
the allocator now gives that function's name and signature
documentation a priority floor in the budgeted packer, even when the
probes suggest the model already knows it.  On the same
OpenDSS items, accuracy rises by $26$ points for Qwen2.5-32B and $31$
points for DeepSeek-Flash; the Llama-70B change remains unresolved.
All $18$ Qwen2.5-32B control-device programs then execute, and the
recalibrated prompts use $27$--$29\%$ of the context used by broad
retrieval.  This same-pilot test shows that the diagnosed allocation
error was actionable, while leaving transfer of the adjustment to
other APIs open.

Together, the two backend pilots support a staged onboarding
sequence: start with broad retrieval; use outcomes on complete programs
to calibrate how profile state maps to injection; and switch to
selective injection once it has been shown, on that library's own
programs, to match retrieval at lower prompt cost.

\subsection{Do Matching Results Come from the Requested Analysis?}
\label{sec:exp:validity_audit}

Most matching results come from the requested analysis, and the full
method raises this share.  A matching number cannot reveal whether the
program used the intended network, changes, solver, and result table, so
we audit the workflow directly.

The clearest comparison uses the same model and benchmark item and
considers only cases in which both methods returned a matching number.
Across these $170$ pairs, the requested analysis was performed by
$87.6\%$ of the full-method programs, compared with $65.9\%$ of the
baseline programs.  Within this comparison, the full method avoided
nearly two thirds of the baseline programs that numerical scoring
accepted even though they followed the wrong analysis.
The difference is $21.8$ percentage points ($95\%$ CI $14.7$--$28.8$); in the vocabulary of the audit, false acceptance falls from $34.1\%$ to $12.4\%$.

This improvement does not make a matching number sufficient for
unattended use.  In a broader sample designed to represent the deployed
open-weight configuration, an estimated $81.5\%$ of matching outputs
performed the requested analysis.  Put more plainly, roughly one in
five still reached the expected number through an unfaithful workflow.
The remaining errors involve wrong network or solver selection, omitted
modifications, index or sign mistakes, and extraction from the wrong
result table or column.  The injected evidence targets these API choices,
which explains why its benefit extends beyond execution and numerical
matching.  Critical outputs still require expert review; two controls
that reduce or route the residual risk are summarised in
Table~\ref{tab:validity_audit_summary}.

\begin{table}[!htbp]
\centering
\footnotesize
\caption{Practical controls supported by the workflow audit.  These
  controls can reduce or route residual risk; neither certifies a
  generated program.  Exact estimates and complete screening results
  are reported in SM Sec.~S7.}
\label{tab:validity_audit_summary}
\setlength{\tabcolsep}{5pt}
\renewcommand{\arraystretch}{1.10}
\begin{tabularx}{\linewidth}{@{}
  >{\raggedright\arraybackslash}p{0.15\linewidth}
  >{\raggedright\arraybackslash}X@{}}
\toprule
Control & What the evidence supports \\
\midrule
Stricter numerical comparison &
When recorded results are rescored with a tenfold tighter tolerance,
about two in five programs that reached a matching number through the
wrong analysis are filtered out, while fewer than two in one hundred
faithful matches are lost.  This is a useful first filter when a trusted
comparison value is available. \\
\addlinespace
Automatic program checks &
The five re-execution checks detect about nine in ten cases in which a
matching number came from the wrong analysis, but fewer than half of
their flags are confirmed as errors.  They
can prioritise expert review, but should not approve or reject programs
automatically. \\
\bottomrule
\end{tabularx}
\end{table}

The stratified draw comprised $610$ numerically accepted outputs and
the paired arm $170$ matched pairs ($340$ outputs, $10$ of which also
belong to the draw).  The checks covered
the network and requested changes, solver behaviour, physical coherence,
result extraction, and whether the printed answer came from the
simulation.  Programs were re-run while these properties were recorded.
Two practising power-systems engineers independently reviewed records
flagged by the automatic checks.
Records that were not flagged were also examined separately rather than
assumed correct.  A complementary audit of executed outputs rejected by
the numerical matcher found that the confirmed misses arose from result
presentation, not from a different physical analysis.  SM Sec.~S7 gives
the complete sampling, review, and statistical procedures.

\FloatBarrier
\section{Discussion}
\label{sec:discussion}

Deployment choices also depend on serving resources, engineering
review, and the limits of the evaluation.

\subsection{Implications for Intelligent Energy Systems}
\label{sec:discussion:implications}

Serving measurements change the model ranking because parameter count
alone does not determine the best local tier.  On the measured subset,
Qwen3-Next matches the $120$B MoE's accuracy on two GPUs instead of
four and uses less than half the energy per successful task.  The
$120$B model has the higher total-token throughput, while the $32$B tier
returns its first token sooner.  Validation-enabled repair raises
accuracy further, but at a much higher energy cost.

The benchmark spans routine power-flow checks, short-circuit analysis,
state estimation, and multi-scenario contingency sweeps.  We measured
token throughput, energy, and aggregate allocated GPU memory directly
for eleven open-weight configurations, and online latency for five
mid-size configurations.  Table~\ref{tab:deployment_guidance}
combines these serving results with the workload envelope and review
requirements.

\begin{figure}[!htbp]
  \centering
  \includegraphics[width=\linewidth]{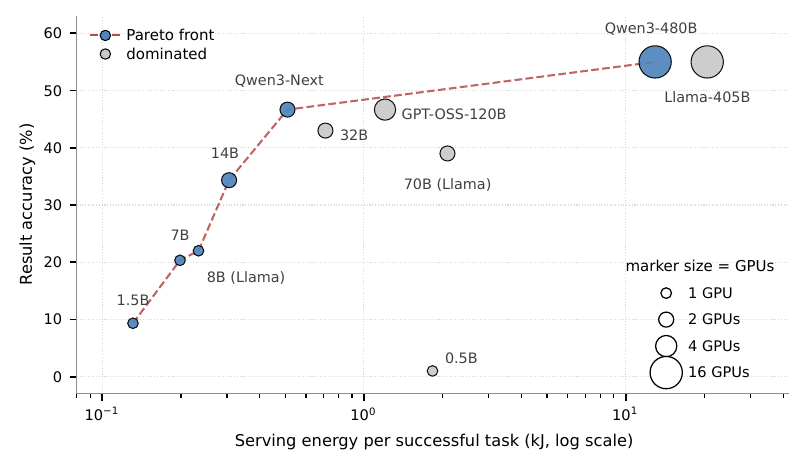}
  \caption{Single-pass condition-C accuracy--energy serving plane for the eleven
    configurations of Table~\ref{tab:serving_metrics} (marker size
    encodes GPU count; energy per successful task on a log axis).
    The dashed line traces the six-point Pareto frontier; grey
    points are dominated.}
  \label{fig:serving_pareto}
\end{figure}

\paragraph{Measured serving footprint}
Table~\ref{tab:serving_metrics} and
Fig.~\ref{fig:serving_pareto} compare accuracy and resource use at the
same operating point, single-pass condition C.  Qwen3-Next reaches
$46.7\%$ subset accuracy at $0.51$\,kJ per successful task.  The
four-GPU $120$B MoE reaches the same accuracy at $1.20$\,kJ, although
it delivers $28\%$ higher total-token throughput.  The $405$B and $480$B
anchors reach $55\%$, but were measured on sixteen allocated GPUs and
with a much larger energy budget.

The complete C+FDRS endpoint occupies a different operating point.
Across the measured tiers of at least $7$B, repair adds $8$--$23$
percentage points of accuracy but uses about $5$--$15$ times the
single-pass energy because it may execute three repair rounds.  The
choice of endpoint therefore depends on whether the additional
accuracy justifies the repair cost.

Interactive serving produces a different ordering.  At
concurrency~$16$, Qwen3-Next has $20\%$ higher output-token throughput than the
two-GPU $32$B tier while using $30\%$ less energy per request and
with $15\%$ lower p95 end-to-end latency.  The $32$B model
nevertheless returns its first token much faster, with a p95 of
$0.09$\,s against $0.70$\,s.
Batch-oriented and first-token-sensitive services may therefore
choose different tiers.

\begin{table}[!htbp]
\centering
\footnotesize
\caption{Measured single-pass serving configurations under condition C
  for eleven open-weight models on an NVIDIA GH200 cluster.  GPU counts
  are the tested tensor-parallel allocations, and VRAM is peak aggregate
  allocated memory under vLLM's $0.9$ GPU-memory-utilisation setting.  The
  offline-batch measurements use a $300$-item subset and report the
  median of three repetitions.  The $0.5$B row is a serving-only floor
  reference outside the accuracy panel.}
\label{tab:serving_metrics}
\setlength{\tabcolsep}{4pt}
\resizebox{\linewidth}{!}{%
\begin{tabular}{llrrrrr}
\toprule
Tier & Model & \makecell{Allocated\\GPUs (TP)} & \makecell{Acc.\\(\%)} &
\makecell{Throughput\\(tok/s)} &
\makecell{Energy\\(kJ/success)} & \makecell{Peak allocated\\VRAM (GB)} \\
\midrule
$0.5$B & Qwen2.5-Coder & 1 (1) & 1.0 & $19{,}350$ & 1.83 & 93.7 \\
$1.5$B & Qwen2.5-Coder & 1 (1) & 9.3 & $41{,}642$ & 0.13 & 94.5 \\
$7$B & Qwen2.5-Coder & 1 (1) & 20.3 & $19{,}312$ & 0.20 & 95.1 \\
$8$B & Llama-3.1 & 1 (1) & 22.0 & $15{,}586$ & 0.23 & 95.3 \\
$14$B & Qwen2.5-Coder & 2 (2) & 34.3 & $14{,}079$ & 0.31 & 190.8 \\
$32$B & Qwen2.5-Coder & 2 (2) & 43.0 & $5{,}512$ & 0.71 & 194.1 \\
$70$B & Llama-3.1 & 2 (2) & 39.0 & $2{,}076$ & 2.09 & 196.6 \\
$80$B & Qwen3-Coder-Next (MoE) & 2 (2) & 46.7 & $5{,}568$ & 0.51 & 203.6 \\
$120$B & gpt-oss (MoE) & 4 (4) & 46.7 & $7{,}138$ & 1.20 & 385.7 \\
$405$B$^{\dag}$ & Llama-3.1 & 16 (16) & 55.0 & $679$ & 20.5 & $1{,}613$ \\
$480$B$^{\dag}$ & Qwen3-Coder (MoE) & 16 (16) & 55.0 & $1{,}132$ & 13.0 & $1{,}606$ \\
\bottomrule
\end{tabular}
}

\smallskip
\begin{minipage}{\linewidth}
\footnotesize
\emph{Note}: measurement protocols and caveats are in SM Sec.~S8.
Throughput counts prompt plus generated tokens per second of
generation wall-clock time.
$^{\dag}$: the $405$B/$480$B rows are single-repetition four-node
anchors on a $100$-item subset.  Peak allocated VRAM is aggregated
across the tested GPUs and largely reflects vLLM's KV-cache reservation
at $0.9$ of each card; neither it nor the tested GPU count is a minimum
hardware requirement.  Latency under the
online-serving protocol is in SM Table~S17.  All displayed accuracy
and energy values use single-pass condition C; C+FDRS is a separate
end-to-end operating point.
\end{minipage}
\end{table}

\subsection{Security Considerations}
\label{sec:discussion:security}

Executing LLM-generated analysis code in an operational pipeline
exposes four distinct threat surfaces.  First, the generated program
is arbitrary Python, and running it confers the privileges of the
execution environment.  The execution harness provides timeouts and
workspace isolation, but only the measures recommended below provide
a security boundary.  Second, executed code can leave side effects on
the file system and process environment (the same two layers).
Third, an accepted output can silently analyse the wrong network,
modification state, or operating point, or be applied outside the
scope for which it was validated (operational misuse).  Such a
workflow may still match the scalar and feed an incorrect operating
decision downstream.  Fourth, under the
injection conditions the retrieved documentation becomes part of the
prompt, so a compromised or stale documentation source is a
prompt-injection channel (addressed by the provenance recommendation
below).

Against the first two surfaces, the harness confines generated code
to a sandbox working directory and redirects file outputs there.  It
also replaces display and browser calls with mocks, strips
interpreter-terminating calls, captures all output streams, and
bounds each run with a timeout and parent watchdog (mechanism inventory
and constants in SM Sec.~S6.1).  The L3 application probes each run
in a fresh single-threaded Python subprocess with its own $30$\,s
bound.  On the OpenDSS arm, where these probes drive native
extensions, a native crash therefore terminates only that probe and is
recorded as a distinct failure type.  Benchmark items execute
in-process under the $60$\,s timeout and the parent watchdog.  These
mechanisms protect the
evaluation process, but they do not form a security boundary.  The
harness provides no network isolation, filesystem allowlist, or
resource quotas.  Its in-process path also runs with the same
interpreter privileges as the harness itself.

The workflow audit quantifies the wrong-workflow part of the third
surface.
Section~\ref{sec:exp:validity_audit} asks how often numerically accepted
programs actually perform the requested analysis and measures the risk
that remains under the full method.  Table~\ref{tab:validity_audit_summary}
shows what tighter numerical comparison and automated re-execution can
support.  They can filter or route outputs for review, but neither can
issue an engineering verdict.

\paragraph{Production deployment recommendations}
A production deployment should not rely on the harness safeguards
alone.  Generated code should execute inside a container or virtual
machine, with network access disabled by default.  A filesystem
allowlist should confine writes to a per-job scratch area, and explicit
quotas should bound CPU, memory, and wall-clock use.  A static
pre-execution screen should reject calls outside the target library's
documented API allowlist.  The
audited screening layer should route flagged accepted outputs to
expert review rather than reject them autonomously.  Human oversight
should be graded jointly by model tier, task risk, and locally
validated acceptance tests.  This graded oversight is the control
against the third surface, operational misuse included.  After site
validation, routine analyses
can enter a risk-based review queue or a sampled spot-check regime.
Protection-coordination, switching-adjacent, and other operationally
critical analyses require per-output expert review.  Finally, because injected
documentation enters the prompt, a deployment should log the
provenance of every injected snippet (source, library version,
content hash) so that a poisoned or stale documentation source is
detectable and auditable after the fact.

\subsection{Limitations and Future Work}
\label{sec:limitations}

\paragraph{Evaluation scope: library, output format, and probe granularity}
The primary benchmark covers the \pp{} library.  The OpenDSS and PyPSA
pilots extend the probing and proactive-allocation path to two
contrasting API surfaces, three deployment forms, and $90$ items per
backend.  They support an onboarding pattern; full-scale cross-library
performance remains untested.  Evaluation on larger suites for OpenDSS,
PyPSA, MATPOWER, ANDES, or GridCal is the next
step~\cite{zimmerman2011matpower,cui2021andes}.

Within \pp{}, each task produces one scalar answer.  This supports
reproducible execution and matching, but leaves full reports,
multi-output tables, interactive scheduling, graphical analysis, and
matrix-valued trajectories for future extensions.  The probes also
operate at function level; parameter-level probes could add finer
diagnostic resolution.

PowerCodeBench is generated from schemas, whereas production
traffic would sample real requests.  The $80$-item human-authored, operator-style holdout reduces
template regularity, while larger production-collected holdouts would
test a broader range of wording and intent.

\paragraph{Calibration across API surfaces}
Profile-conditioned allocation depends on how probe scores map to the
target API surface.  The OpenDSS accessor interface required more
name- and signature-level evidence than the initial allocation
provided.  The targeted adjustment improved the same pilot; evaluation
on other APIs with hierarchical getter/setter bindings and weakly
informative signatures is needed to validate this calibration rule.

\paragraph{Residual risk after numerical matching}
Even under the full method, roughly one in five numerically accepted
outputs in the deployed open-weight sample does not perform the
requested analysis.  Tighter numerical comparison and the automatic
screen can reduce the review load, but operationally critical outputs
still require expert review.  The shared rubric cannot represent the
full range of judgement across institutions.  Deterministic
re-execution also resolves only physics-checkable questions.

\paragraph{Deployment-side assumptions: documentation source and demand drift}
The experiments use official documentation matched to the executed
library version.  A deployment can break this alignment when its
documentation lags the live software, omits APIs, or leaves parameter
semantics ambiguous.  Robustness to such source errors is not varied
in the present ablations.

Role-frequency reweighting likewise assumes an unlabelled query corpus
that represents the intended workload.  The adapted and unadapted
results show that this choice matters when the analysis and
construction mix changes.  A streaming deployment should therefore
monitor query drift and refit offline as needed; the appropriate
cadence and corpus size require production study.

\paragraph{Comparison and panel scope}
Sub-2B models remain below usable capacity in this setting, so the
method targets the
$7$B-and-above tier, where demand-aware injection lifts a
$0.4$--$15\%$ R0 baseline to $33$--$69\%$ Round-3 accuracy.  The
closed-source panel covers four representative mid-tier vendors on the
full release plus two reasoning-tier APIs on a stratified $200$-item
subset (Sec.~\ref{sec:exp:cross_vendor_reasoning}); broader reasoning-tier
coverage and larger reasoning samples are a natural follow-up.
Deployment-time injection is also complementary to
parameter fine-tuning and multi-agent orchestration, which occupy
different cost axes.  Fine-tuning pays an up-front training cost that
repeats as the library evolves~\cite{cheng2025advanceddispatch,liu2024codeupdatearena};
agentic orchestration adds inference-time
compute~\cite{jin2025gridmind,zhang2025poweragent}.  Direct comparisons
with fine-tuned small models, multi-step agents, and newer demand
estimators on the same benchmark remain future work.
\section{Conclusion}
\label{sec:conclusion}
PowerCodeBench treats power-system code generation as a boundary
between natural-language intent, a versioned simulation API, and an
executable workflow.  The measured results answer the three research
questions posed in Sec.~\ref{sec:intro}.

\emph{RQ1: where does API knowledge break down?}  The breakdown is
layered rather than monolithic.  Comprehension (L2) occupies a high,
narrow band for eight of the nine deployable open-weight models, while
recognition (L0) and executable application (L3) remain the sharper
bottlenecks.  Signature recall (L1) is the strongest cross-model
predictor of first-pass accuracy ($\rho_s\!=\!0.93$;
Secs.~\ref{sec:probing} and~\ref{sec:exp:per_task}).  Because
application-layer competence
does not scale smoothly with parameter count, the profiles complement
model size as a selection diagnostic and provide the state needed to
calibrate selective injection for each backend.

\emph{RQ2: can model-side weakness and query-side demand be
reconciled at deployment time?}  On the $2{,}000$-item benchmark, the
validation-enabled C+FDRS endpoint improves every
evaluated model of at least $7$B and each mid-tier API by $32$--$56$
percentage points over an unassisted first pass.  Its reference-free
repair path uses execution feedback alone; where an independent
acceptance signal is available, value-error repair adds a second
route.  Selective injection approaches the full-layer reference
accuracy with $41\%$ of its prompt tokens.  The workflow audit shows
that the gain reaches the underlying analysis: among paired outputs
that both methods pass numerically, $88\%$ under the full method perform
the requested analysis, compared with $66\%$ under plain repair.  A
broader audit of the deployed open-weight configuration nevertheless
finds that roughly one in five matching outputs follows the wrong workflow.

The combined workload and review envelope is summarised in
Table~\ref{tab:deployment_guidance}.

\begin{table}[!htbp]
\centering
\footnotesize
\caption{Workload-aware deployment guidance distilled from the
  measured envelope, serving footprint, and validity audit
  (Secs.~\ref{sec:exp:cross_vendor}--\ref{sec:exp:validity_audit},
  Table~\ref{tab:serving_metrics}).  Envelope cells read the C+FDRS
  endpoint on the full $2{,}000$-item benchmark
  (Fig.~\ref{fig:per_task_heatmap}(b)): a task family is listed when
  every configuration in the tier reaches at least $70\%$ on it, and as
  partial when every configuration reaches at least $50\%$ but not all
  reach $70\%$.  All modes assume the safeguards of
  Sec.~\ref{sec:discussion:security}, including locally validated
  acceptance tests, and differ only in review intensity, which
  concentrates where wrong-workflow risk is highest or decisions are
  operationally critical.  Parenthetical GPU counts identify the tested
  allocations rather than minimum hardware requirements.  Modes are
  recommended configurations, not measured conditions.}
\label{tab:deployment_guidance}
\setlength{\tabcolsep}{4pt}
\begin{tabular}{@{}>{\raggedright\arraybackslash}p{0.16\linewidth}>{\raggedright\arraybackslash}p{0.42\linewidth}>{\raggedright\arraybackslash}p{0.38\linewidth}@{}}
\toprule
Model tier & Workload envelope in this study & Deployment mode \\
\midrule
Sub-$7$B local &
None at usable accuracy (capacity floor) &
Not recommended \\
\addlinespace
$7$B--$14$B local (one--two GPUs) &
Single-network sequential workflows; routine power-flow and DC
power-flow checks from $14$B &
Demand-guided injection with routed repair; per-output expert review
of accepted outputs \\
\addlinespace
$32$B local (two GPUs) &
Routine power-flow and DC power-flow checks, multi-scenario sweeps,
and sequential workflows; OPF and time-series partial &
Demand-guided injection with routed repair; risk-based review queue,
with per-output expert review for operationally critical analyses \\
\addlinespace
$70$B--$120$B local (Llama-$70$B, Qwen3-Next: two GPUs; $120$B MoE:
four GPUs) &
Power flow, DC power flow, OPF, and sequential workflows at aggregate
mid-tier-API accuracy; multi-scenario sweeps and contingency
screening partial; short-circuit partial at best, and state estimation
only on the $120$B MoE (partial) &
Demand-guided injection with routed repair; sampled spot checks, with
per-output expert review for protection-coordination and
switching-adjacent analyses \\
\addlinespace
$405$B--$480$B local (sixteen GPUs) &
Broadest coverage: adds multi-scenario sweeps in full, with state
estimation at the $70\%$ line and short-circuit partial; contingency
correction and semantically grounded queries partial at best &
Demand-guided injection with routed repair; risk-based review queue,
with per-output expert review for operationally critical analyses \\
\addlinespace
Mid-tier hosted APIs &
Routine power-flow and DC power-flow checks and sequential workflows;
OPF and time-series partial; multi-scenario sweeps below the local
tiers of at least $32$B &
Demand-guided injection with routed repair; risk-based review queue,
with per-output expert review for operationally critical analyses \\
\bottomrule
\end{tabular}
\end{table}

\emph{RQ3: which locally deployable tiers reach hosted
mid-tier-API accuracy, and on which workloads?}  Under matched
no-tool conditions, the $70$B--$120$B open-weight models enter the
accuracy range observed across the four mid-tier APIs.  This
comparison concerns the matched no-tool endpoints.  In a separate
stratified check, paid reasoning raises both tested API endpoints, but
the intervention remains beneficial.  At smaller scales the answer is
workload-specific.  Routine power-flow and DC power-flow checks are
in range from $14$B, and the explicit-parameter tier, multi-scenario
sweeps included, from $32$B.  Harder specialised tasks and
semantically grounded queries favour the larger open-weight or hosted
tiers (Sec.~\ref{sec:exp:r0}).  The measured serving footprint
(throughput, latency, energy per successful task, allocated GPU memory) qualifies
the smaller tiers and makes tier choice a joint accuracy--serving
decision (Sec.~\ref{sec:discussion:implications}).

Deployment should therefore combine workload-aware model selection with
local acceptance tests, isolation, and risk-based expert review.

Because the workflow modifies no weights, the serving tier and
evidence policy can evolve with the available hardware and the
calibration of each connected library.  The next step is full-scale
evaluation on further backends and production-collected query
distributions.
\FloatBarrier
\section*{Acknowledgements}
This paper is partially supported by the SAINTES project funded by the ARIA Safeguarded AI programme, as well as the Innovate UK funded TRIAGE project.

The authors acknowledge the use of resources provided by the Isambard-AI National AI Research Resource (AIRR). Isambard-AI is operated by the University of Bristol and is funded by the UK Government's Department for Science, Innovation and Technology (DSIT) via UK Research and Innovation; and the Science and Technology Facilities Council [ST/AIRR/I-A-I/1023].
\section*{Declaration of Competing Interest}
The authors declare that they have no known competing financial interests or
personal relationships that could have appeared to influence the work reported
in this paper.
\section*{Data Availability}
PowerCodeBench is available at \url{https://github.com/huiwxing/PowerCodeBench}
(CC~BY~4.0); the full evaluation/intervention code is released in the same repository.

\section*{CRediT Authorship Contribution Statement}
\textbf{Hui Wu}: Writing -- review \& editing, Writing -- original draft,
Visualization, Validation, Software, Methodology, Investigation, Formal analysis,
Data curation, Conceptualization.
\textbf{Xiaoyang Wang}: Writing -- review \& editing, Supervision, Methodology.
\textbf{Zhong Fan}: Writing -- review \& editing, Supervision, Funding acquisition.

\clearpage
\section*{Supplementary Material}
\addcontentsline{toc}{section}{Supplementary Material}
\setcounter{section}{0}
\setcounter{subsection}{0}
\setcounter{subsubsection}{0}
\setcounter{table}{0}
\setcounter{figure}{0}
\setcounter{equation}{0}
\setcounter{algorithm}{0}
\renewcommand{\thesection}{S\arabic{section}}
\renewcommand{\thesubsection}{\thesection.\arabic{subsection}}
\renewcommand{\thesubsubsection}{\thesubsection.\arabic{subsubsection}}
\renewcommand{\thetable}{S\arabic{table}}
\renewcommand{\thefigure}{S\arabic{figure}}
\renewcommand{\theequation}{S\arabic{equation}}
\renewcommand{\thealgorithm}{S\arabic{algorithm}}
\renewcommand{\theHsection}{supp.\arabic{section}}
\renewcommand{\theHsubsection}{supp.\arabic{section}.\arabic{subsection}}
\renewcommand{\theHsubsubsection}{supp.\arabic{section}.\arabic{subsection}.\arabic{subsubsection}}
\renewcommand{\theHtable}{supp.table.\arabic{table}}
\renewcommand{\theHfigure}{supp.figure.\arabic{figure}}
\renewcommand{\theHequation}{supp.equation.\arabic{equation}}

\section*{Overview}
\label{sm:overview}

\enlargethispage{24pt}
This Supplementary Material (SM) carries the evidence behind the main
manuscript's claims.  Each section settles one question; the roadmap
below gives that question, the part of the main text it serves, and
where its evidence sits.

\begin{table}[!h]
\centering
\footnotesize
\renewcommand{\arraystretch}{0.95}
\setlength{\tabcolsep}{3pt}
\begin{tabular}{@{}p{0.037\linewidth}>{\raggedright\arraybackslash}p{0.485\linewidth}p{0.15\linewidth}>{\raggedright\arraybackslash}p{0.235\linewidth}@{}}
\toprule
SM & Question it settles & Main text & Tables$/$figures \\
\midrule
S1 & Is the released benchmark a controlled sample? & Sec.~3.3 &
     Table~S1, Fig.~S1 \\
S2 & Does the probe corpus cover the APIs the benchmark uses? &
     Sec.~4.2 & Tables~S2--S3 \\
S3 & Were the intervention constants and routes fixed in advance? &
     \mbox{Secs.~4.4--4.5} & Eq.~(S1) \\
S4 & Does the demand predictor transfer, and when does adaptation
     backfire? & \mbox{Secs.~4.3, 5.5} & Tables~S4--S6, Alg.~S1 \\
S5 & Do the full breakdowns overturn the main ordering? &
     \mbox{Secs.~5.2--5.5} & Tables~S7--S11, Fig.~S2 \\
S6 & What must be specified to reproduce a cell? & Sec.~5.1 &
     Table~S12 \\
S7 & Is the false-acceptance drop the method's or the adjudicator's? &
     Sec.~5.7 & Tables~S13--S15 \\
S8 & How were energy and latency measured, and how far do they
     carry? & Sec.~6.1 & Tables~S16--S17 \\
S9 & Why did C not beat budget-matched retrieval at cold start? &
     Sec.~5.6 & Tables~S18--S19 \\
\bottomrule
\end{tabular}
\end{table}

\paragraph{Division of labour with the artifact repository}
The main manuscript and this SM together contain the study design,
protocols, principal numerical results, and the detail needed to
interpret them.  The PowerCodeBench repository
(\url{https://github.com/huiwxing/PowerCodeBench}) adds released inputs,
item-level outputs, audit records, and reproduction scripts for
reanalysis; it is not needed to follow the argument.
All section, table, figure, and equation numbers in this document carry
an ``\textbf{S}'' prefix.

For readability, shortened model labels are used outside the exact-ID
table: Qwen2.5-$x$B denotes Qwen2.5-Coder-$x$B, Qwen3-Next and
Qwen3-480B denote the corresponding Qwen3-Coder models, and
Llama-$x$B denotes Llama-3.1-$x$B.  Table~\ref{tab:sm_model_identifiers}
gives the full deployment identifiers.
\section{Benchmark engineering details}
\label{sm:benchmark}

The frozen $2{,}000$-item release is a controlled sample rather than
an ad-hoc collection: its generation ratios, its coverage of networks,
task families and query patterns, and its rejection diagnostics are
all auditable in Table~\ref{tab:sm_benchmark_protocol}, and
task-specific feasibility filters and uniqueness checks reject failed
or duplicate instantiations before freezing.

\begin{table}[!htbp]
\centering
\caption{PowerCodeBench generation protocol and quality diagnostics.
  Ratios describe the intended generation protocol; item-level counts reflect
  the final validated benchmark used in all experiments.}
\label{tab:sm_benchmark_protocol}
\footnotesize
\setlength{\tabcolsep}{4pt}
\begin{tabularx}{\linewidth}{p{0.26\linewidth} X}
\toprule
Aspect & Evidence in the benchmark release \\
\midrule
Difficulty protocol &
2{,}000 items with a fixed D1/D2/D3/D4 split of
30\%/30\%/20\%/20\% (600/600/400/400). \\
Task coverage &
15 task families: 8 single-step analysis families and 7 multi-step workflow
families, including OPF, contingency analysis, time series, state estimation,
short-circuit studies, sequential diagnosis, and power-flow/short-circuit composition. \\
Query coverage &
17 query-target patterns, including scalar point values, extrema, argmax/argmin,
totals, threshold checks, time-series values, and multi-step workflow outputs. \\
Network coverage &
39 standard \pp{} networks.  The sampling pool explicitly includes both
fast/small networks and large transmission grids (PEGASE, RTE, Polish, GB,
and Oberrhein cases); task-specific feasibility filters determine the final
per-network item counts. \\
Natural-language diversity &
Template-based rendering with LLM-expanded paraphrases; the active single-task
and multi-task template keys used by the release have at least 9 surface
variants per key after validation. \\
Semantic setup coverage &
12 rule-grounded setup rules spanning demand response, line outage and
reinforcement, generator trip/redispatch, shunt support/absorption, and
transformer tap adjustment. \\
Ground-truth types &
1{,}848 float targets, 108 integer targets, and 44 Boolean targets, all
extracted from executable reference solutions. \\
Uniqueness diagnostics &
2{,}000/2{,}000 unique item IDs, 1{,}999/2{,}000 unique natural-language
queries, and 1{,}995/2{,}000 unique scenario serialisations. \\
Semantic diagnostics &
800/800 D3--D4 items contain the intended rule-grounded semantic phrase;
release diagnostics scan the semantic setup span for explicit target indices
or hidden numeric values before freezing the benchmark. \\
\bottomrule
\end{tabularx}
\end{table}

\paragraph{Quality control and semantic-setup audit}
\label{sm:benchmark:quality}
Beyond the three hard admission checks of main-text Sec.~3.4, the
semantic-setup audit verifies the intended rule-grounded phrase in
every D3$/$D4 query.  It scans \emph{only} the semantic-setup span
for explicit target indices or numeric values, so a legitimate number
elsewhere in the query text is never read as a leak.

\begin{figure}[!htbp]
  \centering
  \includegraphics[width=\linewidth]{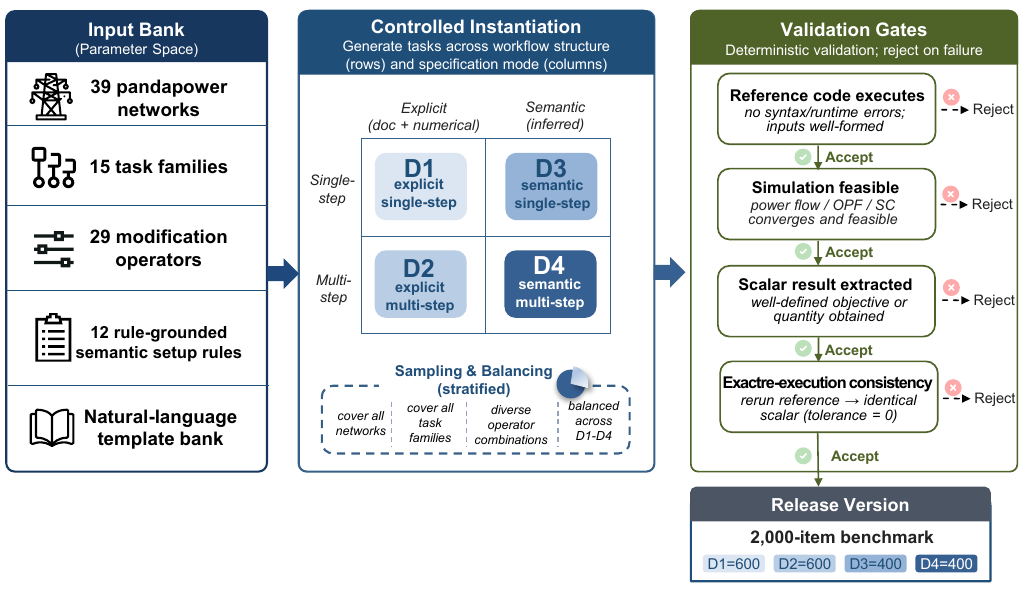}
  \caption{PowerCodeBench generation and validation pipeline. Candidate tasks are sampled from the configurable design space into D1--D4 at the fixed $30\%/30\%/20\%/20\%$ ratio (seed~$42$). Reference-code execution, simulation validity, scalar extraction, re-execution consistency, and semantic-setup checks yield the $2{,}000$-item benchmark.}
  \label{fig:sm_benchmark_pipeline}
\end{figure}

\paragraph{Five-stage construction pipeline}
\label{sm:benchmark:pipeline}
The $2{,}000$ items are produced by the five-stage pipeline summarised
in Figure~\ref{fig:sm_benchmark_pipeline}; per-difficulty,
per-family, and per-network composition statistics of the
release beyond Table~\ref{tab:sm_benchmark_protocol} are plotted in
the PowerCodeBench repository (\texttt{benchmark/}).

\section{Probe corpus engineering details}
\label{sm:probe}

Table~\ref{tab:sm_probe_protocol} records the corpus-level and
layer-level detail not carried by main-text Sec.~4.2, together with
the benchmark-linkage audit.  Table~\ref{tab:sm_probe_results}
then reports the full per-model L0--L3 profiles.

\begin{table}[!htbp]
\centering
\caption{L0--L3 probe-corpus construction protocol and coverage
  diagnostics for the probe release.}
\label{tab:sm_probe_protocol}
\footnotesize
\setlength{\tabcolsep}{4pt}
\begin{tabularx}{\linewidth}{>{\raggedright\arraybackslash}p{0.19\linewidth} >{\raggedright\arraybackslash}X}
\toprule
Aspect & Evidence in the probe corpus \\
\midrule
Corpus and layer detail &
$274$ of the $275$ documented entries include at least one example
snippet available for L3-oriented injection; the L2 layer splits into
$275$ function-purpose, $535$ parameter-meaning, and $170$
return-information multiple-choice items instantiated from non-empty
documentation fields, and the L3 probes require no example snippet.
Entry typing, the total probe count, and the per-layer counts are in
main-text Sec.~4.2. \\
Benchmark linkage &
AST auditing of PowerCodeBench reference code verifies that the probed corpus
covers the core API primitives used by benchmark solutions; non-probed
calls are generic language/library accessors or network-fixture loaders. \\
\bottomrule
\end{tabularx}
\end{table}

\paragraph{Per-model L0--L3 probe scores (full panel)}
\label{sm:probe:scores}
Per-level construction mechanics and scoring rules are specified in
main-text Sec.~4.2; Table~\ref{tab:sm_probe_results} resolves those
scores to every panel model, which main-text Fig.~3 visualises as a
knowledge-profile heatmap.

\begin{table}[!htbp]
\centering
\caption{Mean L0--L3 knowledge-probe scores across all $275$ \pp{} API
  entries, for every open-weight model in the evaluation panel and the
  four closed-source API models from the cross-vendor evaluation of the
  main text.  L0: recognition accuracy; L1: the equally weighted mean
  of path correctness and required-parameter Jaccard similarity; L2:
  comprehension accuracy; L3: application execution success.}
\label{tab:sm_probe_results}
\footnotesize
\begin{tabular}{lcccc}
\toprule
Model & L0 (Recog.) & L1 (Recall) & L2 (Comp.) & L3 (Apply) \\
\midrule
Qwen2.5-1.5B  & 0.244 & 0.384 & 0.357 & 0.036 \\
Llama-8B        & 0.276 & 0.359 & 0.755 & 0.102 \\
Qwen2.5-7B    & 0.362 & 0.653 & 0.904 & 0.156 \\
Qwen2.5-14B   & 0.407 & 0.733 & 0.950 & 0.324 \\
Qwen2.5-32B   & 0.424 & 0.744 & 0.960 & 0.425 \\
Qwen3-Next    & 0.393 & 0.664 & 0.924 & 0.258 \\
Llama-70B       & 0.396 & 0.746 & 0.951 & 0.200 \\
GPT-OSS-120B        & 0.509 & 0.756 & 0.949 & 0.353 \\
Llama-405B      & 0.536 & 0.750 & 0.962 & 0.385 \\
Qwen3-480B    & 0.411 & 0.817 & 0.963 & 0.495 \\
\midrule
\multicolumn{5}{l}{\textit{Closed-source API models (cross-vendor reference)}} \\
Gemini-2.5-Flash    & 0.353 & 0.741 & 0.898 & 0.753 \\
DeepSeek-V4-Flash   & 0.462 & 0.771 & 0.942 & 0.367 \\
Claude-Haiku-4-5    & 0.585 & 0.778 & 0.954 & 0.389 \\
GPT-5.4-mini        & 0.615 & 0.786 & 0.952 & 0.564 \\
\bottomrule
\end{tabular}
\end{table}
\section{Intervention engineering details}
\label{sm:intervention}

Every constant and routing rule of the intervention layer was fixed
before the experiments and reads only the query text and versioned
library artifacts, never a reference answer.
Sec.~\ref{sm:intervention:proactive_group} gives the proactive
injector's constants and selection rules;
Sec.~\ref{sm:intervention:reactive_group} gives the reactive
corrector's routing logic.

\subsection{Proactive injector internals}
\label{sm:intervention:proactive_group}

\paragraph{Layer constants: values and rationale}
\label{sm:intervention:constants}
The deficit floor $\tau_\ell(f,q)$ in the injection score of
main-text Sec.~4.4 lifts the deficit term in three
deterministic high-confidence cases (a query-anchored function
match, a high-confidence demand prediction, and a boundary-card
hit), so the candidate remains in contention even when the probe
deficit alone is small:
\begin{align}
  \tau_\ell(f, q) = \max\!\bigl(&
    \tau^{\text{anchor}}_\ell \, \mathbb{1}[f \in \mathrm{anchor}(q)],\;
    \tau^{\text{intf}}_\ell \, \mathbb{1}[\hat{d}(f,q) \geq d^*], \nonumber\\
    &\tau^{\text{bnd}}_\ell \, \mathbb{1}[f \in \mathrm{boundary}(q)]
  \bigr),
  \label{eq:sm_tau_floor}
\end{align}
with $d^*\!=\!0.85$ the high-confidence demand threshold,
$\mathrm{anchor}(q)$ the set of functions matched by high-precision
query anchors (network-loader names, construction-API substrings),
and $\mathrm{boundary}(q)$ the set of functions tied to a
boundary-card contract triggered by $q$.  The $\tau$ constants
above, together with the $\mathrm{base}$ and $\mathrm{cap}$
constants of the task-need multiplier $a_\ell(q)$
(defined below), take the following
values throughout the paper:
{\small
\[
\setlength{\arraycolsep}{6pt}
\begin{array}{l|cccc}
\ell                       & \text{L0} & \text{L1} & \text{L2} & \text{L3} \\
\hline
\tau^{\text{anchor}}_\ell  & 0.45 & 0.40 & 0    & 0    \\
\tau^{\text{intf}}_\ell    & 0.16 & 0.14 & 0    & 0    \\
\tau^{\text{bnd}}_\ell     & 0    & 0.22 & 0.18 & 0    \\
\mathrm{base}_\ell         & 0.70 & 0.95 & 0.70 & 0.75 \\
\mathrm{cap}_\ell          & 1.05 & 1.15 & 1.30 & 1.40
\end{array}
\]}
These are fixed layer-role priors, not tuned quantities.  L0 carries
the highest anchor threshold because its probe most directly predicts
hallucinated \texttt{AttributeError}; L1 carries floors at both
triggers, consistent with signature recall being the strongest
cross-model predictor of R0 ($\rho_s\!=\!0.93$, $n\!=\!10$); L3
carries the largest ceiling as the most actionable layer; L2 takes
small floors because capable models already score highly on it.  The
$\tau$, $\mathrm{base}$, and $\mathrm{cap}$ constants of this table
were not sensitivity-scanned.  The sensitivity grid of
Fig.~\ref{fig:sm_robustness}(a) scans a different constant family,
the L0--L3 layer weights $w_\ell$ of the risk score, over which R0
moves by at most $1.30$pp pairwise.

\paragraph{Layer-need rules: signal-to-increment mapping}
\label{sm:intervention:layerneed}
The task-need multiplier $a_\ell(q)$ in the injection score of
main-text Sec.~4.4 is a query-derived depth
adjustment.  It starts from $\mathrm{base}_\ell$ (above) and
accumulates layer-specific increments triggered by query-derived
binary signals $\sigma_i(q)$, then is clipped to the per-layer
ceiling $\mathrm{cap}_\ell$:
\[
  a_\ell(q) \;=\; \min\!\Bigl(\,
    \mathrm{base}_\ell + \sum_{i}
    \Delta^{(\ell)}_i\, \mathbb{1}[\sigma_i(q) = 1],\;
    \mathrm{cap}_\ell\Bigr).
\]
All seven signals are computed from $q$ alone by regex or intent
matching; none reads the reference code, the difficulty label, or any
execution feedback.  Five describe the query wording
(\textsl{workflow\_intent}, \textsl{sequential},
\textsl{conditional}, \textsl{aggregation},
\textsl{multi\_modification}) and two the candidate set or query
length (\textsl{constructor\_or\_executor\_role},
\textsl{long\_query} at $\geq\!35$ tokens); the matching vocabulary of
each is listed in \texttt{code/intervention/README.md} of the
repository.  The increment rule $\Delta^{(\ell)}_i$ assigns:
{\small
\[
\setlength{\arraycolsep}{4pt}
\begin{array}{l|cccc}
\text{trigger} & \Delta^{(\mathrm{L0})} & \Delta^{(\mathrm{L1})} & \Delta^{(\mathrm{L2})} & \Delta^{(\mathrm{L3})} \\
\hline
\textsl{constructor\_or\_executor\_role}  & 0    & 0.10 & 0    & 0    \\
\textsl{conditional} \lor \textsl{aggregation} & 0 & 0 & 0.22 & 0 \\
\textsl{workflow\_intent}                 & 0    & 0    & 0.10 & 0.22 \\
\textsl{sequential}                       & 0    & 0    & 0    & 0.22 \\
\textsl{multi\_modification}              & 0    & 0    & 0.10 & 0.12 \\
\textsl{long\_query}                      & 0    & 0    & 0.05 & 0.05 \\
\end{array}
\]}
The number of L3 examples retained by the greedy selector is
additionally capped at $1$, $2$, or $3$ by the count of active
procedural triggers, preventing long L3 snippets from exhausting the
budget on procedurally light queries; the canonical implementation
is \texttt{code/intervention/proactive.py}.

\paragraph{Query anchors, intent-consistency filter, and DataFrame boundary cards}
\label{sm:intervention:proactive}
Three artifact classes serve the proactive injector.  \emph{Query
anchors} fire on explicit network-loader and construction-API
substrings in $q$, triggering the $\tau^{\mathrm{anchor}}_\ell$
component of Eq.~(\ref{eq:sm_tau_floor}).  An
\emph{intent-consistency filter} over the mutually exclusive workflow
executors (time-series, short-circuit, contingency, OPF, DC power
flow) stops the recall-oriented demand predictor from promoting a
low-confidence tail candidate merely because that API is risky for the
target model.  Its exclusion sets come from API-intent fields in the
spec, not from benchmark-specific rules.  \emph{DataFrame boundary
cards} cover recurrent conventions not tied to one function
(e.g.\ \texttt{net.res\_line['loading\_percent']},
\texttt{net.bus['min\_vm\_pu']}), stating the general contract instead
of leaving the model to hallucinate a result or network attribute.
All three are enumerated in \texttt{code/intervention/README.md}.

\subsection{Reactive router internals}
\label{sm:intervention:reactive_group}

\paragraph{Reactive code-error router: exception-class clusters}
\label{sm:intervention:codeerror}
A fix attempt is routed to one of the three code-error routes of
main-text Sec.~4.5 by its exception class and the symbols on the
failing line alone, never by benchmark-specific patterns; row order
in the table below is the order of precedence.

\begin{table}[!ht]
\centering
\footnotesize
\setlength{\tabcolsep}{4pt}
\begin{tabular}{@{}>{\raggedright\arraybackslash}p{0.17\linewidth}
  >{\raggedright\arraybackslash}p{0.44\linewidth}
  >{\raggedright\arraybackslash}p{0.31\linewidth}@{}}
\toprule
Route & Triggering exception classes & Injected artifact \\
\midrule
\textbf{basic\_fix} &
\texttt{SyntaxError}, \texttt{IndentationError}, timeouts, and any
exception carrying no concrete API symbol on the failing line &
None (error feedback only) \\
\addlinespace[2pt]
\textbf{api\_doc} &
\texttt{AttributeError}, \texttt{TypeError}, \texttt{NameError},
\texttt{ImportError}, \texttt{IndexError}, and contract-style failures
(\texttt{ValueError}, \texttt{KeyError},
\texttt{LoadflowNotConverged}) when they implicate a concrete API &
Compact L2 call contract; higher layers suppress redundant lower ones
for the same function \\
\addlinespace[2pt]
\textbf{boundary\_} \textbf{contract} &
Failing line on a known API boundary with no single function
implicated &
Boundary card instead of per-API documentation \\
\bottomrule
\end{tabular}
\end{table}

\paragraph{Value-error branch, output-format sub-path, and fix prompt}
\label{sm:intervention:valueerror}
The numerical-mismatch sub-path extracts an AST trace of the
generated program (ordered target-library calls, pre-analysis table
writes, result-table reads in the final \texttt{print}, and
code-only invariants on the closing arithmetic) and attaches
interface contracts for exactly the APIs and tables the trace
touches.  For each of the eight analysis routines it may encounter,
it adds a \emph{function-output} contract naming the canonical
\texttt{net.res\_*} columns that routine writes.  The mapping is
library-derived, curated once in the library-knowledge artifact, and
independent of any benchmark item.  The output-format sub-path is a
short library-agnostic instruction, emitted when the printed payload
will not parse as the expected scalar, that the last line must print
exactly one numeric scalar.  Together the two sub-paths carry the
\texttt{semantic\_fix} route whose per-model share is reported in
Table~\ref{tab:sm_router_per_model}, and which is the only route the
FDRS step adds over FDR.  The fix prompt is a structured
``\textsc{task}~$/$~\textsc{previous
code}~$/$~\textsc{failure}~$/$~\textsc{relevant doc}~$/$
\textsc{instruction}'' user turn with the system prompt unchanged
across rounds; under a basic-fix decision the relevant-doc block is
empty and the prompt reduces to the standard error-feedback
baseline.  Verbatim templates and the full trace-to-contract
specification: \texttt{docs/PROMPTS.md} in the repository.
\section{Demand modelling: corpus, estimators, and external-query
  evaluation}
\label{sm:demand}

The demand predictor transfers API-role structure across query styles.
Hybrid TF-IDF gives the strongest deployment-style recall, while
role-frequency adaptation helps analysis-heavy workloads and can hurt
construction-heavy ones.  The evidence below separates corpus
independence, estimator choice, adaptation, and external-query
behaviour.

\subsection{Corpus and independence audit}
\label{sm:demand:corpus_group}

\paragraph{Augmented corpus construction and overlap audit}
\label{sm:demand:corpus}
Training pairs are retained only when the reference code executes and
the function list is AST-verified.  Paraphrases of a common seed remain
in one base group, and all splits are group-level.  The corpus trains
only the demand predictor; PowerCodeBench remains the downstream
target.

The augmented corpus and PowerCodeBench share part of the API surface
but not query phrasing.  No benchmark query is a paraphrase or
near-clone of an augmented query, so the downstream benchmark tests
transfer across independently worded tasks.  The overlap audit
separates shared functions from lexical similarity:

\begin{table}[!ht]
\centering
\footnotesize
\setlength{\tabcolsep}{5pt}
\begin{tabular}{@{}>{\raggedright\arraybackslash}p{0.37\linewidth}
  >{\raggedright\arraybackslash}p{0.22\linewidth}
  >{\raggedright\arraybackslash}p{0.32\linewidth}@{}}
\toprule
Comparison & Result & Interpretation \\
\midrule
Reference-function sets & Jaccard $0.25$ &
$15/50$ benchmark functions ($30\%$) are benchmark-only, mainly
large-grid loaders. \\
Query vocabularies & $350/1{,}822$ shared; Jaccard $0.19$ &
Limited token overlap despite the common domain. \\
Per-query best lexical match & mean $0.25$; maximum $0.41$ &
No sampled query pair exceeds the pre-set $0.5$ near-clone threshold. \\
\bottomrule
\end{tabular}
\end{table}
\FloatBarrier

\subsection{Estimators and operating point}
\label{sm:demand:estimators_group}

\paragraph{Six demand-estimator detailed formulations}
\label{sm:demand:estimators}
\emph{(1)~Zero-shot TF-IDF}: top-$k$ functions by TF-IDF cosine
between the query and each function's documentation string.
\emph{(2)~Pairwise TF-IDF + LogReg}: logistic regression on
(query, function-card) pairs, positives from the augmented dataset,
random negatives at ratio 8:1.
\emph{(3)~Hybrid TF-IDF}: linear combination of (1) and (2), mixing
weight $\alpha$ auto-selected for recall@10 on a held-out augmented
validation split.
\emph{(4)~Zero-shot SBERT}: sentence-BERT
(\texttt{all-MiniLM-L6-v2}) embedding cosine.
\emph{(5)~Zero-shot Cross-Encoder}: (1) reranked by
\texttt{cross-encoder/ms-marco-MiniLM-L-12-v2}.
\emph{(6)~Hybrid Cross-Encoder}: (3) with the same reranker.

\paragraph{Six-estimator recall@10/hit@10 results}
\label{sm:demand:estimator_results}
Hybrid TF-IDF is the only estimator that leads on the deployment-style
Bench-ref split while remaining near the best Aug-test recall.  In
Table~\ref{tab:sm_demand_results}(a), the Bench-ref column determines
the deployment choice; the Aug-test column shows its source-style
cost.  Reweighting changes only the supervised component of the three
pairwise or hybrid estimators, and each hybrid arm selects its mixing
weight on its own development split.  Panel~(b) then shows the central
trade: adaptation moves recall toward the deployment split and away
from the source-style split.

\begin{table}[!htbp]
\centering
\caption{Task-demand modelling. (a)~Recall@10$/$hit@10 for six
  estimators on the augmented split (\textit{Aug-test}), the original
  queries (\textit{Aug-orig}), and benchmark-reference labels
  (\textit{Bench-ref}; never trained on).  (b)~Hybrid TF-IDF
  recall@10 before and after role-frequency reweighting.}
\label{tab:sm_demand_results}
\footnotesize
\textbf{(a)} Six estimators (recall@10 / hit@10)\\[2pt]
\begin{tabular}{lccc}
\toprule
Model & Aug-test & Aug-orig & Bench-ref \\
\midrule
Zero-shot TF-IDF         & 0.344 / 0.806 & 0.258 / 0.652 & 0.605 / 0.893 \\
Pairwise TF-IDF+LogReg   & 0.307 / 0.795 & 0.402 / \textbf{1.000} & 0.489 / 0.946 \\
Hybrid TF-IDF            & 0.471 / 0.937 & \textbf{0.443} / 0.957 & \textbf{0.718} / \textbf{0.983} \\
Zero-shot SBERT          & 0.342 / 0.842 & 0.258 / 0.783 & 0.331 / 0.659 \\
Zero-shot Cross-Encoder  & 0.383 / 0.888 & 0.319 / 0.870 & 0.397 / 0.723 \\
Hybrid Cross-Encoder     & \textbf{0.480} / \textbf{0.962} & 0.371 / 0.913 & 0.432 / 0.774 \\
\bottomrule
\end{tabular}
\\[6pt]
\textbf{(b)} Role-frequency reweighting effect (recall@10; $\Delta$ in pp)\\[2pt]
\setlength{\tabcolsep}{5pt}
\begin{tabular}{lrrr|rrr}
\toprule
\multirow{2}{*}{Demand model} &
\multicolumn{3}{c|}{Aug-test recall@10} &
\multicolumn{3}{c}{Bench-ref recall@10} \\
 & before & after & $\Delta$ & before & after & $\Delta$ \\
\midrule
Hybrid TF-IDF (proposed) & $0.522$ & $0.471$ & $-5.1$ & $0.512$ & $0.718$ & $\mathbf{+20.7}$ \\
\bottomrule
\end{tabular}
\end{table}

\paragraph{Recall@$k$ operating point}
\label{sm:demand:recall_k}
Recall flattens near ten candidates on both the augmented held-out and
benchmark-reference splits.  We therefore freeze $k=10$ for
documentation injection; Table~\ref{tab:sm_demand_results}(a) gives
the operating-point recall.

\subsection{Scenario adaptation}
\label{sm:demand:adaptation_group}

\paragraph{Role-frequency reweighting}
\label{sm:demand:reweight}
Role-frequency reweighting addresses a mismatch between a
construction-heavy source corpus and a more analysis-oriented
deployment mix.  The $275$ documented entry points are assigned to
six roles: network construction, network loading, analysis execution,
time-series control, plotting, and other.  The target role shares are
estimated from unlabelled queries, so the procedure changes training
weights without requiring target-domain labels.

Let $\hat{p}_S(r)$ denote the role share in the labelled source corpus
$\mathcal{D}_{\text{train}}$ and $\hat{p}_T(r)$ the role share predicted
on $\mathcal{Q}_T$ by the unadapted ranker.  For each training pair
$(q_i, f_i, y_i)$ with $f_i$ in role $r(f_i)$ we set
\[
  w_i \;=\; \min\!\left(\frac{\hat{p}_T\bigl(r(f_i)\bigr)}{\hat{p}_S\bigl(r(f_i)\bigr)},\;
  w_{\max}\right),
\]
followed by a global normalisation $w_i \leftarrow w_i / \max_j w_j$ so
that the largest weight equals $1$.  Algorithm~\ref{alg:sm_reweight}
states the complete procedure.

\begin{algorithm}[t]
\caption{Scenario-adaptive role-frequency reweighting for the pairwise
demand ranker.}
\label{alg:sm_reweight}
\footnotesize
\begin{algorithmic}[1]
  \State \textbf{Input:} labelled training set
    $\mathcal{D}_{\text{train}} = \{(q_i, f_i, y_i)\}$,
    unlabelled target queries $\mathcal{Q}_T$,
    role assignment $r:\mathcal{F}\to\mathcal{R}$
    derived from the function cards.
  \State Compute source role distribution
    $\hat{p}_S(r) \gets$ training share of role $r$ over
    $\mathcal{D}_{\text{train}}$.
  \State Compute target role distribution
    $\hat{p}_T(r) \gets$ predicted share of role $r$ over $\mathcal{Q}_T$
    (via the unadapted ranker's top-$k$ outputs).
  \For{each role $r \in \mathcal{R}$}
    \State $\tilde{w}(r) \gets \min\bigl(\hat{p}_T(r)/\hat{p}_S(r),\,
    w_{\max}\bigr)$
    \Comment{cap $w_{\max}\!=\!3$ to stabilise training}
  \EndFor
  \State Normalise $w(r) \gets \tilde{w}(r) / \max_{r'} \tilde{w}(r')$.
  \State Assign $w_i \gets w(r(f_i))$ to each training pair.
  \State Refit pairwise logistic regression with sample weights $\{w_i\}$;
         all other pipeline components remain unchanged.
  \State \textbf{Output:} scenario-adapted pairwise ranker.
\end{algorithmic}
\end{algorithm}

\paragraph{Cross-style held-out generalisation of the adaptation}
\label{sm:demand:holdout_styles}
Weights learned from either query style reproduce the matched-style
endpoint within $0.29$pp.  The comparison holds the evaluation items
fixed because the explicit and semantic partitions have different
unweighted baselines.

\begin{center}
\footnotesize
\setlength{\tabcolsep}{3pt}
\begin{tabularx}{\linewidth}{@{}>{\raggedright\arraybackslash}p{0.24\linewidth}
  >{\centering\arraybackslash}p{0.09\linewidth}
  >{\centering\arraybackslash}p{0.12\linewidth}
  >{\centering\arraybackslash}p{0.14\linewidth}
  >{\raggedright\arraybackslash}X@{}}
\toprule
Evaluation split & $n$ & Matched & Cross-style &
\makecell[l]{$\Delta$ cross$-$matched\\(pp [$95\%$ CI])} \\
\midrule
D1+D2 explicit & $1{,}200$ & $0.6862$ & $0.6869$ &
$+0.07\,[-0.12,+0.27]$ \\
D3+D4 semantic & $800$ & $0.6619$ & $0.6590$ &
$-0.29\,[-0.62,0.00]$ \\
\bottomrule
\end{tabularx}

\smallskip
{\scriptsize The unweighted baselines are $0.5625$ and $0.4777$,
respectively.  Paired-query bootstrap: $B=10{,}000$, seed $22$.}
\end{center}

\subsection{External-query evaluation: construction protocols and full metrics}
\label{sm:demand:external_group}

The estimator generalises to all three external sources, but
the value of adaptation depends on their API-role mix.
Table~\ref{tab:sm_external_protocols} documents how the sets were
constructed and screened; main-text Table~7 gives the two fixed-arm
endpoints, and Table~\ref{tab:sm_external_queries} shows the
role-level mechanism and exploratory selector.  Every item has an
executable reference program, and oracle functions are extracted with
the same engine and function universe used for the benchmark.

\paragraph{Adaptation as a deployment-mix-specific knob (mechanism)}
\label{sm:demand:ext_mechanism}
The role columns explain the mixed result.  Adaptation consistently
favours analysis execution and often network loading, while
construction recall pays the cost.  Analysis-heavy sets therefore
improve, whereas the construction-heavy community set loses recall
even as its hit rate rises.  The same role weights are used for
all four adapted runs.  This pattern is consistent with differences in
workload mix; no set-specific refitting is performed.

\paragraph{Exploratory query-conditional selector}
\label{sm:demand:conditional}
Only the community set provides a non-degenerate test of conditional
routing.  The rule sends analysis-style queries to the adapted
arm and construction-style queries to the unadapted arm.  The other
three sets are routed wholly to one arm and therefore reproduce that
arm by construction.  On the small cross-platform set every query is
routed to the unadapted arm, so the retrained selector mean falls
below the fixed arm by exactly that arm's retraining variation
($3.2$pp).  The selector remains exploratory until tested on a new
mixed-role distribution.

\paragraph{Evaluation protocol}
\label{sm:demand:ext_protocol}
The estimator and the adaptation target $\mathcal{Q}_T$ were fixed
before any external query was collected.  For the external evaluation
the ranker was retrained on the full source corpus and its mixing
weight re-selected on the internal split ($0.6$ unadapted, $0.5$
adapted) before any external query was scored.  Each external set,
including its query text, reference code, and oracle API labels, was
frozen before evaluation and used only for evaluation.  Table~\ref{tab:sm_external_queries}(a) gives the
query-conditional selector; panel~(b) gives the role-level recall
behind the mechanism reading above.

\begin{table}[!htbp]
\centering
\caption{Construction protocols and screening provenance for the
  three external query sets of main-text Sec.~5.5.
  All reference programs execute to completion with zero silently
  dropped items ($n_{\mathrm{eval}}=n$ for every set).}
\label{tab:sm_external_protocols}
\begingroup
\footnotesize
\renewcommand{\arraystretch}{1.02}
\setlength{\tabcolsep}{4pt}
\begin{tabularx}{\linewidth}{@{}>{\raggedright\arraybackslash}p{0.18\linewidth}
  >{\raggedright\arraybackslash}X@{}}
\toprule
Aspect & Protocol record \\
\midrule
\multicolumn{2}{@{}l}{\textbf{(a) Set 1: community-sourced controlled
rewrites ($n{=}84$)}} \\[1pt]
Source &
\pp{} issue tracker and Discussions via the public REST$/$search API,
over $29$ search routes fixed in advance; titles, excerpts, URLs and
fetch dates only. \\
\addlinespace[2pt]
Screening funnel &
$448\!\to\!362$ deduplicated; Engineer~1 first-pass
$80/137/145$ include$/$borderline$/$exclude; Engineer~2 blindly
re-screened the $217$ include$\cup$borderline pool; joint review
downgraded $26$ includes to exclude and promoted $35$ borderlines to
include $\to$ $89$ eligible; $5$ near-identical cross-posts dropped
at authoring $\to$ $n{=}84$. \\
\addlinespace[2pt]
Rewriting &
Engineer~1 rewrote under fixed rules: network, operation, analysis
type and target quantity never altered, no solution steps or function
names added; Engineer~2 verified semantic preservation. \\
\addlinespace[2pt]
Composition &
$12$ families; $6/39/39$ D1$/$D2$/$D3; $50$ oracle functions;
$55/84$ ($65\%$) networks outside the PowerCodeBench pool. \\
\midrule
\multicolumn{2}{@{}l}{\textbf{(b) Set 2: cross-platform supplement
($n{=}10$)}} \\[1pt]
Source &
Declared extension round over a multi-source pool: Discussions
($6$), Stack Overflow ($2$), openmod ($2$); screened separately, never
merged into the $n{=}84$ set. \\
\addlinespace[2pt]
Screening funnel &
$595$ triaged on metadata$/$excerpt under the same criteria;
Engineer~2 independently reviewed the $69$-item full-text shortlist;
$8$ built directly and the two engineers jointly admitted $2$ of $6$
hard disagreements $\to$ $n{=}10$. \\
\addlinespace[2pt]
Rewriting &
Same two-engineer inclusion$/$exclusion, rewriting, and verification
protocol as Set~1. \\
\addlinespace[2pt]
Composition &
$4$ D2 $/$ $6$ D3; $7/10$ non-standard networks. \\
\midrule
\multicolumn{2}{@{}l}{\textbf{(c) Set 3: engineer-written ($n{=}20$)}} \\[1pt]
Source &
Two practising power engineers; their approved edition is the
authoritative text (corrections: bus numbering, N$-$1 boundaries,
loading criteria, short-circuit conditions, measurement sets,
stopping rules). \\
\addlinespace[2pt]
Screening funnel &
None.  One item required implementing the Baran--Wu $69$-bus feeder
from published data, checked against published base-case values. \\
\addlinespace[2pt]
Rewriting &
None: query text enters evaluation byte-identical; network notes and
expected-quantity fields hidden from the ranker. \\
\addlinespace[2pt]
Composition &
$7/6/7$ D1$/$D2$/$D3; $24$ oracle functions; $19$ float and $1$
integer ground truths. \\
\bottomrule
\end{tabularx}
\endgroup
\end{table}

\begin{table}[!htbp]
\centering
\caption{External-query evaluation of demand ranking.
  (a)~Exploratory query-conditional selection on three external sets
  ($n=114$) and the held-out internal set ($n=80$): share routed to the
  adapted arm and recall@10 ($\%$), including the mean [min, max] over
  five retraining seeds. (b)~Per-role recall@10 ($\%$), unadapted
  $\to$ adapted; ``---'' marks a role absent from the set, and PCB
  denotes PowerCodeBench.}
\label{tab:sm_external_queries}
\footnotesize
\textbf{(a)} Query-conditional selector\\[2pt]
\setlength{\tabcolsep}{4pt}
\begin{tabular}{lrrrr}
\toprule
 & & \multicolumn{3}{c}{conditional selector} \\
Query set & $n$ &
\makecell{routed to\\adapted} & cond. &
\makecell{cond., $5$ seeds\\mean [min, max]} \\
\midrule
Internal holdout (PCB scenarios reused) & 80 & $80/80$ & 68.94 & $71.74$ $[70.92, 72.17]$ \\
Community-sourced controlled rewrites   & 84 & $35/84$ & 62.06 & $62.39$ $[57.07, 65.07]$ \\
Cross-platform supplement               & 10 & $0/10$ & 64.72 & $61.51$ $[58.14, 64.72]$ \\
Engineer-written                        & 20 & $20/20$ & 64.52 & $65.45$ $[64.52, 67.26]$ \\
\bottomrule
\end{tabular}
\\[6pt]
\textbf{(b)} Per-role recall@10 ($\%$, unadapted $\to$ adapted)\\[2pt]
\setlength{\tabcolsep}{4pt}
\resizebox{\linewidth}{!}{%
\begin{tabular}{lccccc}
\toprule
Query set & \makecell{Analysis\\executor} & \makecell{Network\\loader} &
\makecell{Construction} & \makecell{Controller$/$\\time-series} & Other \\
\midrule
Internal holdout ($n{=}80$)      & $83.3 \to 96.1$ & $18.2 \to 48.5$ & $38.5 \to 3.8$  & $100.0 \to 100.0$ & $40.0 \to 50.0$ \\
Community-sourced rewrites ($n{=}84$) & $77.2 \to 100.0$ & $44.0 \to 46.4$ & $66.1 \to 39.2$ & $79.2 \to 83.3$   & $0.0 \to 14.3$ \\
Cross-platform ($n{=}10$)        & $90.0 \to 100.0$ & $50.0 \to 50.0$ & $68.8 \to 31.3$ & $66.7 \to 77.8$   & --- \\
Engineer-written ($n{=}20$)      & $81.8 \to 95.5$ & $10.0 \to 45.0$ & $53.8 \to 38.5$ & ---               & --- \\
\bottomrule
\end{tabular}
}

\smallskip
\parbox{0.96\linewidth}{\footnotesize\raggedright All items were evaluated
($n_{\mathrm{eval}}=n$). Each selector run uses its retrained ranker;
across the five runs, the re-evaluated fixed arms differ from the
fixed unadapted$/$adapted arms by up to $3.2$pp in five-seed mean,
the largest being the
unadapted arm on the $n{=}10$ set. The $n{=}84$ set includes two plotting items with role
recall $0$ in both arms; plotting has adaptation weight $0$, and
``Other'' spans the residual utility roles. The adapted-role
weights for analysis, loading, controller$/$time-series, construction,
other, and plotting are $1.0/0.564/0.293/0.083/0.082/0.0$. On the
community set, overall hit@$10$ changes from $83.33$ to $97.62$, and
the conditional selector's $62.06\%$ recall is $6.99$pp above the
better fixed arm ($55.07\%$).\par}
\end{table}
\section{Full experimental results}
\label{sm:full_results}

The detailed results preserve the main-text ordering.  Plain repair
has a low and front-loaded ceiling, semantically grounded tasks remain
hard after injection, robustness perturbations are small beside the
intervention effect, and measured-cost matching does not rescue dense
retrieval.

\subsection{No-intervention baselines}
\label{sm:full_results:nobaseline_group}

\paragraph{Per-model fix-round trajectory (R0--R3) without intervention}
\label{sm:full_results:fix_trajectory}
Execution feedback alone has a low ceiling, and for most models the
largest gain arrives in the first repair round.  The early columns of
Table~\ref{tab:sm_fix_trajectory} show that the R0--R1 increment
exceeds each later increment for six of the ten models; the smallest
models remain near their starting point.

\begin{table}[!htbp]
  \centering
  \caption{Accuracy progression ($\%$) across fix rounds (R0--R3) on
    PowerCodeBench, no intervention applied; this is the evidence
    behind the front-loading claim and the $1.5$B zero line of
    main-text Sec.~5.4.  $\Delta$ denotes the
    total R0$\to$R3 uplift attributable to fix rounds, computed from
    unrounded values (so it may differ from the rounded-column
    difference by $\le 0.1$pp).}
  \label{tab:sm_fix_trajectory}
  \footnotesize
  \setlength{\tabcolsep}{4pt}
  \begin{tabular}{lccccc}
    \toprule
    Model & R0 & R1 & R2 & R3 & $\Delta$(R0$\to$R3) \\
    \midrule
    \multicolumn{6}{l}{\textit{Sub-10B local models}} \\
    Qwen2.5-1.5B  & 0.0 & 0.0 & 0.0 & 0.0 & +0.0 \\
    Qwen2.5-7B    & 0.4 & 0.4 & 0.4 & 0.5 & +0.1 \\
    Llama-8B        & 0.5 & 0.7 & 1.1 & 1.3 & +0.8 \\
    \midrule
    \multicolumn{6}{l}{\textit{Mid-size local models}} \\
    Qwen2.5-14B   & 1.6 & 3.2 & 3.7 & 3.8 & +2.2 \\
    Qwen2.5-32B   & 3.8 & 8.6 & 10.2 & 11.2 & +7.4 \\
    \midrule
    \multicolumn{6}{l}{\textit{Large local models}} \\
    Qwen3-Next    & 4.0 & 5.2 & 5.9 & 7.0 & +3.0 \\
    Llama-70B       & 5.0 & 8.5 & 12.9 & 15.0 & +10.0 \\
    GPT-OSS-120B        & 14.1 & 19.4 & 21.7 & 22.9 & +8.8 \\
    Llama-405B      & 13.5 & 21.7 & 26.3 & 28.5 & +15.1 \\
    Qwen3-480B    & 15.0 & 22.0 & 25.3 & 26.5 & +11.5 \\
    \bottomrule
  \end{tabular}
\end{table}

\paragraph{Per-difficulty accuracy at R3 (without intervention)}
\label{sm:full_results:diff_r3}
Semantic grounding is already the limiting factor without
intervention.  Across the panel, D3/D4 reach roughly half the
execution-conditioned accuracy of D1/D2.  Llama-8B's zero accuracy on
D3/D4 accompanies an execution rate below $2\%$, so the failure occurs
before value matching.  The full no-intervention R3 matrix is released
with the item-level results; Table~\ref{tab:sm_pertask_compact} gives
the corresponding proactive Round-0 view.

\subsection{Proactive base-C breakdowns}
\label{sm:full_results:basec_group}

\paragraph{Per-model reactive ablation under proactive base C}
\label{sm:full_results:reactive_pairwise_C}
Under proactive base C, the value-error branch is the only added step
that is non-negative for every model.  Routed documentation is close
to always-on documentation overall, while its negative deltas on the
largest models are consistent with C having already supplied useful
API evidence.  The three delta columns of
Table~\ref{tab:sm_reactive_pairwise_C} separate always-on
documentation, routed documentation, and value-error repair.

\begin{table}[!htbp]
\centering
\caption{Per-model accuracy ($\%$) at Round-3 under proactive base C,
  ablation deltas in pp.  The \textit{Mean} row equals the C row of
  the main-text reactive summary.  Deltas: FD--FX, always-on
  documentation over plain fix; FDR--FD, demand-routed versus
  always-on; FDRS--FDR, the value-error branch.}
\label{tab:sm_reactive_pairwise_C}
\footnotesize
\setlength{\tabcolsep}{4pt}
\resizebox{\linewidth}{!}{%
\begin{tabular}{lrrrr|rrr}
\toprule
Model & C+FX & C+FD & C+FDR & C+FDRS &
\makecell{$\Delta$\\FD--FX} & \makecell{$\Delta$\\FDR--FD} & \makecell{$\Delta$\\FDRS--FDR} \\
\midrule
Qwen2.5-1.5B   &  9.90 &  9.40 &  9.55 &  9.60 & $-0.50$ & $+0.15$ & $+0.05$ \\
Qwen2.5-7B     & 29.55 & 31.30 & 31.25 & 32.80 & $+1.75$ & $-0.05$ & $+1.55$ \\
Qwen2.5-14B    & 42.55 & 46.45 & 46.10 & 48.60 & $+3.90$ & $-0.35$ & $+2.50$ \\
Qwen2.5-32B    & 50.75 & 52.20 & 52.75 & 52.75 & $+1.45$ & $+0.55$ & $+0.00$ \\
Llama-8B         & 28.95 & 31.05 & 34.40 & 36.30 & $+2.10$ & $+3.35$ & $+1.90$ \\
Llama-70B        & 49.00 & 51.00 & 51.45 & 57.45 & $+2.00$ & $+0.45$ & $+6.00$ \\
Llama-405B       & 64.75 & 66.45 & 64.75 & 68.70 & $+1.70$ & $-1.70$ & $+3.95$ \\
Qwen3-Next     & 55.00 & 57.85 & 57.95 & 59.55 & $+2.85$ & $+0.10$ & $+1.60$ \\
Qwen3-480B     & 63.80 & 65.45 & 63.75 & 65.90 & $+1.65$ & $-1.70$ & $+2.15$ \\
GPT-OSS-120B         & 58.40 & 59.80 & 58.35 & 60.25 & $+1.40$ & $-1.45$ & $+1.90$ \\
\midrule
\textit{Mean}        & 45.27 & 47.10 & 47.03 & 49.19 & $+1.83$ & $-0.07$ & $+2.16$ \\
\bottomrule
\end{tabular}
}
\end{table}

\paragraph{Per-task and per-difficulty Round-0 accuracy under proactive base C}
\label{sm:full_results:per_task_r0}
Semantically grounded tasks remain the limiting difficulty even at the
largest scales.  Table~\ref{tab:sm_pertask_compact} makes the contrast
visible within each row: D1/D2 remain easier than D3/D4.  Main-text
Fig.~6 gives the companion task-family view.

\begin{table}[!htbp]
\centering
\caption{Per-difficulty Round-0 accuracy ($\%$) under proactive base C
  (no reactive correction) on the $2{,}000$-item benchmark: nine
  open-weight models (by parameter count) and four closed-source API
  vendors.  Per-family resolution in main-text Fig.~6; full matrices
  in \texttt{results/full\_matrices/}.}
\label{tab:sm_pertask_compact}
\footnotesize
\setlength{\tabcolsep}{4pt}
\begin{tabular}{lrrrr}
\toprule
 & \multicolumn{4}{c}{Per difficulty} \\
Model & D1 & D2 & D3 & D4 \\
\midrule
\multicolumn{5}{l}{\textit{Open-weight (locally deployable)}} \\
Qwen2.5-7B    & 30.7 & 30.3 & 11.0 & 13.3 \\
Llama-8B        & 20.2 & 35.8 &  4.8 & 11.8 \\
Qwen2.5-14B   & 39.3 & 40.2 & 26.0 & 25.8 \\
Qwen2.5-32B   & \textbf{52.0} & \textbf{51.0} & 25.3 & 33.8 \\
Llama-70B       & 44.0 & 48.0 & 25.0 & 27.8 \\
Qwen3-Next    & 48.3 & 57.5 & 39.5 & 37.8 \\
GPT-OSS-120B        & 54.8 & 49.2 & 39.5 & 32.8 \\
Llama-405B      & 68.3 & 50.0 & 46.0 & 33.8 \\
Qwen3-480B    & 67.0 & 61.3 & 50.0 & 40.8 \\
\midrule
\multicolumn{5}{l}{\textit{Closed-source API (mid-tier)}} \\
Claude-Haiku-4-5    & 39.8 & 44.5 & 25.8 & 32.3 \\
Gemini-2.5-Flash    & 47.3 & 43.8 & 33.8 & 29.5 \\
GPT-5.4-mini        & 60.5 & 61.2 & 40.8 & 29.0 \\
DeepSeek-V4-Flash   & 54.7 & 40.2 & 36.0 & 25.0 \\
\bottomrule
\end{tabular}
\end{table}

\subsection{Robustness and variance checks}
\label{sm:full_results:robustness_group}

\paragraph{Naturalistic-query holdout (per-model)}
\label{sm:full_results:holdout}
The naturalistic holdout preserves the broad model ordering, although
the C+FDRS endpoints are lower than on the full benchmark and the
bare-prompt endpoints are higher.  The per-model
columns of Table~\ref{tab:sm_holdout} show both the endpoint shift and
the intervention gain; they do not by themselves identify why a model
is more sensitive to the change in wording.

\begin{table}[!htbp]
\centering
\caption{Naturalistic-query holdout versus the full $2{,}000$-item
  PowerCodeBench (PCB), open-weight panel ($\geq\!7$B): per-model
  evidence for the broad tier ordering preserved in main-text Sec.~5.5.
  The holdout is $80$ human-authored queries ($\pm 5$pp one-SE band);
  its C+FDRS endpoints average $6.78$pp below PCB and its A endpoints
  $5.79$pp above.}
\label{tab:sm_holdout}
\footnotesize
\setlength{\tabcolsep}{4pt}
\begin{tabular}{lrrrr|rr}
\toprule
                       & \multicolumn{2}{c}{Holdout ($n\!=\!80$)} & \multicolumn{2}{c|}{PCB ($n\!=\!2{,}000$)} & \multicolumn{2}{c}{$\Delta$ FDRS--A} \\
Model                  & A     & C+FDRS & A     & C+FDRS & Hold     & PCB      \\
\midrule
Qwen2.5-7B       &  0.00 & 28.75 &  0.40 & 32.80 & $+28.75$ & $+32.40$ \\
Llama-8B           &  1.25 & 20.00 &  0.50 & 36.30 & $+18.75$ & $+35.80$ \\
Qwen2.5-14B      &  6.25 & 42.50 &  1.60 & 48.60 & $+36.25$ & $+47.00$ \\
Qwen2.5-32B      &  7.50 & 42.50 &  3.80 & 52.75 & $+35.00$ & $+48.95$ \\
Qwen3-Next       &  7.50 & 51.25 &  4.00 & 59.55 & $+43.75$ & $+55.55$ \\
Llama-70B          & 15.00 & 61.25 &  5.00 & 57.45 & $+46.25$ & $+52.45$ \\
GPT-OSS-120B           & 18.75 & 56.25 & 14.05 & 60.25 & $+37.50$ & $+46.20$ \\
Llama-405B         & 30.00 & 66.25 & 13.45 & 68.70 & $+36.25$ & $+55.25$ \\
Qwen3-480B       & 23.75 & 52.50 & 15.05 & 65.90 & $+28.75$ & $+50.85$ \\
\midrule
\textit{Panel mean}    & 12.22 & 46.81 &  6.43 & 53.59 & $+34.58$ & $+47.16$ \\
\bottomrule
\end{tabular}
\end{table}

\begin{figure}[!htbp]
  \centering
  \includegraphics[width=\linewidth]{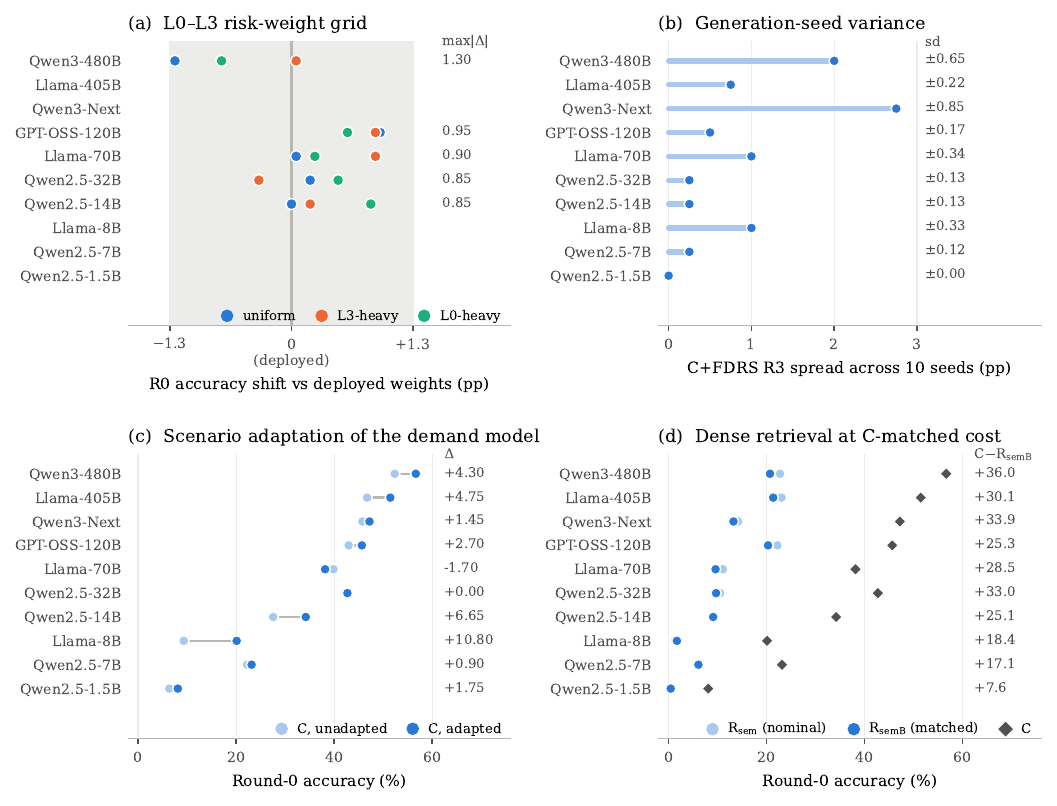}
  \caption{Four robustness checks on a shared model axis.
    \textbf{(a)}~Risk-weight grid, as the shift from the deployed
    mild-L3 setting; shaded band $\pm1.3$pp.
    \textbf{(b)}~Seed spread of C+FDRS R3 over $10$ vLLM seeds.
    \textbf{(c)}~Condition C, unadapted versus role-reweighted demand
    predictor.
    \textbf{(d)}~Dense retrieval at nominal budget (R$_{\text{sem}}$)
    versus C-matched measured cost (R$_{\text{semB}}$).}
  \label{fig:sm_robustness}
\end{figure}

\paragraph{Risk-score weighting sensitivity}
\label{sm:full_results:risk_weight}
The risk-layer weights are not load-bearing.  Across the grid in
Fig.~\ref{fig:sm_robustness}(a), the largest accuracy shift is
$1.30$pp, and only three of the thirty item-paired contrasts differ
from zero.  All three involve the L3-heavy corner:

\begin{center}
\footnotesize
\setlength{\tabcolsep}{6pt}
\begin{tabular}{@{}lrr@{}}
\toprule
Contrast & $\Delta$ accuracy (pp) & Exact McNemar $p$ \\
\midrule
Llama-70B: deployed $\to$ L3-heavy & $+0.90$ & $0.022$ \\
Llama-70B: uniform $\to$ L3-heavy & $+0.85$ & $0.030$ \\
Qwen3-480B: uniform $\to$ L3-heavy & $+1.30$ & $0.033$ \\
\bottomrule
\end{tabular}
\end{center}

The four weight settings are registered in
Sec.~\ref{sm:repro:prompt}.

\paragraph{Profile-conditioning isolation at a shared nominal budget}
\label{sm:full_results:profile_isolation}
Profile conditioning changes the evidence supplied but produces no
detectable accuracy increment in this same-nominal-budget scan.  The paired
Round-0 design replaces the per-model layer deficits with the maximal
constant deficit $r_{\mathcal{M},\ell}(f)=1$, while holding the demand model,
renderer, selection rules, and nominal budget fixed.  Across
Qwen2.5-14B, Llama-70B, and Qwen3-480B on the same $400$ stratified
items, prompt length changes on $33$--$64\%$ of items, confirming that
the manipulation is active.

\begin{center}
\footnotesize
\setlength{\tabcolsep}{7pt}
\begin{tabular}{@{}rrr@{}}
\toprule
Budget $B$ & Profile$-$constant accuracy (pp) & Paired $95\%$ CI (pp) \\
\midrule
$600$     & $-1.17$ & $[-2.83,+0.50]$ \\
$800$     & $+1.08$ & $[-1.00,+3.25]$ \\
$1{,}000$ & $+0.58$ & $[-1.33,+2.42]$ \\
$2{,}000$ & $-0.08$ & $[-1.92,+1.67]$ \\
\bottomrule
\end{tabular}
\end{center}

All four panel-mean and twelve per-model intervals cover zero.  The
$41\%$ prompt-cost result belongs to condition C as a complete
selective pipeline, not to the profile in isolation.

\paragraph{Demand-adaptation robustness (per-model)}
\label{sm:full_results:adapt_robustness}
Most of the proactive gain survives without adaptation.  The
panel-mean gains are $27.79$pp over A and $22.59$pp over R, and both
contrasts are positive for every model.  Adaptation adds $3.16$pp,
about $10\%$ of the complete lift.  Figure~\ref{fig:sm_robustness}(c)
gives the per-model Round-0 comparison.

\paragraph{Multi-seed generation variance}
\label{sm:full_results:multiseed}
Run-to-run jitter is small beside the effect it qualifies.  Across ten
independent vLLM launches, the panel-median standard deviation is
$0.20$pp and the worst cell spans $2.75$pp, compared with
$32$--$56$pp intervention gains; Fig.~\ref{fig:sm_robustness}(b)
shows the per-model spread.  Greedy decoding removes sampling but not
exact-logit ties, parallel-kernel variation, or continuous-batching
order.  Closed-source APIs are excluded because their generation
seeds are not exposed.

\subsection{Dense retrieval at matched measured cost: budget calibration and full panel}
\label{sm:full_results:semantic_rag}

At equal \emph{measured} cost the dense-retrieval gap does not close;
it stems from selection quality, not from token budget.  This
subsection defines the equal-cost condition R$_{\text{semB}}$,
records how its budget was calibrated, and reports the full panel in
panel~(d) of Fig.~\ref{fig:sm_robustness}.

\paragraph{R$_{\text{semB}}$ definition}
\label{sm:full_results:rsemb_def}
Equal nominal budgets do not equalise measured prompt cost.
R$_{\text{semB}}$ therefore changes only the nominal budget of
R$_{\text{sem}}$ to match C's realised tokens.  Both dense conditions
use the same $275$ documentation cards, top-$10$ depth, renderer, and
greedy packer; they omit the demand model, risk gating, and boundary
cards.  The scorer is Sentence-BERT bi-encoder cosine
(\texttt{all-MiniLM-L6-v2}) without reranking.

\paragraph{Full-panel result}
\label{sm:full_results:rsemb_result}
Cost matching usually lowers dense-retrieval accuracy and often
widens its gap to C; only the floor-regime models remain effectively
unchanged.  Figure~\ref{fig:sm_robustness}(d) gives both variants for
all ten models.  The persistent gap points to selection quality, not
token volume.

\paragraph{$B^{*}$ calibration protocol}
\label{sm:full_results:rsemb_calibration}
A single budget was selected offline on a stratified calibration set,
fixed before the full-panel run, and applied without changing the
retriever, renderer, or ranking model.  The compact record below
separates nominal limits from measured tokens.

\begin{table}[!ht]
\centering
\footnotesize
\setlength{\tabcolsep}{5pt}
\begin{tabular}{@{}>{\raggedright\arraybackslash}p{0.215\linewidth}
  >{\raggedright\arraybackslash}p{0.26\linewidth}
  >{\raggedright\arraybackslash}p{0.435\linewidth}@{}}
\toprule
Quantity & Value & Reading \\
\midrule
Nominal budget & $2{,}000$ &
C realises about $1{,}450$ tokens; R$_{\text{sem}}$ realises
$3{,}639$--$3{,}715$, and BM25 R realises $4{,}202$ (per-model mean
of realised prompt tokens; main-text Table~6 reports the pooled mean
$4{,}189$). \\
Target window & $1{,}450\pm10\%$; $[1{,}305,1{,}595]$ &
$B^{*}=600$ is the unique grid hit ($+0.7\%$); neighbouring budgets
miss by $-16.5\%$ and $+18.1\%$. \\
Production means & R$_{\text{semB}}$: $1{,}429$--$1{,}474$;
C: $1{,}415$--$1{,}504$ &
One $B^{*}$ is used for all ten models. \\
Calibration check & $n=100$, seed $22$;
strata $30/30/20/20$ &
Runtime C mean $1{,}464.8$; reconstructed token counts match
$100/100$. \\
\bottomrule
\end{tabular}
\end{table}
\FloatBarrier

\subsection{Reactive-router destinations per model}
\label{sm:full_results:router_per_model}

Adding the value-error route changes where repair attempts go:
executed but numerically wrong programs move from basic repair to
\texttt{semantic\_fix}.  Table~\ref{tab:sm_router_per_model} makes the
shift visible in the smaller \texttt{basic\_fix} share and the new
semantic share.  Shares are computed over fix attempts and sum to
$100\%$ within each run, up to rounding.

\begin{table}[!htbp]
\centering
\caption{Reactive-router destinations per model: share ($\%$) of fix
  attempts per route under proactive base C, for the C+FDR and
  C+FDRS runs.  These are the per-route, per-model shares cited in
  main-text Sec.~5.4.
  Shares are computed from the per-route attempt
  counts of the released run logs; each row sums to $100\%$ within a
  run up to rounding.}
\label{tab:sm_router_per_model}
\footnotesize
\setlength{\tabcolsep}{4pt}
\begin{tabular}{lrrr|rrrr}
\toprule
 & \multicolumn{3}{c|}{C+FDR} & \multicolumn{4}{c}{C+FDRS} \\
Model & \makecell{basic\\fix} & \makecell{api\\doc} &
\makecell{bound.\\contr.} & \makecell{basic\\fix} &
\makecell{api\\doc} & \makecell{bound.\\contr.} &
\makecell{semantic\\fix} \\
\midrule
Qwen2.5-1.5B & 22.4 & 45.4 & 32.1 &  3.4 & 46.0 & 31.9 & 18.7 \\
Qwen2.5-7B   & 23.8 & 52.9 & 23.3 &  2.1 & 53.2 & 22.9 & 21.9 \\
Qwen2.5-14B  & 36.4 & 53.8 &  9.9 &  3.6 & 53.6 &  9.7 & 33.1 \\
Qwen2.5-32B  & 34.7 & 55.3 &  9.9 &  3.3 & 52.8 &  7.6 & 36.3 \\
Llama-8B       & 20.1 & 53.8 & 26.1 &  3.4 & 52.9 & 25.7 & 18.1 \\
Llama-70B      & 52.6 & 35.3 & 12.2 & 12.2 & 36.8 & 10.6 & 40.4 \\
Llama-405B     & 49.7 & 37.2 & 13.2 &  4.9 & 36.4 & 11.9 & 46.9 \\
Qwen3-Next   & 44.1 & 48.6 &  7.3 &  4.9 & 48.3 &  6.6 & 40.2 \\
Qwen3-480B   & 60.1 & 31.1 &  8.8 &  5.1 & 30.6 &  8.5 & 55.8 \\
GPT-OSS-120B       & 45.5 & 50.2 &  4.3 &  5.1 & 49.6 &  3.6 & 41.7 \\
\bottomrule
\end{tabular}
\end{table}
\section{Reproducibility configuration}
\label{sm:reproducibility}

This section is the lookup panel for the fixed configuration values
used across the experiments.  Three families of constants live
elsewhere and are not repeated here: the injector constants in
Sec.~\ref{sm:intervention:constants}, the benchmark generation
parameters in Sec.~\ref{sm:benchmark}, and the audit protocol in
Sec.~\ref{sm:validity}.  The system prompt,
the user-turn templates of each condition, and the fix-round template
are included in the PowerCodeBench repository
(\texttt{docs/PROMPTS.md}) instead of being reproduced here.

\subsection{Decoding and execution}
\label{sm:repro:prompt}

\begin{table}[!htbp]
\centering
\footnotesize
\setlength{\tabcolsep}{4pt}
\begin{tabularx}{\linewidth}{@{}>{\raggedright\arraybackslash}p{0.18\linewidth}
  >{\raggedright\arraybackslash}X@{}}
\toprule
Setting & Value \\
\midrule
System prompt &
A role statement for the target library, a four-step instruction
(load the network, apply the modifications, run the analysis, print
the final numeric value), and an output-only-code rule.  Documentation
snippets enter through the user turn under the injection conditions;
fix rounds use a separate debugger system prompt. \\
Decoding &
Greedy, temperature $0$, every model.  Max new tokens $4{,}096$
(first pass) and $2{,}048$ (fix rounds). \\
Serving (open weight) &
vLLM, BF16, no quantisation; NVIDIA GH200, four GPUs per node, tensor
parallelism scaled to model size. \\
Sandbox &
$60\,$s execution timeout with a $75\,$s parent watchdog that recovers
stuck workers; execution harness as below. \\
API policy (closed source) &
Same max new tokens; temperature $0$ where the vendor exposes it;
reasoning-mode features and tool$/$function calling disabled. \\
Access windows &
Four-vendor API panel of main-text Sec.~5.2: 6--10~May~2026.
DeepSeek cells of the transfer pilot (Sec.~\ref{sm:transfer}):
26--27~July~2026. \\
\bottomrule
\end{tabularx}
\end{table}

Two settings need explanatory notes.  The tool- and reasoning-mode
switches are off so that the four-vendor panel stays inside the
mid-tier deployment regime evaluated in the main text; vendor agent
scaffolding is outside this comparison.  The \emph{execution harness}
does more than impose a timeout: each run executes with the harness
sandbox directory as its working directory and file output paths
redirected into it; display, figure, and browser calls are replaced
by mocks (matplotlib on a non-interactive backend);
interpreter-terminating calls are stripped; and
\texttt{stdout}$/$\texttt{stderr} are captured by the harness.  The
harness applies no filesystem allowlist.  Benchmark items execute
in-process under the $60\,$s timeout and $75\,$s watchdog.  On every
backend, L3 application probes run in a fresh single-threaded Python
subprocess with its own $30\,$s bound.

\paragraph{Seeds and determinism}
All experiments use fixed seeds and nominally greedy decoding: the
PowerCodeBench item generator uses seed $42$
(Sec.~\ref{sm:benchmark:pipeline}); the demand-predictor train/test
group split uses seed $22$; and vLLM open-weight inference uses
sampling seed $22$.  The multi-seed variance check of
Sec.~\ref{sm:full_results:multiseed} runs on a stratified $400$-item
subset (\texttt{sample\_seed}$=$$22$, strata $=$ difficulty) under
generation seeds $\{22,33,44,55,66,77,88,99,111,222\}$.  Temperature
$0$ removes stochastic sampling but does not make parallel GPU
serving bitwise deterministic; the independent-run spread is reported
in Sec.~\ref{sm:full_results:multiseed}.  The released reproduction
pipeline pins the L0--L3 layer weights to the deployed mild-L3
setting $(0.20, 0.25, 0.20, 0.35)$ (the \texttt{RISK\_WEIGHTS}
constant of the proactive-injector module).  The alternatives in the
sensitivity grid (Fig.~\ref{fig:sm_robustness}(a)), namely uniform
$(0.25, 0.25, 0.25, 0.25)$, L3-heavy $(0.0, 0.20, 0.20, 0.60)$, and
inverted L0-heavy $(0.40, 0.30, 0.20, 0.10)$, are explicit opt-in
overrides.

\begin{table}[!htbp]
\centering
\setlength{\tabcolsep}{2pt}
\footnotesize
\caption{Model panel of main-text Sec.~5.1 plus the $0.5$B
  serving-only reference of Table~11: identifiers, parameter
  counts, serving configuration.  Open-weight models run locally on
  vLLM at BF16; closed-source models use vendor APIs under the
  no-tool, temperature-$0$ regime of \S\ref{sm:repro:prompt}.  Params
  is total/active where reported for MoE entries.}
\label{tab:sm_model_identifiers}
\begin{tabularx}{\linewidth}{@{}>{\raggedright\arraybackslash}p{0.24\linewidth}
  >{\raggedright\arraybackslash}X
  >{\centering\arraybackslash}p{0.085\linewidth}
  >{\raggedright\arraybackslash}p{0.16\linewidth}@{}}
\toprule
Model & Identifier & Params (B) & Serving \\
\midrule
\multicolumn{4}{@{}l}{\textit{Open-weight (HuggingFace \texttt{repo/name})}} \\
Qwen2.5-Coder-0.5B & \texttt{Qwen/Qwen2.5-Coder-0.5B-Instruct} & 0.5 & vLLM, BF16 (serving only) \\
Qwen2.5-Coder-1.5B & \texttt{Qwen/Qwen2.5-Coder-1.5B-Instruct} & 1.5 & vLLM, BF16 \\
Qwen2.5-Coder-7B   & \texttt{Qwen/Qwen2.5-Coder-7B-Instruct}   & 7   & vLLM, BF16 \\
Qwen2.5-Coder-14B  & \texttt{Qwen/Qwen2.5-Coder-14B-Instruct}  & 14  & vLLM, BF16 \\
Qwen2.5-Coder-32B  & \texttt{Qwen/Qwen2.5-Coder-32B-Instruct}  & 32  & vLLM, BF16 \\
Qwen3-Coder-Next   & \texttt{Qwen/Qwen3-Coder-Next}            & 80  & vLLM, BF16 \\
Qwen3-Coder-480B   & \texttt{Qwen/Qwen3-Coder-480B-A35B-Instruct} & 480/35 & vLLM, BF16 \\
Llama-3.1-8B       & \texttt{meta-llama/Llama-3.1-8B-Instruct}   & 8   & vLLM, BF16 \\
Llama-3.1-70B      & \texttt{meta-llama/Llama-3.1-70B-Instruct}  & 70  & vLLM, BF16 \\
Llama-3.1-405B     & \texttt{meta-llama/Llama-3.1-405B-Instruct} & 405 & vLLM, BF16 \\
GPT-OSS-120B       & \texttt{openai/gpt-oss-120b}                & 120 & vLLM, BF16 \\
\midrule
\multicolumn{4}{@{}l}{\textit{Closed-source mid-tier API (vendor model ID)}} \\
Claude-Haiku-4-5  & \texttt{claude-haiku-4-5}  & --- & API, no-tool \\
Gemini-2.5-Flash  & \texttt{gemini-2.5-flash}  & --- & API, no-tool \\
GPT-5.4-mini      & \texttt{gpt-5.4-mini}      & --- & API, no-tool \\
DeepSeek-V4-Flash & \texttt{deepseek-v4-flash} & --- & API, no-tool \\
\bottomrule
\end{tabularx}
\end{table}

\subsection{Vanilla BM25 (condition R) configuration}
\label{sm:repro:bm25}

The vanilla BM25 retrieval baseline (condition R) indexes the same
$275$-entry documentation corpus used by the L0--L3 probe
generator and the demand predictor.  Each documented entry is
flattened to ``name + full path + category + signature + description
+ parameter names and descriptions + return information''
(the \texttt{examples} field is excluded so that BM25 is not biased
toward construction-heavy demonstration patterns) and tokenised with
a regular-expression identifier tokeniser
(\texttt{[A-Za-z][A-Za-z0-9\_]*}, lower-cased, preserving
\texttt{snake\_case}).  Retrieval uses
\texttt{rank\_bm25.BM25Okapi} at the library defaults
$(k_1{=}1.5, b{=}0.75, \varepsilon{=}0.25)$, and emits the
top-$10$ functions, which are then rendered through the same
per-function snippet builder used by C and by X, the demand-hit
full-layer reference condition of main-text Sec.~5.1 (no boundary
cards, no query anchors, no model-side risk gating).

\subsection{Demand predictor training and adaptation}
\label{sm:repro:demand}

The corpus, the $8{:}1$ negative--positive sampling ratio, the
group-level split, and the reweighting algorithm are specified in
main-text Sec.~4.3 and Secs.~\ref{sm:demand:corpus}
and~\ref{sm:demand:reweight}.  Four deployment values do not follow
from them and are registered here.  The pairwise component is fitted
with \texttt{scikit-learn} L2 logistic regression.  The hybrid mixing
weight $\alpha$ is auto-selected on the held-out augmented
development split by maximising recall@$10$ over the grid
$\alpha \in \{0.0, 0.1, \ldots, 1.0\}$, and the two rankers of
the end-to-end ablation select $\alpha=0.7$ (unadapted) and
$\alpha=0.4$ (role-reweighted).  The adaptation target
$\mathcal{Q}_T$ is the PowerCodeBench natural-language query pool.
The undeployed demand-model variants (zero-shot SBERT, zero-shot
cross-encoder, hybrid cross-encoder) follow the
\texttt{sentence-transformers} defaults documented alongside the
artifact and are not part of the deployed C pipeline.

\subsection{Software versions}
\label{sm:repro:software}

The experimental environment uses Python $3.11$, \pp{} $3.4.0$
(held fixed for the experimental window), NumPy $2.2.6$, pandas $2.3.3$,
\texttt{vllm} $0.15.1$, \texttt{transformers} $4.57.6$,
\texttt{sentence-transformers} $5.2.3$, \texttt{scikit-learn} $1.6.0$,
and \texttt{rank-bm25} $0.2.2$.  The OpenAI and DeepSeek endpoints
use the OpenAI-compatible client in \texttt{openai} $2.26.0$;
Anthropic and Gemini calls use their HTTPS REST interfaces through
Python's standard-library client.  All client versions are pinned in
the released environment specification.
\section{Engineering-validity audit of the scalar acceptance protocol}
\label{sm:validity}

Scalar matching is informative but does not establish that the
requested physical workflow was followed.  The paired audit measures
how the intervention changes false acceptance for the same item and
model, whereas the deployed-frame audit estimates the remaining risk
among accepted outputs in operation
(Tables~\ref{tab:sm_e1p_mcnemar_full}
and~\ref{tab:sm_far_stratified}).

Throughout, the \emph{false-acceptance rate} (FAR) is the fraction of
scalar-accepted outputs whose final ruling is contaminated because the
program did not realise the queried physical scenario.  Item-defect
rulings are removed from both numerator and denominator.

\subsection{What was audited and how}
\label{sm:validity:sampling}

The audit examines $610$ outputs that passed scalar matching.  Five
instrumented checks create a high-recall review queue but never decide
validity.  Flagged records then pass through a fixed decision
calculus, independent review by two engineers, and adversarial
refutation; the review also includes every unflagged record.
Because every successful challenge removes a contamination ruling,
this chain is deliberately biased against overstating contamination.

\paragraph{Sampling frames}
All frames were drawn with fixed seeds before adjudication from the
scalar-accepted records of the $2{,}000$-item benchmark run.
The unpaired frames estimate FAR within endpoint populations; the
paired frame estimates the within-item, within-model difference on
common support.

\begin{table}[!htbp]
\centering
\footnotesize
\setlength{\tabcolsep}{4pt}
\begin{tabularx}{\linewidth}{@{}>{\raggedright\arraybackslash}p{0.14\linewidth}>{\raggedright\arraybackslash}p{0.18\linewidth}>{\raggedright\arraybackslash}p{0.24\linewidth}>{\raggedright\arraybackslash}p{0.08\linewidth}>{\raggedright\arraybackslash}X@{}}
\toprule
Frame & Endpoint & Models & $n$ & Strata \\
\midrule
\textbf{main\_P} & \emph{C+FDRS} (deployed) & Open-weight panel &
$400$ & Tier $\times$ family \\
\textbf{api\_T5} & \emph{C+FDRS} (deployed) & Four closed-source API
models & $60$ & Tier $\times$ family \\
\textbf{contrast\_S} & \emph{A+FX} (plain-fix baseline) & Open-weight
panel, T2--T4 & $150$ & Tier $\times$ family, T2--T4 only$^{\ast}$ \\
\textbf{paired} & Both arms on the same (model, item) &
Common support: $2{,}106$ combinations accepted under both arms &
$170$ pairs & $120$-pair base draw plus a pre-specified $50$-pair T4
extension (Sec.~\ref{sm:validity:paired}) \\
\bottomrule
\end{tabularx}

\smallskip
\parbox{0.96\linewidth}{\footnotesize\raggedright $^{\ast}$The \textbf{contrast\_S} accepted pool at T1
is empty, so that frame structurally contains no T1 stratum.
T1--T4 are capability tiers of the open-weight panel: T1 sub-$7$B,
T2 $7$--$32$B, T3 $70$--$120$B, and T4 the upper group
(Llama-405B, Qwen3-480B, and Qwen3-Next), which is enumerated by
model rather than by a parameter range.  \textbf{api\_T5} is a frame
label for the closed-source panel, not a fifth capability tier.\par}
\end{table}

An independent $20$-item calibration pilot is excluded from every
reported frame.  Design-weighted results use the sampler's allocation
cells as strata, with a H\'ajek ratio estimator and stratified
nonparametric bootstrap intervals that are mildly conservative by
construction.

\paragraph{Automatic checks}
\label{sm:validity:gate}

The automatic layer is sensitive enough to prioritise review but not
precise enough to decide validity.  Any non-pass creates an
adjudication flag; contamination is determined only by the
review pipeline.  Table~\ref{tab:sm_gate_operating}(a) reports its
operating characteristics.

The reference program must first reproduce the recorded ground truth
in the released environment.  This entry gate passes for every item in
the $610$-record round.  Each
generated program then passes through five instrumented checks:

(1)~\emph{Network and modification state}: structural signature and
component-level modification fingerprint, taken before the first
solve for multi-solve families.
(2)~\emph{Solver identity and convergence} takes a trace over the
\pp{} solver entry points and checks it against a per-family
expectation table with the reference run's own trace as authority.
(3)~\emph{Power balance and sanity}: injection$-$load$-$losses
residual within $\max(0.5\,\mathrm{MW},\,0.5\%$ of load$)$, voltages
in a hard $[0.5, 1.5]$\,p.u.\ band, OPF feasibility, short-circuit
result tables present.
(4)~\emph{Result extraction} AST-parses the printed extraction
expression on both sides and aligns it via a fixed table of
mathematically equivalent idioms.
(5)~\emph{Output provenance} combines a static scan for ground-truth
literals without a preceding solver call with a perturbation
re-execution verifying the printed scalar responds to the network
(provably perturbation-insensitive ``stiff'' quantities exempted).

\begin{table}[!htbp]
\centering
\caption{Screening and tolerance sensitivity.
  (a)~Five-check layer versus the $610$ audit rulings (item defects
  excluded; flag $=$ non-pass).  (b)~Tolerances from
  $10^{-2}/10^{-3}$ to $10^{-3}/10^{-4}$; ``cont.\ rej.''$/$``legit.\
  lost'' are shares of false acceptances rejected$/$legitimate matches
  lost.}
\label{tab:sm_gate_operating}
\footnotesize
\textbf{(a)} Five-check automatic layer\\[2pt]
\setlength{\tabcolsep}{4pt}
\begin{tabular}{lrrrrrrr}
\toprule
Frame & $n$ & Prev. & Flag & Sens. & Spec. & PPV & NPV \\
\midrule
main\_P       & 375 & 22.40 & 49.87 & 89.29 & 61.51 & 40.11 & 95.21 \\
api\_T5       &  53 & 13.21 & 45.28 & 85.71 & 60.87 & 25.00 & 96.55 \\
C combined    & 428 & 21.26 & 49.30 & 89.01 & 61.42 & 38.39 & 95.39 \\
contrast\_S   & 144 & 36.11 & 61.11 & 92.31 & 56.52 & 54.55 & 92.86 \\
\midrule
\textbf{All frames} & \textbf{572} & \textbf{25.00} & \textbf{52.27} & \textbf{90.21} & \textbf{60.37} & \textbf{43.14} & \textbf{94.87} \\
\bottomrule
\end{tabular}
\\[2pt]
\parbox{0.96\linewidth}{\footnotesize\raggedright Pooled: TP$=129$, FP$=170$, FN$=14$, TN$=259$;
balanced accuracy $75.29$.  Wilson $95\%$ on the pooled row: %
sens.\ $[84.24, 94.08]$, spec.\ $[55.67, 64.89]$,
PPV $[37.65, 48.81]$, NPV $[91.58, 96.92]$.
Of the $332$ flagged records, $33$ were later ruled item defects.
PPV$/$NPV $=$ positive$/$negative predictive value; Prev.\ $=$
prevalence.\par}
\\[6pt]
\textbf{(b)} Matching-tolerance tightening\\[2pt]
\setlength{\tabcolsep}{3.5pt}
\begin{tabular}{lcccc}
\toprule
Frame & FAR$_{\mathrm{unw}}$ (\%) & FAR$_{\mathrm{w}}$ (\%) [boot $95\%$] &
cont.\ rej.\ (\%) & legit.\ lost (\%) \\
\midrule
main\_P      & $22.40 \to 14.88$ & $18.51 \to 11.96\,{\scriptstyle[9.02, 15.38]}$  & 40.48 & 1.72 \\
api\_T5      & $13.21 \to 11.76$ & $12.65 \to 11.74\,{\scriptstyle[3.69, 22.02]}$  & 14.29 & 2.17 \\
C combined   & $21.26 \to 14.47$ & $16.69 \to 11.89\,{\scriptstyle[8.38, 15.80]}$  & 38.46 & 1.78 \\
contrast\_S  & $36.11 \to 27.78$ & $33.51 \to 25.45\,{\scriptstyle[17.53, 33.98]}$ & 32.69 & 1.09 \\
\bottomrule
\end{tabular}
\end{table}

\paragraph{Decision calculus}
\label{sm:validity:pipeline}

The governing meta-rule, fixed before the formal round, is: an
accepted output is a false acceptance if and only if at least one of
checks (1)--(5) fails in a way not covered by the \emph{equivalence
exception}, which applies only when (a) the deviation is a pure
implementation$/$idiom difference, a mathematically equivalent
computation, or a method choice the query does not constrain,
\emph{and} (b) the resulting physical scenario and the extracted
quantity coincide with the reference intent.  Ten boundary cases
were fixed before the calibration pilot:

\begin{table}[!htbp]
\centering
\footnotesize
{\emph{Boundary cases (equivalent versus contaminated), fixed
before the calibration pilot.}}\\[3pt]
\begin{tabular}{@{}p{0.36\linewidth}p{0.60\linewidth}@{}}
\toprule
\emph{Equivalent} & \emph{Contaminated} \\
\midrule
Loader aliasing &
Solver substitution changing a query-constrained physical model (AC
for DC) \\
Naming-only differences &
Omission of a query-specified modification, even when the queried
scalar is insensitive to it \\
Indexing idioms &
Compensating unit$/$scaling errors \\
Mathematically identical aggregations &
Hard-coded outputs \\
Superset analyses with correct sub-results &
A different short-circuit fault case \\
\bottomrule
\end{tabular}
\end{table}

\emph{Worked example.}  A generation answers a DC power-flow query
with \texttt{runpp(method='dc')}.  The oracle shows this keyword is
silently absorbed (the call is bit-identical to plain AC
\texttt{runpp} and differs from \texttt{rundcpp}), so the executed
physics is AC where the query constrains the model to DC.  Check (2)
fails, the equivalence exception's clause (a) is not met (the query
\emph{does} constrain the method), and the side is ruled
contaminated even when the printed scalar lands inside the matching
tolerance.

A $20$-item calibration pilot ($\kappa=0.80$) yielded seven
rule amendments (full statements in \texttt{audit/}).  Two of them cover
underdetermined items: R2 (the only deviation is random-number
consumption shape on a dimension the query does not pin down, and
every faithful reading lands within tolerance $\rightarrow$ item
defect) and R3 (invariance oracle proves the queried scalar
unchanged $\rightarrow$ equivalent; a faithful legal reading
$\rightarrow$ item defect; no legal reading produces the accepted
value $\rightarrow$ contaminated).  R5 fixes the refuter's three
attack heuristics: internal-consistency checks within the accepted
program, dialect checks against reference-corpus API usage, and
no-modification oracle contrasts.

\paragraph{Review chain}
Every flagged record was labelled independently by two practising
power-systems engineers under the decision calculus.  The reported
$\kappa=0.8817$ applies to the $332$ flagged records, a subset enriched
for difficult cases; it is not an agreement estimate for the full
$610$-record population.

Every surviving contamination ruling was challenged under the three R5
tests, and every reversal reduced the contamination count.  The $278$
unflagged records were censused rather than assumed clean.

Both arms within each paired sampling stage shared one instrument, so
mid-chain instrument fixes do not affect within-stage contrasts,
although automatic-check distributions should not be compared cell by
cell across rounds.

Disagreements were resolved under the decision calculus using
deterministic re-execution in a pinned environment whenever the
disputed claim was executable.

The same chain ran on all three rounds:

\begin{table}[!ht]
\centering
\footnotesize
\setlength{\tabcolsep}{4pt}
\begin{tabularx}{\linewidth}{@{}>{\raggedright\arraybackslash}p{0.31\linewidth}>{\raggedright\arraybackslash}p{0.21\linewidth}>{\raggedright\arraybackslash}p{0.20\linewidth}>{\raggedright\arraybackslash}X@{}}
\toprule
Chain stage &
\makecell[l]{$610$-item round\\($610$ items)} &
\makecell[l]{Paired base\\($232$ sides)} &
\makecell[l]{T4 extension\\($98$ sides)} \\
\midrule
Flagged for review & $332$ & $137$ & $63$ \\
Reviewer $\kappa$ (binary) & $0.8817$ & $0.860$ & $1.000$ \\
Reviewer $\kappa$ (four class) & --- & $0.868$ & $0.955$ \\
Adjudicated disagreements & $19$ & $12$ & $2$ \\
Refutation upheld$/$overturned$/$defect &
$129/7/2$ of $138$ flagged; $14/0/0$ of $14$ census & $56/0/0$ of $56$ & $19/0/0$ of $19$ \\
Census additions &
$+14$ contaminated, $+8$ equivalent, $+5$ item defect ($27$
reclassifications) &
$+1$ cont.\ (\emph{C+FDRS} side) &
$+1$ cont.\ (\emph{A+FX} side) \\
\bottomrule
\end{tabularx}

\smallskip
\parbox{0.96\linewidth}{\footnotesize\raggedright The four classes are contaminated, equivalent, item
defect, and unresolved; this $\kappa$ describes only the paired review
stages.  Adjudicated disagreements are counted four-class in the
paired base and T4-extension columns and binary (contaminated versus
not) in the $610$-item column.  The $14$ contaminated census additions
are the FN cell of Table~\ref{tab:sm_gate_operating}(a).  For the
$610$-item round, the flagged-set binary table
(both contaminated $/$ Reviewer-1 only $/$ Reviewer-2 only $/$
neither) is $126/16/3/187$, giving $313/332$ agreement.  Ten paired
sides reuse rulings from that round; the other $232+98$ sides
were reviewed afresh.\par}
\end{table}

\subsection{Results}
\label{sm:validity:far}

The two audit designs give complementary answers.  In the deployed
C+FDRS frame, the design-weighted FAR is $18.51\%$; on common support,
FAR falls from $34.12\%$ under A+FX to $12.35\%$ under C+FDRS for the
same item and model.

\paragraph{Residual risk in deployed frames}
The deployed-frame residual remains material but declines with model
tier.  Table~\ref{tab:sm_far_stratified} separates the raw sample rates
from the design-weighted population estimates.

\begin{table}[!htbp]
\centering
\caption{Stratified false-acceptance rates for the $610$-item audit
  of main-text Sec.~5.7.
  T4 is the upper group: Llama-405B, Qwen3-480B, and Qwen3-Next.
  $n_{\mathrm{eff}}$ excludes item-defect rulings; ``cont.''\ counts
  final contaminated rulings surviving adversarial refutation.
  Unweighted $=$ sample rate with Wilson $95\%$ CI; weighted $=$
  design-weighted population projection with stratified-bootstrap
  $95\%$ CI ($B=2{,}000$).}
\label{tab:sm_far_stratified}
\footnotesize
\setlength{\tabcolsep}{4pt}
\begin{tabular}{lrrcc}
\toprule
Stratum & $n_{\mathrm{eff}}$ & cont. &
FAR$_{\mathrm{unw}}$ (\%) [Wilson $95\%$] &
FAR$_{\mathrm{w}}$ (\%) [boot $95\%$] \\
\midrule
main\_P (all tiers) & 375 & 84 & $22.40\,{\scriptstyle[18.47, 26.89]}$ & $18.51\,{\scriptstyle[15.13, 21.95]}$ \\
\quad T1 (sub-7B)    &  60 & 26 & $43.33\,{\scriptstyle[31.57, 55.90]}$ & $44.95\,{\scriptstyle[34.21, 55.48]}$ \\
\quad T2 (7--32B)    & 105 & 23 & $21.90\,{\scriptstyle[15.06, 30.73]}$ & $21.27\,{\scriptstyle[15.15, 27.92]}$ \\
\quad T3 (70--120B)  &  96 & 18 & $18.75\,{\scriptstyle[12.20, 27.70]}$ & $18.54\,{\scriptstyle[12.39, 24.96]}$ \\
\quad T4               & 114 & 17 & $14.91\,{\scriptstyle[9.52, 22.59]}$  & $14.66\,{\scriptstyle[9.31, 20.49]}$ \\
api\_T5              &  53 &  7 & $13.21\,{\scriptstyle[6.55, 24.84]}$  & $12.65\,{\scriptstyle[4.72, 22.90]}$ \\
C endpoint combined  & 428 & 91 & $21.26\,{\scriptstyle[17.65, 25.38]}$ & $16.69\,{\scriptstyle[13.20, 20.41]}$ \\
contrast\_S (T2--T4) & 144 & 52 & $36.11\,{\scriptstyle[28.72, 44.22]}$ & $33.51\,{\scriptstyle[25.86, 41.84]}$ \\
\quad T2             &  48 & 26 & $54.17\,{\scriptstyle[40.29, 67.42]}$ & $56.08\,{\scriptstyle[42.82, 69.05]}$ \\
\quad T3             &  47 & 15 & $31.91\,{\scriptstyle[20.40, 46.17]}$ & $35.33\,{\scriptstyle[23.00, 49.30]}$ \\
\quad T4             &  49 & 11 & $22.45\,{\scriptstyle[13.02, 35.88]}$ & $26.51\,{\scriptstyle[14.21, 39.07]}$ \\
\bottomrule
\end{tabular}

\smallskip
\parbox{0.96\linewidth}{\footnotesize Within main\_P tiers, weighted
and unweighted estimates differ by at most $0.7$pp except T1
($1.6$pp).  T1 accounts for $15.0\%$ of the sample but $2.0\%$ of its
accepted population; api\_T5 accounts for $13.0\%$ of the sample and
$32.2\%$ of the combined accepted population.  This allocation lowers
the main\_P estimate by $3.89$pp and the combined-C estimate by
$4.57$pp.  Unweighted per-model API FARs are
$7.69\%$, $8.33\%$, $13.33\%$, and $23.08\%$, with
$n_{\mathrm{eff}}=12$--$15$ each; they are qualitative only.}
\end{table}

Weighting lowers the combined-C estimate because the high-FAR lowest
tier was oversampled and the lower-FAR API frame was undersampled.

The baseline-versus-endpoint difference is a \emph{separately sampled
contrast}, contrast\_S $-$ main\_P.  All three point estimates are
positive:

\begin{center}
\footnotesize
\setlength{\tabcolsep}{4pt}
\begin{tabular}{@{}p{0.42\linewidth}p{0.26\linewidth}p{0.26\linewidth}@{}}
\toprule
Framing & Unweighted (pp) & Weighted (pp) \\
\midrule
All sampled tiers &
$13.71\,{\scriptstyle[6.47, 20.94]}$ &
$15.00\,{\scriptstyle[6.10, 24.34]}$ \\
\addlinespace[2pt]
Tier-aligned T2--T4 subset &
$17.70\,{\scriptstyle[10.58, 25.13]}$ &
$15.58\,{\scriptstyle[6.90, 25.09]}$ \\
\addlinespace[2pt]
Common-support population, directly standardised to its tier mix &
$9.28\,{\scriptstyle[-2.69, 20.61]}$ &
$13.28\,{\scriptstyle[0.26, 25.37]}$ \\
\bottomrule
\end{tabular}
\end{center}

The unweighted common-support interval includes zero.  These estimates
compare separately drawn endpoint samples and are neither paired nor
causal.

\paragraph{Paired common-support result}
\label{sm:validity:paired}

On common support, the full method has a substantially lower
false-acceptance rate for the same item and model
(Table~\ref{tab:sm_e1p_mcnemar_full}).  The primary and secondary
versions, tier breakdown, and discordant-pair power gate were fixed
before the final labels existed.

The census found one additional contaminated side in each paired
sampling stage.  The base-round miss occurred on the C+FDRS side and
was therefore counted against the intervention; the T4-extension miss
occurred on the A+FX side.

\begin{table}[!htbp]
\centering
\caption{Paired false-acceptance results on the $170$-pair
  common-support sample, with exact McNemar tests.}
\label{tab:sm_e1p_mcnemar_full}
\footnotesize
\setlength{\tabcolsep}{2pt}
\begin{tabular}{lrcccccc}
\toprule
Version & $n$ & b$/$C$/$A$/$n & FAR$_{\mathrm{C}}$ &
FAR$_{\mathrm{A}}$ & $\Delta$(A$-$C) pp [$95\%$ CI] & disc. & exact $p$ \\
\midrule
Primary            & 170 & $17/4/41/108$ & 12.35 & 34.12 & $21.76\,{\scriptstyle[14.71, 28.82]}$ & 45 & $9.3\times10^{-9}$ \\
Defect-excluded    & 158 & $17/3/38/100$ & 12.66 & 34.81 & $22.15\,{\scriptstyle[15.19, 29.75]}$ & 41 & $1.0\times10^{-8}$ \\
\midrule
\quad T2 (7--32B)   &  30 & $2/0/14/14$  &  6.67 & 53.33 & $46.67\,{\scriptstyle[30.00, 63.33]}$ & 14 & $1.2\times10^{-4}$ \\
\quad T3 (70--120B) &  40 & $3/1/12/24$  & 10.00 & 37.50 & $27.50\,{\scriptstyle[12.50, 42.50]}$ & 13 & $3.4\times10^{-3}$ \\
\quad T4               & 100 & $12/3/15/70$ & 15.00 & 27.00 & $12.00\,{\scriptstyle[4.00, 20.00]}$  & 18 & $7.5\times10^{-3}$ \\
\bottomrule
\end{tabular}

\smallskip
\parbox{0.96\linewidth}{\footnotesize\raggedright
Each pair compares C (C+FDRS) with A (A+FX) for the same item and model
after both arms produced scalar-accepted outputs.  The sample combines
the $120$-pair base draw and the pre-specified $50$-pair T4 extension.
FAR is reported in percent and $\Delta$(A$-$C) in percentage points;
``b$/$C$/$A$/$n'' counts pairs false-accepted in both arms, C only, A
only, or neither.  The primary analysis treats a side ruled
suspected-item-defect as clean; the secondary analysis excludes the
$12$ pairs touching such a side.  T4 comprises Llama-405B, Qwen3-480B,
and Qwen3-Next.  The pre-specified power gate (discordant $\geq 15$)
passes for both pooled versions and, after the extension, for T4; T2
and T3 are direction-consistent but individually under the gate.  The
T4 $p$-value shown is unadjusted and becomes approximately $0.045$
under the conservative six-look Bonferroni correction.\par}
\end{table}

The pooled reduction survives the defect-excluded analysis and has the
same direction in every tier.  The T4 extension was frozen after the
stratified interim look and before any new data were generated.  Its
point estimate remains $12.0$pp from the $n=50$ interim sample to the
merged $n=100$ sample.  The unadjusted $p=0.0075$ remains significant
at approximately $0.045$ under the conservative six-look Bonferroni
correction.

The two designs address different populations: the paired design
estimates the intervention effect on common support, the separately
sampled contrast describes FAR levels in the deployed endpoint
populations.

\subsection{Sensitivity and errata}

\paragraph{Matching-tolerance sensitivity}
\label{sm:validity:tolerance}

Tightening the scalar tolerance removes a meaningful share of false
acceptances while discarding few legitimate matches
(Table~\ref{tab:sm_gate_operating}(b)).  It is therefore a useful
first filter when a trusted comparison value exists, but the remaining
FAR shows that tolerance alone cannot validate the workflow.

\paragraph{Executed outputs rejected by scalar matching}
\label{sm:validity:unmatched}

One additional frame tests what scalar rejection might hide:
confirmed misses are rare and arise from presentation.  In a separate
stratified sample of $100$ executed-but-unmatched outputs, eight are
presentation-layer misses ($95\%$ Wilson CI $4.1$--$15.0\%$), three
are specification ambiguities, and the remaining $89$ are genuine
errors.

This frame is a mechanism diagnostic, not a population-level
correction.  It is not self-weighting and therefore supports no
adjustment to any reported accuracy or FAR estimate.

\paragraph{Item-defect errata}
\label{sm:validity:errata}

A \emph{suspected-item-defect} ruling is a judgment about a
benchmark item, not about a model: the item's text under-determines
the queried answer (R2) or admits multiple legal readings whose
ground truths straddle the tolerance (R3).  Such rulings are removed
from both numerator and denominator of every FAR ($38$ of $610$:
$25$ in main\_P, $7$ in api\_T5, $6$ in contrast\_S); in the paired
round the twelve pairs touching an item-defect side are dropped in
the secondary McNemar version.  Excluding the $31$ unique benchmark
items implicated by these record-level rulings changes the archived
primary full-suite accuracy cells by at most $0.54$ percentage points
and leaves the paired-equivalence decisions unchanged; the affected
items and this recomputation are listed in the repository's
\texttt{audit/item\_defect\_errata.json}.
When two audit rounds touch the same item, the earlier ruling is
carried rather than re-judged, and the errata channel never alters
any final model-side ruling.
\section{Deployment serving metrics: measurement protocols and full tables}
\label{sm:serving}

The serving measurements distinguish three practical operating
points.  Qwen3-Next provides an efficient high-accuracy point within
one node, dense tiers recover throughput under batching, and
validation-enabled repair trades substantially more energy for
additional accuracy.  Energy is read directly from per-GPU NVML
cumulative counters on exclusively allocated nodes.  The two full
records are in Tables~\ref{tab:sm_serving_protocol_a}
and~\ref{tab:sm_serving_latency}.

\subsection{Measurement protocols and scope}
\label{sm:serving:protocols}

\label{sm:serving:protocol_a}
Protocol A compares single-pass C with end-to-end C+FDRS.
\label{sm:serving:protocol_b}
Protocol B supplies the request-level latency and concurrency record.

\begin{table}[!htbp]
\centering
\footnotesize
\setlength{\tabcolsep}{4pt}
\begin{tabularx}{\linewidth}{@{}>{\raggedright\arraybackslash}p{0.15\linewidth}
  >{\raggedright\arraybackslash}p{0.36\linewidth}
  >{\raggedright\arraybackslash}X@{}}
\toprule
Protocol & Workload and window & Repetition and instrumentation \\
\midrule
A: offline batch &
$300$ items, stratified by difficulty and task family at seed $22$;
C covers generation only, while C+FDRS includes execution and up to
three repairs with $16$ workers. &
Three repetitions share one model load after eight uncounted warmup
requests; medians are reported, and peak allocated VRAM is the maximum
across repetitions.  NVML records only the
engine-occupied GPUs (TP $1/2/4$, or all $16$ GPUs for the anchors),
with zero counter or sampling errors. \\
\addlinespace[2pt]
B: online serving &
Five configurations, concurrency $1/4/16$; $200$ prompts of about
$1.4$k tokens and fixed $512$-token completions with
\texttt{--ignore-eos}. &
Full-pool warmup gives all levels the same prefix-cache state; client
warmup is outside the energy window.  All $15$ cells complete
$200/200$ requests, with zero counter or sampling errors. \\
\bottomrule
\end{tabularx}
\end{table}

\paragraph{Measurement caveats (as-measured provenance)}
\label{sm:serving:caveats}
Four caveats bound the measurements; each either understates
efficiency or leaves the latency result unchanged.

\begin{table}[!htbp]
\centering
\footnotesize
\renewcommand{\arraystretch}{1.00}
\setlength{\tabcolsep}{3pt}
\begin{tabularx}{\linewidth}{@{}>{\raggedright\arraybackslash}p{0.12\linewidth}>{\raggedright\arraybackslash}p{0.50\linewidth}>{\raggedright\arraybackslash}X@{}}
\toprule
Caveat & Observed check & Consequence for interpretation \\
\midrule
Single-repetition anchors &
Two confirmatory runs reproduce the anchor throughput within
$+2.8\%$; the counter window is at least as long as generation.  For
the $480$B anchor the cluster power is averaged over per-node windows
of unequal length ($154$--$180$\,s), so its printed power differs by
$4\%$ from the product of energy per item and items per second. &
Any excess window duration remains in the energy total, so energy can
only be overstated and efficiency understated. \\
\addlinespace[2pt]
Protocol-A settle &
The counter continues for about $2.7$\,s after generation. &
Small-tier throughput is conservatively understated. \\
\addlinespace[2pt]
Protocol-B client startup &
A fixed client-startup cost is included in the energy window and can
raise J$/$request by up to $1.5\times$. &
Table~\ref{tab:sm_serving_latency} gives the energy-window value and a
service-window lower bound; the true value lies between them.
Client-timed latency is unaffected. \\
\addlinespace[2pt]
Repetition dispersion &
The $70$B repetitions divide into a lower-energy first run and a
higher-energy steady-state pair. &
The median retains the higher-energy steady-state side; the full range
in Table~\ref{tab:sm_serving_protocol_a} makes that choice visible and
excludes first-run kernel-autotune effects. \\
\bottomrule
\end{tabularx}
\end{table}

\subsection{Full serving tables}
\label{sm:serving:tables}

Table~\ref{tab:sm_serving_protocol_a} is the full protocol-A record
behind main-text Table~11: per-tier hardware configuration, item
throughput, and energy at
both operating points, including the end-to-end$/$single-pass energy
ratio E(C+FDRS)$/$E(C).  Table~\ref{tab:sm_serving_latency} is the
full protocol-B concurrency gradient with time-to-first-token (TTFT)
and time-per-output-token (TPOT) columns and the per-concurrency
energy split.

\begin{table}[!htbp]
\centering
\caption{Protocol-A offline-batch results on the stratified $300$-item
  subset (median of three repetitions): hardware, single-pass C
  accuracy, throughput, and energy, and end-to-end C+FDRS performance.
  Panel~(a) reports peak allocated VRAM per GPU under vLLM's $0.9$
  GPU-memory-utilisation setting.  In panel~(b), Acc and J$/$succ.\ are medians taken per column over
  the three repetitions.}
\label{tab:sm_serving_protocol_a}
\footnotesize
\renewcommand{\arraystretch}{1.00}
\setlength{\tabcolsep}{2.5pt}
\textbf{(a) Hardware, node load, and condition C (single pass)}
\par\smallskip
\begin{tabular*}{\linewidth}{@{\extracolsep{\fill}}lrrrrrrr@{}}
\toprule
Tier & \makecell{GPUs$\times$nodes\\(TP)} &
\makecell{VRAM$/$GPU\\(GB)} &
\makecell{Acc.\\(\%)} &
\makecell{items$/$s} & \makecell{J$/$item\\{[}min--max{]}} &
\makecell{J$/$succ.} &
\makecell{Power$/$util.\\(W$/$\%)} \\
\midrule
$0.5$B & $1\times1$ (1) & 93.7 & 1.0 & 11.9 & 18.3 [18.1--20.1] & $1{,}828$ & 217 $/$ 73.6 \\
$1.5$B & $1\times1$ (1) & 94.5 & 9.3 & 25.5 & 12.2 [12.2--16.2] & 131 & 312 $/$ 65.8 \\
$7$B & $1\times1$ (1) & 95.1 & 20.3 & 11.9 & 40.3 [40.2--56.5] & 198 & 479 $/$ 85.7 \\
$8$B & $1\times1$ (1) & 95.3 & 22.0 & 9.16 & 51.3 [51.3--101] & 233 & 470 $/$ 88.5 \\
$14$B & $2\times1$ (2) & 95.4 & 34.3 & 8.27 & 105 [105--149] & 305 & 867 $/$ 91.1 \\
$32$B & $2\times1$ (2) & 97.0 & 43.0 & 3.21 & 306 [304--307] & 712 & 984 $/$ 96.4 \\
$70$B & $2\times1$ (2) & 98.3 & 39.0 & 1.26 & 813 [530--814] & $2{,}085$ & $1{,}023$ $/$ 98.5 \\
$80$B & $2\times1$ (2) & 101.8 & 46.7 & 3.17 & 238 [236--266] & 510 & 754 $/$ 95.1 \\
$120$B & $4\times1$ (4) & 96.4 & 46.7 & 2.84 & 562 [554--589] & $1{,}203$ & $1{,}610$ $/$ 96.0 \\
$405$B$^\dagger$ & $16\times4$ (16) & 100.9 & 55.0 & 0.416 & $11{,}269$ & $20{,}489$ & $4{,}720$ $/$ 91.4 \\
$480$B$^\dagger$ & $16\times4$ (16) & 100.5 & 55.0 & 0.555 & $7{,}125$ & $12{,}955$ & $4{,}119$ $/$ 78.4 \\
\bottomrule
\end{tabular*}

\par\medskip
\textbf{(b) C+FDRS (end-to-end)}
\par\smallskip
\begin{tabular*}{\linewidth}{@{\extracolsep{\fill}}lrrrr@{}}
\toprule
Tier & \makecell{Acc\\(\%)} & \makecell{J$/$item} &
\makecell{J$/$succ.\\task} & \makecell{E ratio\\vs.\ C} \\
\midrule
$0.5$B & 1.0 & 152 & $15{,}246$ & $8.3\times$ \\
$1.5$B & 10.7 & 191 & $1{,}791$ & $15.6\times$ \\
$7$B & 31.3 & 412 & $1{,}343$ & $10.2\times$ \\
$8$B & 36.0 & 772 & $2{,}132$ & $15.0\times$ \\
$14$B & 47.7 & $1{,}248$ & $2{,}600$ & $11.9\times$ \\
$32$B & 51.3 & $2{,}615$ & $5{,}110$ & $8.5\times$ \\
$70$B & 61.7 & $3{,}757$ & $6{,}093$ & $4.6\times$ \\
$80$B & 57.7 & $2{,}301$ & $4{,}134$ & $9.7\times$ \\
$120$B & 59.0 & $4{,}209$ & $7{,}193$ & $7.5\times$ \\
$405$B$^\dagger$ & --- & --- & --- & --- \\
$480$B$^\dagger$ & --- & --- & --- & --- \\
\bottomrule
\end{tabular*}

\smallskip
\parbox{\linewidth}{\footnotesize\raggedright
All C results are single-pass.  Acc is matched accuracy on this subset
and may differ from full-benchmark values through sampling.  C
J$/$item gives the repetition [min--max].  E ratio is
E(C+FDRS)$/$E(C), and node power sums the engine-occupied GPUs.  The
$405$B and $480$B rows marked $\dagger$ are single-run C anchors on
$100$ items using $16$ GPUs across four nodes; C+FDRS and Protocol B
were not run.  The $0.5$B row is a serving-only floor outside the
accuracy panel.  Exact model identities appear in
Table~\ref{tab:sm_model_identifiers}.\par}
\end{table}

\begin{table}[!htbp]
\centering
\caption{Protocol-B online serving (\texttt{vllm bench serve}), five
  configurations at concurrency $1/4/16$.  All $15$ cells completed
  $200/200$ requests with $512$-token outputs and
  \texttt{--ignore-eos}; latencies are median$/$p95.  J$/$request is
  bracketed by the energy-window and service-window bounds.
  TTFT $=$ time to first token, TPOT $=$ time per output token, ITL
  $=$ inter-token latency, e2e $=$ end-to-end request latency.}
\label{tab:sm_serving_latency}
\scriptsize
\renewcommand{\arraystretch}{1.00}
\setlength{\tabcolsep}{1.5pt}
\begin{tabular*}{\linewidth}{@{\extracolsep{\fill}}lrrrrrrrrr@{}}
\toprule
\makecell{Tier\\(TP)} & conc & \makecell{TTFT\\med $/$ p95\\(ms)} &
\makecell{TPOT$/$ITL\\med $/$ p95\\(ms)} &
\makecell{e2e\\med $/$ p95\\(s)} & req$/$s &
\makecell{out\\tok$/$s} & \makecell{J$/$req\\(window)} &
\makecell{J$/$req\\(lower\\bd.)} &
\makecell{Power (W)\\$/$ util\\(\%)} \\
\midrule
$7$B (1) & 1 & 14.7 $/$ 15.7 & 5.14 $/$ 5.23 & 2.64 $/$ 2.65 & 0.378 & 194 & $1{,}320$ & $1{,}272$ & 481 $/$ 95.6 \\
 & 4 & 24.1 $/$ 25.8 & 5.25 $/$ 5.32 & 2.70 $/$ 2.72 & 1.478 & 757 & 356 & 311 & 460 $/$ 86.8 \\
 & 16 & 33.2 $/$ 70.2 & 5.83 $/$ 6.00 & 3.01 $/$ 3.08 & 5.129 & $2{,}626$ & 117 & 78 & 400 $/$ 65.5 \\
\midrule
$14$B (2) & 1 & 17.2 $/$ 18.7 & 6.33 $/$ 6.49 & 3.25 $/$ 3.26 & 0.307 & 157 & $2{,}652$ & $2{,}580$ & 793 $/$ 96.9 \\
 & 4 & 26.2 $/$ 32.3 & 6.58 $/$ 6.75 & 3.39 $/$ 3.42 & 1.179 & 603 & 724 & 649 & 765 $/$ 88.4 \\
 & 16 & 30.7 $/$ 73.7 & 7.67 $/$ 7.93 & 3.97 $/$ 4.02 & 3.921 & $2{,}008$ & 248 & 180 & 706 $/$ 70.6 \\
\midrule
$32$B (2) & 1 & 22.2 $/$ 24.3 & 12.10 $/$ 12.15 & 6.21 $/$ 6.22 & 0.161 & 82 & $5{,}868$ & $5{,}784$ & 932 $/$ 98.3 \\
 & 4 & 37.4 $/$ 48.3 & 12.43 $/$ 12.55 & 6.39 $/$ 6.43 & 0.625 & 320 & $1{,}548$ & $1{,}458$ & 912 $/$ 93.8 \\
 & 16 & 54.5 $/$ 93.4 & 13.96 $/$ 14.31 & 7.18 $/$ 7.34 & 2.155 & $1{,}103$ & 489 & 410 & 883 $/$ 83.0 \\
\midrule
$70$B (2) & 1 & 214.1 $/$ 300.1 & 22.54 $/$ 22.62 & 11.74 $/$ 11.85 & 0.085 & 44 & $12{,}543$ & $12{,}442$ & $1{,}060$ $/$ 99.1 \\
 & 4 & 788.5 $/$ 972.5 & 22.97 $/$ 23.07 & 12.57 $/$ 12.75 & 0.318 & 163 & $3{,}357$ & $3{,}259$ & $1{,}037$ $/$ 96.8 \\
 & 16 & $1{,}674.1$ $/$ $2{,}813.7$ & 24.80 $/$ 25.17 & 15.82 $/$ 16.04 & 0.984 & 504 & $1{,}115$ & $1{,}017$ & $1{,}001$ $/$ 90.6 \\
\midrule
$80$B (2) & 1 & 181.8 $/$ 258.6 & 5.49 $/$ 5.57 & 2.99 $/$ 3.08 & 0.329 & 169 & $1{,}593$ & $1{,}542$ & 508 $/$ 94.1 \\
 & 4 & 381.7 $/$ 418.2 & 6.99 $/$ 7.13 & 3.96 $/$ 4.02 & 1.010 & 517 & 638 & 572 & 578 $/$ 87.5 \\
 & 16 & 586.6 $/$ 702.3 & 10.53 $/$ 11.20 & 6.06 $/$ 6.21 & 2.590 & $1{,}326$ & 341 & 272 & 704 $/$ 76.3 \\
\bottomrule
\end{tabular*}
\end{table}

\subsection{Accuracy--energy Pareto reading}
\label{sm:serving:pareto}

Four additional readings sharpen the deployment interpretation.

First, the two four-node anchors reach the same $55.0\%$ subset
accuracy, but the $480$B tier uses $0.63\times$ the energy and is
therefore the more efficient of the two.

Second, the measurement-subset tie between Qwen3-Next and the $120$B
model does not hold on the full benchmark.  Qwen3-Next has the higher
condition-C accuracy there ($47.20\%$ vs.\ $45.65\%$), although the
$120$B model retains $28\%$ higher total-token throughput.

Third, batching substantially improves the dense tiers.  At
concurrency~$16$, they deliver $11.6$--$13.6\times$ their
concurrency-$1$ request throughput while using roughly one tenth of
the energy per request.  The $70$B tier nevertheless reaches
$2.81$\,s p95 time to first token, so batching efficiency does not
imply comparable interactive responsiveness.

Fourth, the four-node anchors draw $4.1$--$4.7$\,kW and use
$7.1$--$11.3$\,kJ per attempt.  This places them outside the resource
range considered for routine local service despite their higher
accuracy.
\section{Cross-backend transfer at library cold start}
\label{sm:transfer}

The workflow can be instantiated for a new library, and selective
injection usually improves on a bare prompt; in this pilot, however,
its advantage over retrieval did not survive library cold start.  The
failure analysis links part of that gap to a mismatch between isolated
API recall and accessor use inside complete programs.

\subsection{Setup}

For each library, we regenerated only the backend-facing evidence and
tasks; the workflow itself remained unchanged.

\paragraph{Benchmarks and self-contained prompts}
\label{sm:transfer:bench}

The transfer prompts are self-contained and show no leakage from the
original backend.  Each prompt embeds its self-built network setup and
names the target library; a scan of all $24$ result files
(\texttt{transfer/import\_scan\_v2\_postfix.json}) finds zero \pp{}
imports across all $2{,}160$ generations.

Each backend receives a $90$-item mini-benchmark comprising
five task families with $18$ items each, split evenly between
D1\_basic and D3\_semantic.  Every item contains a natural-language
query, executable reference code, and scalar ground truth produced in
an isolated subprocess.  Admission requires convergence, physical
sanity, and sensitivity to the requested modification.  The
documentation specifications contain $76$ curated OpenDSS entries and
$67$ PyPSA entries, with every usage example execution-verified.

\paragraph{Conditions and budget-matched retrieval calibration}
\label{sm:transfer:conditions}

On each mini-benchmark, every backend--model cell is evaluated at
Round~$0$ under four conditions: bare prompting A, selective injection
C, broad dense retrieval R$_{\text{sem}}$, and the same retriever with a
backend-level budget matched to C's realised injection,
R$_{\text{semB}}$.

The nominal comparison is not cost-matched: R$_{\text{sem}}$ injects
$3{,}854$--$4{,}154$ documentation tokens per prompt, compared with
$313$--$556$ for C.  R$_{\text{semB}}$ uses one budget per backend,
calibrated to C's realised backend means of $459.5$ tokens for OpenDSS
and $507.1$ for PyPSA.  Budgets of $250$ and $270$ produce realised
injection within $3.5\%$ of these backend targets.  This is a
backend-level cost match, not an exact per-model match.

\subsection{Results: cold start and recalibration}
\label{sm:transfer:results}

Across the six backend--model cells, selective injection raises
accuracy above the bare prompt in five, but it does not reproduce the
mature-\pp{} advantage over retrieval.  In the two cells with resolved
gains, targeted OpenDSS recalibration recovers $50\%$ and $82\%$ of
the original broad-retrieval gap, indicating that part of the deficit
came from a correctable allocation error.

\paragraph{Cold-start comparison}

Broad retrieval has the higher point estimate in five of six cells.  After its
prompt budget is matched to C, two OpenDSS cells still favour
retrieval, while the other four comparisons show no detected
difference at $n=90$.  Context volume therefore does not explain the
cold-start result by itself; which evidence is selected also matters.
Table~\ref{tab:sm_transfer_cells} gives the full accuracy and execution
matrix together with the paired contrasts.

\paragraph{Targeted recalibration (C-recal)}
\label{sm:transfer:crecal}

The targeted adjustment is useful but incomplete.  It uses
$27.0$--$28.7\%$ of broad retrieval's prompt tokens.  The two local
models remain below broad retrieval, whereas the DeepSeek comparison
is unresolved at $n=90$.  PyPSA does not share this accessor
organisation and is not included in the arm.

C-recal was defined from the frozen base results before any
recalibrated output was generated.  It changes condition C in one
respect only: the name- and signature-layer deficits of demand-hit
functions receive a fixed minimum, which raises the priority of the
corresponding documentation during packing.  The table below reports
the result on the same $90$-item OpenDSS pilot.

\begin{table}[!ht]
\centering
\footnotesize
\setlength{\tabcolsep}{2pt}
\begin{tabular}{@{}p{0.20\linewidth}p{0.10\linewidth}p{0.24\linewidth}p{0.22\linewidth}p{0.19\linewidth}@{}}
\toprule
Model & \makecell[l]{C-recal\\acc.\ (\%)} &
\makecell[l]{$\Delta$(C-recal$-$C)\\(pp [$95\%$ CI])} &
\makecell[l]{$\Delta$(C-recal$-$R$_{\text{sem}}$)\\(pp [$95\%$ CI])} &
\makecell[l]{Prompt tokens\\C-recal $/$ R$_{\text{sem}}$} \\
\midrule
Qwen2.5-32B & $41.1$ &
$+25.6\,{\scriptstyle[+14.4,+36.7]}$, $p=1.2\times10^{-4}$ &
$-25.6\,{\scriptstyle[-38.9,-12.2]}$ &
$1{,}243 / 4{,}604$ \\
\addlinespace[2pt]
Llama-70B & $44.4$ &
$+7.8\,{\scriptstyle[-1.1,+16.7]}$, n.s. &
$-18.9\,{\scriptstyle[-31.1,-6.7]}$ &
$1{,}220 / 4{,}489$ \\
\addlinespace[2pt]
DeepSeek-V4-Flash & $68.9$ &
$+31.1\,{\scriptstyle[+17.8,+44.4]}$, $p=2.5\times10^{-5}$ &
$-6.7\,{\scriptstyle[-15.6,+2.2]}$, n.s. &
$1{,}315 / 4{,}580$ \\
\bottomrule
\end{tabular}
\end{table}
\begin{table}[!htbp]
\centering
\caption{Transfer pilot.  (a)~Round-$0$ accuracy (Wilson $95\%$ CI)
  and indented execution rate, all $24$ cells.  (b)~Paired
  $\Delta$(C$-$R$_{\text{semB}}$) in pp, $95\%$ item-bootstrap CIs
  ($10{,}000$ resamples of $n=90$ pairs).  A star marks a CI excluding
  zero; a contained zero means no detected difference.}
\label{tab:sm_transfer_cells}
\footnotesize
\textbf{(a)} Full $24$-cell matrix (accuracy [Wilson $95\%$ CI]; execution rate)\\[2pt]
\setlength{\tabcolsep}{3pt}
\begin{tabular}{llcccc}
\toprule
Backend & Model & A & R$_{\text{sem}}$ & R$_{\text{semB}}$ & C \\
\midrule
\multirow{6}{*}{OpenDSS}
 & Qwen2.5-32B & $17.8\,{\scriptstyle[11.2,26.9]}$ & $66.7\,{\scriptstyle[56.4,75.5]}$ & $41.1\,{\scriptstyle[31.5,51.4]}$ & $15.6\,{\scriptstyle[9.5,24.4]}$ \\
 & \quad\emph{exec.} & $33.3$ & $78.9$ & $65.6$ & $34.4$ \\
 & Llama-70B & $3.3\,{\scriptstyle[1.1,9.3]}$ & $63.3\,{\scriptstyle[53.0,72.5]}$ & $38.9\,{\scriptstyle[29.5,49.2]}$ & $36.7\,{\scriptstyle[27.5,47.0]}$ \\
 & \quad\emph{exec.} & $13.3$ & $84.4$ & $64.4$ & $78.9$ \\
 & DeepSeek-V4-Flash & $6.7\,{\scriptstyle[3.1,13.8]}$ & $75.6\,{\scriptstyle[65.8,83.3]}$ & $60.0\,{\scriptstyle[49.7,69.5]}$ & $37.8\,{\scriptstyle[28.5,48.1]}$ \\
 & \quad\emph{exec.} & $28.9$ & $84.4$ & $77.8$ & $51.1$ \\
\midrule
\multirow{6}{*}{PyPSA}
 & Qwen2.5-32B & $40.0\,{\scriptstyle[30.5,50.3]}$ & $54.4\,{\scriptstyle[44.2,64.3]}$ & $53.3\,{\scriptstyle[43.1,63.3]}$ & $57.8\,{\scriptstyle[47.5,67.5]}$ \\
 & \quad\emph{exec.} & $45.6$ & $88.9$ & $74.4$ & $75.6$ \\
 & Llama-70B & $5.6\,{\scriptstyle[2.4,12.3]}$ & $60.0\,{\scriptstyle[49.7,69.5]}$ & $51.1\,{\scriptstyle[40.9,61.2]}$ & $50.0\,{\scriptstyle[39.9,60.1]}$ \\
 & \quad\emph{exec.} & $31.1$ & $81.1$ & $74.4$ & $72.2$ \\
 & DeepSeek-V4-Flash & $68.9\,{\scriptstyle[58.7,77.5]}$ & $94.4\,{\scriptstyle[87.6,97.6]}$ & $83.3\,{\scriptstyle[74.3,89.6]}$ & $81.1\,{\scriptstyle[71.8,87.9]}$ \\
 & \quad\emph{exec.} & $71.1$ & $97.8$ & $91.1$ & $97.8$ \\
\bottomrule
\end{tabular}
\\[6pt]
\textbf{(b)} Budget-matched paired item-level accuracy difference (pp [$95\%$ bootstrap CI])\\[2pt]
\setlength{\tabcolsep}{4pt}
\begin{tabular}{llc}
\toprule
Backend & Model & $\Delta$(C$-$R$_{\text{semB}}$) \\
\midrule
\multirow{3}{*}{OpenDSS}
 & Qwen2.5-32B & $-25.6\,{\scriptstyle[-36.7,-14.4]}^{*}$ \\
 & Llama-70B & $-2.2\,{\scriptstyle[-13.3,+8.9]}$ \\
 & DeepSeek-V4-Flash & $-22.2\,{\scriptstyle[-37.8,-6.7]}^{*}$ \\
\midrule
\multirow{3}{*}{PyPSA}
 & Qwen2.5-32B & $+4.4\,{\scriptstyle[-7.8,+16.7]}$ \\
 & Llama-70B & $-1.1\,{\scriptstyle[-13.3,+10.0]}$ \\
 & DeepSeek-V4-Flash & $-2.2\,{\scriptstyle[-13.3,+8.9]}$ \\
\bottomrule
\end{tabular}
\end{table}

\subsection{Mechanism and migration notes}

The failed OpenDSS cells identify a specific portability problem: an
isolated probe can recognise an accessor without showing that the
model can select and reproduce it inside a composed program.  This
distinction explains the recalibration result.

\paragraph{Accessor selection in composed programs}
\label{sm:transfer:mechanism}

\begin{table}[!ht]
\centering
\caption{OpenDSS control-device results for the two local models:
  $76$-entry mean probe profiles, realised injected tokens per prompt,
  and non-execution among $18$ items per condition.}
\label{tab:sm_opendss_control}
\footnotesize
\setlength{\tabcolsep}{6pt}
\begin{tabular}{lcc}
\toprule
 & Qwen2.5-32B & Llama-70B \\
\midrule
L0 recognition                    & $0.618$ & $0.447$ \\
L1 path correctness               & $0.947$ & $0.197$ \\
L1 required-parameter Jaccard     & $0.526$ & $0.145$ \\
L1 composite recall               & $0.737$ & $0.171$ \\
L2 comprehension                  & $0.969$ & $0.978$ \\
L3 application                    & $0.408$ & $0.079$ \\
\midrule
Injected tokens, C                & $393$   & $526$ \\
Injected tokens, R$_{\text{sem}}$ & $3{,}969$ & $3{,}854$ \\
\midrule
Control-device non-executable, A                & $15/18$ & $13/18$ \\
Control-device non-executable, C                & $17/18$ & $2/18$ \\
Control-device non-executable, R$_{\text{sem}}$ & $6/18$  & $1/18$ \\
\bottomrule
\end{tabular}

\smallskip
\parbox{0.96\linewidth}{\footnotesize\raggedright
Profile values are means per function.  L1 composite equally weights
path correctness and required-parameter Jaccard; injection rows report
realised mean tokens per prompt.  For Qwen2.5-32B, $14/15$ condition-A,
$17/17$ condition-C, and $6/6$ R$_{\text{sem}}$ non-executions were
name-level failures.\par}
\end{table}

Table~\ref{tab:sm_opendss_control} shows the reversal.  Qwen2.5-32B
has the stronger isolated L1 profile, so the gate supplies less
documentation, yet non-execution rises under C.  Llama-70B has the
weaker profile, receives more documentation, and shows lower
non-execution under C.

The failure patterns point to missing name- and signature-level
evidence.  Qwen2.5-32B uses
near-miss names such as \texttt{Tap} for \texttt{TapNumber}, and
places calls under the wrong interface, such as \texttt{Bus.kvar}
where \texttt{Circuit.TotalPower} is required for source reactive
power.  These are name- and signature-level errors, the same layers
that isolated probes marked as known.  The profile can therefore
overestimate the model's ability to choose and reproduce an accessor
path inside a composed program, causing the allocator to withhold
useful evidence.

The zero-shot demand ranker may also contribute to the mismatch.
APIs with hierarchical getter$/$setter bindings or weakly informative
signatures may require backend-specific calibration in addition to
regenerating the backend-facing artifacts.

\paragraph{Backend migration scope}
\label{sm:transfer:ledger}
Onboarding a new library requires regenerating only the artifacts that
encode that library's documentation and API conventions; the code that
consumes those artifacts runs unchanged.
The target library must also be installed in the recorded execution
environment, here \texttt{opendssdirect} $0.9.4$ and \texttt{pypsa}
$1.2.3$.
\begin{center}
\enlargethispage{12pt}
\footnotesize
\renewcommand{\arraystretch}{0.86}
\setlength{\tabcolsep}{3pt}
\begin{tabularx}{\linewidth}{@{}>{\raggedright\arraybackslash}p{0.23\linewidth}>{\raggedright\arraybackslash}X>{\raggedright\arraybackslash}p{0.15\linewidth}@{}}
\toprule
Component & What it encodes & Onboarding action \\
\midrule
\multicolumn{3}{@{}l}{\emph{Framework code (library-independent)}} \\
Probe templates & L0--L3 question forms generated from any API specification & Reused \\
Risk aggregation & Eq.~(1): per-function profile from layer scores & Reused \\
Demand-ranking formulation & Query-to-function ranking (run zero-shot on a new library) & Reused \\
Injection scoring and packing & Eq.~(2) and the token-budget packer & Reused \\
Exception routing & Mapping of runtime failures to repair routes & Reused \\
Intervention policy and benchmark chain & Execution, validation, and scoring harness & Reused \\
\addlinespace
\multicolumn{3}{@{}l}{\emph{Documentation-derived artifacts (library-specific)}} \\
API specification and layer renderings & Library entries rendered at L0--L3 with executable examples & Regenerated \\
Role assignments & Which functions load, modify, or analyse networks & Regenerated \\
Boundary cards & Result-table and naming conventions & Regenerated \\
Exception classes & Error types triggering each repair route & Regenerated \\
Result-table and function-output contracts & Where each analysis routine writes its outputs & Regenerated \\
Benchmark feasibility filters & What counts as an admissible task item & Regenerated \\
\bottomrule
\end{tabularx}
\end{center}

\clearpage
\bibliographystyle{elsarticle-num}
\bibliography{references}
\end{document}